\documentclass[aps,prb,twocolumn,superscriptaddress,showpacs,floatfix]{revtex4-1}
\usepackage{graphicx}
\usepackage{bm}
\usepackage{stmaryrd}
\usepackage{latexsym}
\usepackage{amssymb}
\usepackage{amsfonts}
\usepackage{amsmath}
\usepackage{color}
\usepackage{xspace}

\newcommand{\ket}[1]{\ensuremath{|#1\rangle}\xspace}
\newcommand{\bra}[1]{\ensuremath{\langle #1|}\xspace}

\newcommand{\phd}{{\phantom{\dagger}}}

\renewcommand\Im{\operatorname{Im}}

\newcommand{\namedfootnote}[2]{\footnote{#2}\global\edef#1{Note\thefootnote}}

\begin{document}
\title{Statics and dynamics of weakly coupled antiferromagnetic spin-$1/2$ ladders in a magnetic field}
\date{\today}

\author{Pierre Bouillot}
\affiliation{DPMC-MaNEP, University of Geneva, CH-1211 Geneva, Switzerland}
\author{Corinna Kollath}
\affiliation{D\'epartement de Physique
Th\'eorique, University of Geneva, CH-1211 Geneva, Switzerland}
\affiliation{Centre de Physique Th\'eorique, Ecole Polytechnique, CNRS, 91128
 Palaiseau Cedex, France}
\author{Andreas M. L\"auchli}
\affiliation{Max Planck Institut f\"ur Physik komplexer Systeme, D-01187 Dresden, Germany}
\author{Mikhail Zvonarev}
\affiliation{Department of Physics, Harvard University, Cambridge, MA 02138, USA}
\affiliation{LPTMS, CNRS and Universite Paris-Sud, UMR8626, Bat. 100, 91405 Orsay, France}
\author{Benedikt Thielemann}
\affiliation{Laboratory for Neutron Scattering, ETH Zurich and Paul Scherrer Institute, CH-5232 Villigen, Switzerland}
\author{Christian R\"uegg}
\affiliation{London Centre for Nanotechnology and Department of Physics and Astronomy, University College London, London WC1E 6BT, United Kingdom}
\author{Edmond Orignac}
\affiliation{LPENSL, UMR 5672, CNRS, F-69364 Lyon Cedex 07, France}
\author{Roberta Citro}
\affiliation{Dipartimento di Fisica ``E. R. Caianiello'' and Institute CNR-SPIN, Universit\`a degli Studi di Salerno, Via Ponte don Melillo, Fisciano, I-84084 Salerno, Italy}
\author{Martin Klanj\v{s}ek}
\affiliation{Jo\v{z}ef Stefan Institute, Jamova 39, 1000 Ljubljana, Slovenia}
\affiliation{Laboratoire national des Champs Magn\'etiques Intenses, CNRS (UPR3228)
Universit\'e J. Fourier Grenoble I, Universit\'e de Toulouse, UPS, INSA, F-38042 Grenoble Cedex 09, France}
\author{Claude Berthier}
\affiliation{Laboratoire national des Champs Magn\'etiques Intenses, CNRS (UPR3228)
Universit\'e J. Fourier Grenoble I, Universit\'e de Toulouse, UPS, INSA, F-38042 Grenoble Cedex 09, France}
\author{Mladen Horvati\'{c}}
\affiliation{Laboratoire national des Champs Magn\'etiques Intenses, CNRS (UPR3228)
Universit\'e J. Fourier Grenoble I, Universit\'e de Toulouse, UPS, INSA, F-38042 Grenoble Cedex 09, France}
\author{Thierry Giamarchi}
\affiliation{DPMC-MaNEP, University of Geneva, CH-1211 Geneva, Switzerland}

\begin{abstract}
We investigate weakly coupled spin-$1/2$ ladders in a magnetic field. The work is motivated by recent experiments on the compound $\mathrm{(C}_5\mathrm{H}_{12}\mathrm{N)}_2\mathrm{CuBr}_4$ (BPCB). We use a combination of numerical and analytical methods, in particular the density matrix renormalization group (DMRG) technique, to explore the phase diagram and the excitation spectra of such a system. We give detailed results on the temperature dependence of the magnetization and the specific heat, and the magnetic field dependence of the nuclear magnetic resonance (NMR) relaxation rate of single ladders. For coupled ladders, treating the weak interladder coupling within a mean-field or quantum Monte Carlo approach, we compute the transition temperature of triplet condensation and its corresponding antiferromagnetic order parameter. Existing experimental measurements are discussed and compared to our theoretical results. Furthermore we compute,
using time dependent DMRG, the dynamical correlations of a single spin
ladder. Our results allow to directly describe the inelastic neutron scattering cross
section up to high energies. We focus on the evolution of the spectra with the
magnetic field and compare their behavior for different couplings. The
characteristic features of the spectra are interpreted using different
analytical approaches such as the mapping onto a spin chain, a Luttinger
liquid (LL) or onto a t-J model. For values of parameters for which such
measurements exist, we compare our results to inelastic neutron scattering experiments on the
compound BPCB and find excellent agreement. We make additional predictions for the high energy part of the spectrum that are potentially testable in future experiments.
\end{abstract}
\pacs{75.10.Jm, 75.40.Gb, 75.40.Cx, 75.30.Kz}

\maketitle

\section{Introduction}
Many fascinating magnetic properties of solids are related to quantum effects\cite{Auerbach_book_magnetism}. In particular, due to the Pauli principle, the interplay between
interactions and kinetic energy can induce a
strong antiferromagnetic spin exchange. Such exchange
leads to a remarkable dynamics for the spin
degrees of freedom. On simple structures, the antiferromagnetic exchange can stabilize an antiferromagnetic order. By variations in dimensionality and connectivity of the lattice a variety of complex phenomena can arise.

Recently, among those effects two fascinating situations in which the interaction strongly favors the formation of dimers have been explored in detail. The first situation concerns a
high dimensional system in which the
antiferromagnetic coupling can lead to a spin liquid state made of singlets along the dimers. In such a spin liquid the
application of a magnetic field leads to the creation of
triplons which are spin-$1$ excitations. The triplons which
behave essentially like itinerant bosons can condense leading
to a quantum phase transition that is in the universality class
of Bose-Einstein
condensation\cite{giamarchi_ladder_coupled,nikuni00_tlcucl3_bec,ruegg03_tlcucl3}
(BEC). Such transitions have been explored experimentally and theoretically in a
large variety of materials, belonging to different structures
and dimensionalities\cite{giamarchi_BEC_dimers_review}.
On the other hand, low dimensional systems behave quite
differently. Quantum fluctuations
are extreme, and no ordered state is usually possible. In many quasi one-dimensional systems the
ground state properties are described by
Luttinger liquid (LL) physics\cite{giamarchi_book_1d,gogolin_1dbook} that predicts a {\it quasi} long range order. The elementary excitations are spin-$1/2$ excitations (spinons). They behave essentially as interacting spinless fermions. This typical behaviour can be observed in spin ladder systems in the presence of a magnetic field. Although such systems have been studied theoretically intensively for many years in both zero\cite{dagotto_ladder_review,gopalan_2ch,Weihong_spin_ladder,Troyer_thermo_ladder,barnes_ladder,reigrotzki_ladder_field,Zheng_bound_state_ladder,sushkov_ladder_boundstates,knetter_ladder} and finite magnetic field\cite{giamarchi_ladder_coupled,mila_ladder_strongcoupling,orignac_BEC_NMR,chitra_spinchains_field,furusaki_correlations_ladder,Wang_thermo_ladder,wessel01_spinliquid_bec,normand_bond_spinladder,Maeda_spinchain_magnetization,Usami_LL_parameter,hikihara_LL_ladder_magneticfield,Tachiki_spin_ladder}, a {\it quantitive} description of the LL low energy physics remained to be performed specially for a direct comparison with experiments.

Quite recently the remarkable ladder
compound\cite{Patyal_BPCB}
$\mathrm{(C}_5\mathrm{H}_{12}\mathrm{N)}_2\mathrm{CuBr}_4$,
usually called BPCB (also known as (Hpip)$_2$CuBr$_4$), has been investigated.
The compound BPCB
has been identified to be a very good realization of weakly
coupled spin ladders. The fact that the interladder coupling is
much smaller than the intraladder coupling leads to a clear
separation of energy scales. Due to this separation the compound offers
the exciting possibility to study both the phase with Luttinger
liquid properties typical for low dimensional systems \emph{and} the
BEC condensed phase typical for high dimensions. Additionally, the
magnetic field required for the realization of different phases
lies for this compound in the experimentally reachable range.
The LL predictions have been quantitatively tested for magnetization and specific
heat\cite{Ruegg_thermo_ladder},
nuclear magnetic reasonance\cite{Klanjsek_NMR_3Dladder} (NMR) and neutron diffraction\cite{Thielemann_ND_3Dladder}
measurements. Additionally the BEC
transition and its corresponding order parameter have experimentally been
observed by NMR\cite{Klanjsek_NMR_3Dladder} and neutron diffraction
measurements\cite{Thielemann_ND_3Dladder}.

In addition, the excitations of this compound have recently been
observed by inelastic neutron
scattering\cite{Thielemann_INS_ladder,Savici_BPCB_INS} experiments (INS). These are directly related to the dynamical correlations of spin ladders in a magnetic field. These dynamical correlations have hardly been investigated so far. The direct investigation of such excitations is of high interest,
since they not only characterize well the spin
system, but the properties of the triplon/spinon excitations are
also closely related to the properties of some itinerant
bosonic/fermionic systems. Indeed using such mappings \cite{giamarchi_book_1d}
of spin systems to itinerant fermionic or bosonic systems, the quantum spin
systems can be used as quantum simulators to address some of
the issues of itinerant quantum systems.
One of their advantage compared to regular itinerant systems is the
fact that the Hamiltonian of a spin system is in general well characterized, since the spin exchange constants can be
directly measured. The exchange between the spins would
correspond to short range interactions, leading to very good
realization of some of the models of itinerant particles, for which the short range of the
interaction is usually only an approximation. In that respect
quantum spin systems play a role similar to the one of cold
atomic gases \cite{bloch_cold_atoms_optical_lattices_review},
in connection with the question of itinerant interacting systems.

In this paper, we present a detailed calculation of the properties of weakly coupled spin-$1/2$ ladders. We focus in particular on their dynamics and their low energy physics providing a detailled analysis and a {\it quantitative} description necessary for an unbiased comparison with experiments. More precisely, we explore the phase diagram of such a system computing static quantities
(magnetization, specific heat, BEC critical temperature, order
parameter) and the NMR relaxation rate, using a combination of analytic (mostly Luttinger liquid
theory and Bethe ansatz (BA)) and numerical (mostly density matrix renormalization group (DMRG) and quantum Monte Carlo (QMC))
techniques. We compare our results with the various measurements on the
compound BPCB. A short account of some of these
results in connection with measurements on BPCB was previously published in
Refs.~\onlinecite{Ruegg_thermo_ladder,Klanjsek_NMR_3Dladder,Thielemann_ND_3Dladder}.
We here extend these results and give the details on how the theoretical results were obtained.
Motivated by recent experimental measurements, we further investigate the excitation spectra and dynamical correlation functions
at high and intermediate energies, for which a theoretical description is very challenging. We show how for the low
energy part of the spectrum it is possible to use the mapping
to low energy effective theories such as the LL or to a spin chain which can be solved by Bethe ansatz techniques\cite{caux_heisenbergchaindyn,caux_bosegascorrelations}.
Such a technique does not work, however, for energies of the order of the magnetic exchange of the
system. In this manuscript we thus complement such analytical approaches by a DMRG analysis.
We use the recent real-time variant to obtain the
dynamics\cite{Vidal_time_DMRG,white_time_DMRG,daley_time_DMRG,Schollwoeck_tDMRG}
in real time and the dynamical correlation functions. The same
technique can also be used to obtain finite temperature
results\cite{Verstraete_finiteT_DMRG,Zwolak_finiteT_DMRG,White_finT}.
This allows to obtain an accurate computation of the excitation
spectra and correlation functions in the high energy regime. We use different
analytical approaches to interpret our numerical results.

The plan of the paper is as follows.
Sec.~\ref{sec:coupledladder} defines the model of weakly
coupled spin ladders. Its basic excitations and phase diagram
are introduced as well as the spin chain mapping which proves to be very helpful
for the physical interpretations. Sec.~\ref{sec:methods}
briefly recalls the different analytical (LL, BA) and
numerical (DMRG, QMC)
techniques which we used to obtain the results described in the present paper. Sec.~\ref{sec:staticproperties} gives a detailed
characterization of the phase diagram focusing on the static
properties and the NMR relaxation rate.
Sec.~\ref{sec:dynamicalcorrelation} presents the computed
dynamical correlations of a single spin ladder at different magnetic fields and couplings. The numerical calculations are compared to previous
results (link cluster expansion, spin chain mapping, weak
coupling approach) and analytical descriptions (LL, t-J model).
Sec.~\ref{sec:experimental} directly compares the computed
quantities to experimental measurements. In particular the theoretical spectra
are compared to the low energy INS measurements on the compound BPCB. It also provides predictions for the high energy
part of the INS cross section. Finally,
Sec.~\ref{sec:conclusions} summarizes our conclusions and discusses
further perspectives.

\section{Coupled spin-$1/2$ ladders}\label{sec:coupledladder}

In this section we introduce the theoretical model of weakly coupled spin-$1/2$ ladders in a magnetic field. We recall its low-temperature phase diagram, paying
special attention to the regime of strong coupling along
the rungs of the ladder. This regime is particularly interesting
since it is realized in the spin-ladder compound
$\mathrm{(C}_5\mathrm{H}_{12}\mathrm{N)}_2\mathrm{CuBr}_4$,
customarily called BPCB. We discuss briefly the energy scales for BPCB
in the present section, leaving more detailled discussions
for Sec.~\ref{sec:experimental}.

\subsection{Model}\label{sec:model}

The Hamiltonian we consider is
\begin{equation}\label{equ:coupledladdershamiltonian}
H_{\textrm{3D}}=\sum_\mu H_{\mu}+J'\sum\mathbf{S}_{l,k,\mu}\cdot\mathbf{S}_{l',k',\mu'}.
\end{equation}
Here $H_\mu$ is the Hamiltonian of the single ladder $\mu$ and $J^\prime$ is the strength of the interladder coupling. The operator $\mathbf{S}_{l,k,\mu}=(S_{l,k,\mu}^x,S_{l,k,\mu}^y,S_{l,k,\mu}^z)$ acts at the site $l$ ($l=1,2,\ldots,L$) of the leg $k$ ($k=1,2$) of the ladder $\mu$. Often we will omit ladder indices from the subscripts of the operators (in particular, replace $\mathbf{S}_{l,k,\mu}$ with $\mathbf{S}_{l,k}$) to lighten notation. $S^\alpha_{l,k}$ ($\alpha=x,y,z$) are conventional spin-$1/2$ operators with $[S^x_{l,k},S^y_{l,k}]=iS^z_{l,k}$, and $S^\pm_{l,k}=S^x_{l,k}\pm iS^y_{l,k}$.

\begin{figure}
\begin{center}
\includegraphics[width=0.58\linewidth]{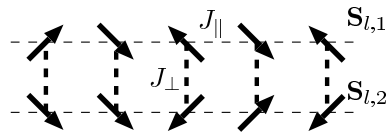}
\end{center}
\caption{Single ladder structure: $J_ \perp$ ($J_ \parallel$) is the coupling along the rungs (legs) represented by thick (thin) dashed lines and ${\bf S}_{l,k}$ are the spin operators acting on the site  $l$ of the leg $k=1,2$.
\label{fig:singleladder}}
\end{figure}

The Hamiltonian $H_\mu$  of the spin-$1/2$ two-leg ladder illustrated in Fig.~\ref{fig:singleladder} is
\begin{equation} \label{equ:spinladderhamiltonian}
H_\mu= J_\perp H_\perp + J_\parallel H_\parallel
\end{equation}
where $J_{\perp}$ ($J_{\parallel}$) is the coupling constant along the rungs (legs) and
\begin{gather}
H_\perp=\sum_{l}\mathbf{S}_{l,1}\cdot\mathbf{S}_{l,2}-h^z J_\perp^{-1} M^z \label{equ:Hperp}\\
H_\parallel=\sum_{l,k} \mathbf{S}_{l,k}\cdot\mathbf{S}_{l+1,k} \label{equ:Hparallel}
\end{gather}
The magnetic field, $h^z,$ is applied in the $z$ direction, and $M^z$ is the $z$-component of the total spin operator $\mathbf{M}=\sum_{l}(\mathbf{S}_{l,1} +\mathbf{S}_{l,2})$. Since $H_\mu$ has the symmetry $h^z\rightarrow-h^z$, $M^z\rightarrow-M^z$, we only consider $h^z\ge0$. The relation between $h^z$ and the physical magnetic field in experimental units is given in Eq.~\eqref{equ:experimentalhz}.

\subsection{Energy scales}
In the present paper we focus on the case of spin-$1/2$ antiferromagnetic
ladders weakly coupled to one another.
This means that the interladder coupling $J^\prime>0$ is much smaller than the intraladder couplings  $J_\parallel$ and $J_\perp$, i.e.
\begin{equation}
0<J^\prime \ll J_\parallel \text{ and } J_\perp. \label{equ:Jprimecond}
\end{equation}
As we will show, the model~\eqref{equ:coupledladdershamiltonian} accurately
describes the magnetic properties of the compound BPCB. Its detailed description is given in Sec.~\ref{sec:experimental}. The couplings have been experimentally determined to be\cite{Klanjsek_NMR_3Dladder,Thielemann_ND_3Dladder}
\begin{equation}\label{equ:0jprime}
 J'\approx20-100~\text{mK}
\end{equation}
and\cite{Klanjsek_NMR_3Dladder}
\begin{equation}\label{equ:0couplings}
J_{\parallel}\approx3.55~\mathrm{K},\quad J_\perp\approx12.6~\mathrm{K}.
\end{equation}
More details about the determination of the couplings for the compound BPCB are given in Sec.~\ref{sec:experimental}.

\subsection{Spin ladder to spin chain mapping} \label{sec:spinchainmap}

\begin{figure}
\begin{center}
\includegraphics[width=\linewidth]{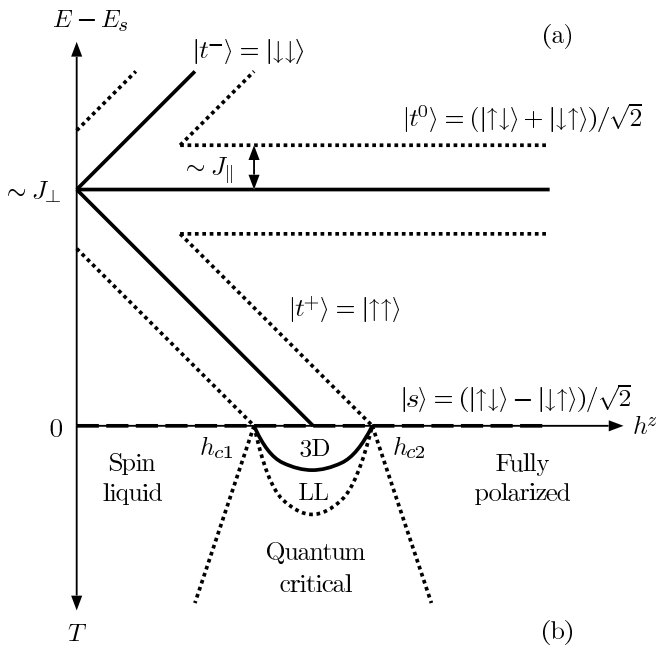}
\end{center}
\caption{(a) Energy of the triplets $|t^+\rangle$, $|t^0\rangle$, $|t^-\rangle$ (solid lines) and singlet $|s\rangle$ (dashed line)
versus the applied magnetic field in the absence of an interrung coupling ($J_\parallel=0$). The dotted lines represent the limits of the triplets
excitation band when $J_\parallel\neq0$. (b) Phase diagram of weakly coupled spin ladders: crossovers (dotted lines)
and phase transition (solid line) that only exists in the presence of an interladder coupling are sketched.\label{fig:phasediagramm}}
\end{figure}

The physical properties of a single ladder \eqref{equ:spinladderhamiltonian} are defined by the value of the dimensionless coupling
\begin{equation}\label{equ:couplingratio}
\gamma=\frac{J_\parallel}{J_\perp}.
\end{equation}
In the limit $J_\parallel=0$ (therefore $\gamma=0$) the rungs of the ladder are decoupled. We denote this \textit{decoupled bond limit} hereafter. The four eigenstates of each decoupled rung are: the
singlet state
\begin{equation}
\ket{s}=\frac{\vert{\uparrow \downarrow}\rangle-\vert{\downarrow \uparrow}\rangle}{\sqrt{2}}
\label{equ:smult}
\end{equation}
with the energy $E_s=-3J_\perp/4,$ spin $S=0,$ and $z$-projection of the spin $S^z=0$, and three triplet states
\begin{equation}
\ket{t^+}=\vert{\uparrow\uparrow}\rangle, \quad
\ket{t^0}=\frac{\vert{\uparrow \downarrow}\rangle+\vert{\downarrow
\uparrow}\rangle}{\sqrt{2}}, \quad
\ket{t^-}=\vert{\downarrow\downarrow}\rangle
\label{equ:tmult}
\end{equation}
with $S=1,$ $S^z=1,0,-1$, and energies $E_{t^+}=J_\perp/4-h^z$,
$E_{t^0}=J_\perp/4$, $E_{t^-}=J_\perp/4+h^z$, respectively. The ground state is $\ket s$ below the critical value of the magnetic field, $h_c^{\rm DBL}= J_\perp$, and $\ket{t_+}$ above. The dependence of the energies on the magnetic field is shown in Fig.~\ref{fig:phasediagramm}.a.

A small but finite $\gamma>0$ delocalizes triplets and creates bands of excitations with a bandwidth $\sim J_\parallel$ for each triplet branch. This leads to three distinct phases in the ladder system~\eqref{equ:spinladderhamiltonian} depending on the magnetic field:
\begin{itemize}
\item[(i)] {\it Spin liquid phase}\footnote{This phase is also called {\it quantum disordered} \cite{giamarchi_BEC_dimers_review}}, which is characterized by a spin-singlet
ground state (see Sec.~\ref{sec:criticalfields}) and a gapped excitation spectrum (see Sec.~\ref{sec:spinliquidexcitations}). This phase appears for magnetic fields ranging from $0$ to $h_{c_1}.$

\item[(ii)] {\it Gapless phase,} which is characterized by a gapless excitation spectrum. It occurs between the critical fields $h_{c_1}$ and $h_{c_2}$. The ground state magnetization per rung, $m^z=\langle M^z\rangle/L$, increases from $0$ to $1$ for $h^z$ running from $h_{c_1}$ to $h_{c_2}$. The low energy physics can be described by the LL theory (see Sec.~\ref{sec:luttinger_liquid}).

\item[(iii)] {\it Fully polarized phase,} which is characterized by the fully polarized ground state and a gapped excitation spectrum. This phase appears above $h_{c2}$.
\end{itemize}

Besides ladders the transition between (i) and (ii) can occur in several other gapped systems such as Haldane $S=1$ chains or frustrated chains\cite{schu_spins,affleck_field,sachdev_qaf_magfield,chitra_spinchains_field}. In the gapless phase, the distance between the ground state and the bands $\ket{t^0}$ and $\ket{t^-},$ which is of the order of $ J_\perp$, is much larger than the width of the band $\ket{t^+} \sim J_\parallel$, since $\gamma \ll 1$. 

For small $\gamma$ the ladder problem can be reduced to a simpler spin chain problem. The essence of the {\it spin chain mapping}\cite{tachiki_laddermapping,chaboussant_mapping,mila_ladder_strongcoupling,giamarchi_ladder_coupled} is to project out $\ket{t^0}$ and $\ket{t^-}$ bands from the Hilbert space of the model~\eqref{equ:spinladderhamiltonian}. The remaining states $\ket{s}$ and $\ket{t^+}$ are identified with the spin states
\begin{equation}
|\tilde\downarrow\rangle =|s\rangle,  \quad |\tilde\uparrow\rangle = |t^+\rangle. \label{equ:Hsreduced}
\end{equation}
The local spin operators ${\bf S}_{l,k}$ can therefore be identified in the reduced Hilbert space spanned by the states \eqref{equ:Hsreduced} with the new effective spin-$1/2$ operators $\tilde {\bf S}_{l}$:
\begin{equation}\label{equ:spinchainmaping}
\begin{array}{lll}
S^\pm_{l,k} = \frac{(-1)^k}{\sqrt{2}}\tilde S^\pm_l, \quad S^z_{l,k} = \frac{1}{4}\left(1+ 2 \tilde S^z_l\right).
\end{array}
\end{equation}
The Hamiltonian~\eqref{equ:spinladderhamiltonian} reduces to the Hamiltonian of the spin-$1/2$ XXZ Heisenberg chain
\begin{multline}\label{equ:strongcouplinghamiltonian}
H_{\text{XXZ}}=J_\parallel
\sum_{l}\left(\tilde S_l^x\tilde S_{l+1}^x+\tilde S_l^y\tilde S_{l+1}^y+\Delta\tilde S_l^z\tilde S_{l+1}^z\right)\\
-\tilde h^z\tilde M^z +L\left(-\frac{J_\perp}{4}+\frac{J_\parallel}{8}-\frac{h^z}{2}\right).
\end{multline}
Here the pseudo spin magnetization is $\tilde M^z=\sum_{l}\tilde S^z_l$, the magnetic field $\tilde
h^z=h^z-J_\perp-J_\parallel/2$ and the anisotropy parameter
\begin{equation}
\Delta=\frac{1}{2}.
\end{equation}
Note that the spin chain mapping constitutes a part of a more general strong coupling expansion of the model~\eqref{equ:spinladderhamiltonian}, as discussed in the appendix~\ref{sec:tjmodelmapping}.

For the compound BPCB the parameter $\gamma$ is rather small
\begin{equation}\label{equ:couplingratio2}
\gamma\approx\frac{1}{3.55}\approx 0.282.
\end{equation}
and the spin chain mapping~\eqref{equ:strongcouplinghamiltonian} gives the values of many observables decently well. Some important effects are, however, not captured by this approximation. Examples will be given in later sections.

\subsection{Role of weak interladder coupling} \label{sec:weakinterladderc}

Let us now turn back to the more general Hamiltonian~\eqref{equ:coupledladdershamiltonian} and discuss the role of a weak interladder coupling $J^\prime$ (couplings ordered as in Eq.~\eqref{equ:Jprimecond}). The spin liquid and fully polarized phases are almost unaffected by the presence of $J^\prime$ whenever the gap in the excitation spectrum is larger than $J^\prime$ (see, e.g., Ref.~\onlinecite{orignac_BEC_NMR} for more details). However, a new 3D antiferromagnetic order in the plane perpendicular to $h^z$ emerges in the gapless phase for $T~\lesssim J^\prime.$ The corresponding phase, called {\it 3D-ordered}, shows up at low enough temperatures $T_c$ in numerous experimental systems with reduced dimensionality and a
gapless spectrum~\cite{giamarchi_BEC_dimers_review}. For the temperature $T~\gtrsim J^\prime$ the ladders decouple from each other and the system undergoes a deconfinement transition into a Luttinger liquid regime (which will be described in Sec.~\ref{sec:luttinger_liquid}). For $T~\gtrsim J_\parallel$ the rungs decouple from each other and the system becomes a (quantum disordered) paramagnet. All the above mentioned phases are illustrated in Fig.~\ref{fig:phasediagramm}.b.

\section{Methods}\label{sec:methods}

In this section, we present the methods used to study the
ladder system and its mean-field extension to the case of
weakly coupled ladders. We first focus on the so called density
matrix renormalization group (DMRG) or matrix product state
(MPS) methods. These numerical methods allow us to investigate
dynamical correlations at zero and finite temperature. Additionally we discuss the Bethe ansatz used to obtain properties of the system after the spin chain mapping. Furthermore we introduce an analytical low energy description for the gapless phase, the Luttinger liquid theory. This theory in combination with a numerical determination of its parameters (see appendix~\ref{sec:LLappendix}) gives a quantitative description of the low energy physics. Finally, we treat the weak interladder coupling $J'$ by a mean
field approach, both analytically and numerically, and a quantum Monte Carlo (QMC) technique.

\subsection{DMRG}
A numerical method used to determine static and dynamical quantities at zero and finite temperature of a quasi one-dimensional system is the
DMRG. This method was originally introduced by S.R.~White
\cite{white_dmrg} to study static properties of one dimensional
systems. Since
usually the dimension of the total Hilbert space of a many-body
quantum system is too large to be treated exactly, the main idea of the DMRG algorithm is to describe the
important physics using a reduced effective space. This
reduced effective space is chosen optimally by using a
variational principle. The DMRG has been proven very successfully in many situations and has been generalized to compute dynamical properties of quantum systems using different approaches in frequency space \cite{Schollwock_DMRG,Hallberg_rev, Jeckelmann_rev}.
Recently the interest in this method even increased
after a successful generalization to time-dependent phenomena
and finite temperature situations
\cite{Vidal_time_DMRG,daley_time_DMRG,white_time_DMRG,Verstraete_finiteT_DMRG,Zwolak_finiteT_DMRG,White_finT,Schollwoeck_tDMRG}. The real-time calculations give an alternative route to determine dynamical properties of the system\cite{white_time_DMRG} which we use in the following.
An overview of the method, its extensions and its successful
applications to real-time and finite temperature can be found in Refs.~\onlinecite{Schollwock_DMRG,Hallberg_rev}. Further details on the method and its technical aspects are given in the appendix~\ref{sec:DMRG}.

\subsection{Bethe ansatz}

The spin-1/2 XXZ chain~\eqref{equ:strongcouplinghamiltonian} which is obtained after the spin chain mapping of the system is exactly solvable: the so-called Bethe ansatz technique gives explicit analytic expressions for its eigenfunctions and spectrum\cite{gaudin_book,korepin_book}. To convert this information into a practical recipe of calculation of the correlation functions is a highly sophisticated problem. However, a known solution to this problem (Ref.~\onlinecite{caux_heisenbergchaindyn} and references therein) incorporates involved analytics and numerics, the latter limiting the precision of the final results to about the same extent as to-date implementations of the DMRG method. Calculation of the thermodynamic properties of the spin-1/2 XXZ chain model \eqref{equ:strongcouplinghamiltonian} by the Bethe ansatz technique is a simpler, but still non-trivial task, requiring a solution of an infinite set of non-linear coupled integral equations\cite{takahashi_strings}. The solution of such equations can be only found numerically, and already in the 1970s this was done with a high precision\cite{Takahashi_XXZthermodnum72}.

Later on, an alternative to the Bethe ansatz, the quantum transfer matrix method, was used to get the thermodynamics of the XXZ chain in a magnetic field\cite{Takahashi_XXZthermodqtm91}. Within this approach the free energy of the system is expressed through the largest eigenvalue of the transfer matrix. This largest eigenvalue is given by the solution of a set of non-linear equations (Eq.~(66.a) of Ref.~\onlinecite{Takahashi_XXZthermodqtm91}) which are written in a form very suitable for solving them iteratively. In the present paper we followed this route and got the results for the specific heat for all temperatures and various magnetic fields with very high precision, see Sec.~\ref{sec:specificheat}.

\subsection{Luttinger Liquid (LL)}\label{sec:luttinger_liquid}

The Luttinger liquid Hamiltonian governs the dynamics of the free bosonic excitations with linear spectrum and can be written as\cite{giamarchi_book_1d,gogolin_1dbook}
\begin{equation}\label{equ:luttingerliquid}
H_{\textrm{LL}} = \frac{1}{2\pi}\int dx\left[uK\left(\partial_x\theta(x)\right)^2+\frac{u}{K}\left(\partial_x\phi(x) \right)^2\right],
\end{equation}
where $\phi$ and $\theta$ are canonically commuting bosonic fields, $[\phi(x),\partial_{y}\theta(y)]=i\pi\delta(x-y).$ The dimensionless parameter $K$ entering Eq.~\eqref{equ:luttingerliquid} is customarily called the Luttinger parameter, and $u$ is the propagation velocity of the bosonic excitations (velocity of sound). Many 1D interacting quantum systems belong to the Luttinger Liquid (LL) universality class: the dynamics of their low-energy excitations is governed by the Hamiltonian~\eqref{equ:luttingerliquid} and the local operators are written through the free boson fields $\theta$ and $\phi$ (the latter procedure if often called bosonization).

The spin-$1/2$ XXZ Heisenberg chain, Eq.~\eqref{equ:strongcouplinghamiltonian}, in the gapless phase is a well-known example of a model belonging to the LL universality class. Its local operators are expressed through the boson fields as follows\cite{giamarchi_book_1d}:
\begin{multline}\label{equ:luttingeroperator1}
\tilde S^\pm(x) = e^{\mp i\theta(x)}\left[\sqrt{2A_x}(-1)^x \right.\\ \left.+2\sqrt{B_x}\cos(2\phi(x)-2\pi\tilde m^z x)\right]
\end{multline}
and
\begin{multline}\label{equ:luttingeroperator2}
\tilde S^z(x)= \tilde m^z-\frac{\partial_x \phi(x)}{\pi}\\
+\sqrt{2A_z}(-1)^x\cos(2\phi(x)-2\pi \tilde m^zx).
\end{multline}
Here the continuous coordinate $x=la$ is given in units of the lattice spacing~$a,$ $\tilde m^z=\langle\tilde M^z\rangle/L$ is the magnetization per site of the spin chain, and $A_x$, $B_x$ and $A_z$ are coefficients which depend on the parameters of the model~\eqref{equ:strongcouplinghamiltonian}. How to calculate $K,$ $u,$ $A_x$, $B_x,$ and $A_z$ is described in the appendix~\ref{sec:LLparameters}.

The Hamiltonian~\eqref{equ:strongcouplinghamiltonian} is the leading term in the strong coupling expansion (the parameter $\gamma=J_\parallel/J_\perp\ll1,$ see appendix~\ref{sec:tjmodelmapping}) of the model~\eqref{equ:spinladderhamiltonian}. Local operators of the latter model are bosonized by combining Eqs.~\eqref{equ:spinchainmaping}, \eqref{equ:luttingeroperator1}, and \eqref{equ:luttingeroperator2}. The analysis of the model~\eqref{equ:spinladderhamiltonian} suggests\cite{giamarchi_ladder_coupled,furusaki_correlations_ladder, chitra_spinchains_field} that the bosonization of the local spins can be performed for {\it any} values of $J_\perp$ and $J_\parallel$ in the gapless regime. We would like to stress that even for a small $\gamma$ some parameters out of $K,$ $u,$ $A_x$, $B_x,$ and $A_z$ show significant numerical differences if calculated within the spin chain~\eqref{equ:strongcouplinghamiltonian} compared to the spin ladder~\eqref{equ:spinladderhamiltonian}. We discuss this issue in the appendix~\ref{sec:LLparameters}.

\subsection{Mean-field approximation}\label{sec:mean-field}

Up to now, we have presented methods adapted to deal with one
dimensional systems. In real compounds, an interladder coupling
is typically present. As discussed in
Sec.~\ref{sec:weakinterladderc}, in the incommensurate regime
this interladder coupling $J'$ (cf.~Eq.~\eqref{equ:coupledladdershamiltonian}) can lead to a new
three dimensional order (3D-ordered phase in
Fig.~\ref{fig:phasediagramm}.b) at temperatures of the order of
the coupling $J'$. In the case of BPCB the interladder coupling is
much smaller than the coupling inside the ladders, i.e. $J'\ll
J_\perp,J_\parallel$ (Sec.~\ref{sec:bpcb}). Therefore, unless one is extremely close
to $h_{c1}$ or $h_{c2}$ one can treat the interladder coupling
with a standard mean-field approximation. This approach
incorporates all the fluctuations inside a ladder. However, it
overestimates the effect of $J'$ by neglecting quantum
fluctuations between different ladders. Such effects can partly be
taken into account by a suitable change of the interladder
coupling\cite{Thielemann_ND_3Dladder} that will be discussed in Sec.~\ref{sec:coupledladderproperties}. Close to the critical
fields the interladder coupling $J'$ becomes larger than the
effective energy of the one dimensional system. This forces one to
consider a three dimensional approach from the start and brings
the physics of the system in the universality class of
Bose-Einstein condensation
\cite{giamarchi_ladder_coupled,giamarchi_BEC_dimers_review}.
In the following we consider that we are far enough (i.e.~by an
energy of the order of $J'$) away from the critical points so
that we can use the mean-field approximation.

Since the single ladder correlation functions along the magnetic field direction ($z$-axis) decay
faster than the staggered part of the ones in the perpendicular $xy$-plane (see Eqs.~\eqref{equ:xxcorrelationsimplify}~and~\eqref{equ:zzcorrelationsimplify}
for the LL exponent $K$ of the ladder shown in
Fig.~\ref{fig:LLparameter}), the three dimensional order will
first occur in this plane. Thus the dominant order parameter
is the $q=\pi$ staggered magnetization perpendicular to
the applied magnetic field. The mean-field decoupling of the
spin operators of neighboring ladders thus reads
\begin{align}\label{equ:mean-fieldvalue}
S^x_{l,k}&\cong -(-1)^{l+k}m^x_a\Rightarrow m^x_a=-(-1)^{l+k}\langle
S^x_{l,k}\rangle\\
S^z_{l,k}&\cong \frac{m^z}{2}-(-1)^{l+k} m^z_a
\end{align}
We have chosen the $xy$-ordering to be along the $x$-axis and
$m^z_a$ will be very small and therefore neglected.

This approximation applied on the interladder interaction part of the 3D Hamiltonian $H_{\textrm{3D}}$
(Eq.~\eqref{equ:coupledladdershamiltonian}) leads to
\begin{multline}\label{equ:mean-fieldhamiltonian}
H_{\textrm{MF}}=J_{\parallel}H_{\parallel} + J_{\perp} H_{\perp}\\
+\frac{n_cJ'm^z}{4}\sum_{l,k}S_{l,k}^z
+\frac{n_cJ'm^x_a}{2}\sum_{l,k}(-1)^{l+k}S_{l,k}^x.
\end{multline}
Here we assume that the coupling is dominated by $n_c$ neighboring ladders, where $n_c$ is the rung connectivity ($n_c=4$ for the case of BPCB, cf.~Fig. \ref{fig:structure}). This mean-field Hamiltonian
corresponds to a single ladder in a site dependent magnetic
field with a uniform component in the $z$-direction and a staggered component in the $x$-direction. The ground state wave function of the Hamiltonian must be determined fulfilling the self-consistency condition for $m^z$ and $m^x_a$ using numerical or analytical methods. This amounts to minimize the ground state energy of some variational Hamiltonian.

\subsubsection{Numerical mean-field}\label{sec:numericalmean-field}
The order parameters $m^z$ and $m^x_a$ can be computed
numerically by treating the mean-field Hamiltonian $H_{\textrm{MF}}$
self consistently with DMRG. These parameters are evaluated
recursively in the middle of the ladder (to minimize the
boundary effects) starting with $m^z=0$ and $m^x_a=0.5$. An
accuracy of $<10^{-3}$ on these quantities is quickly reached
after a few recursive iterations (typically $\sim 5$) of the DMRG keeping few hundred DMRG states and treating a system of length $L=150$. We
verified by keeping as well the alternating part of the $z$-order
parameter $m^z_a$ that this term is negligible ($<10^{-5}$).

\subsubsection{Analytical Mean-field}

Using the low energy LL description of our ladder system (see
Sec.~\ref{sec:luttinger_liquid}), it is possible to treat the
mean-field Hamiltonian $H_{\textrm{MF}}$ within the bosonization
technique. Introducing the LL operators~\eqref{equ:luttingeroperator1} and~\eqref{equ:luttingeroperator2}  in $H_{\textrm{MF}}$ \eqref{equ:mean-fieldhamiltonian} and keeping only the
most relevant terms leads to the Hamiltonian\cite{schulz_coupled_spinchains,nagaosa_quasi1D_SC}
\begin{multline}\label{equ:sinegordonhamiltonian}
H_{\textrm{SG}} = \frac{1}{2\pi}\int dx\left[uK\left(\partial_x\theta(x)\right)^2+\frac{u}{K}\left(\partial_x\phi(x) \right)^2\right]\\+\sqrt{A_x}n_cJ'm_a^x\int dx\cos(\theta(x))
\end{multline}
where we neglected the mean-field renormalization of $h^z$ in~\eqref{equ:mean-fieldhamiltonian}. This Hamiltonian differs from
the standard LL Hamiltonian $H_{\textrm{LL}}$~\eqref{equ:luttingerliquid} by a cosine term corresponding to
the $x$-staggered magnetic field in~\eqref{equ:mean-fieldhamiltonian}. It is known as the sine-Gordon Hamiltonian\cite{coleman_equivalence,luther_chaine_xyz,giamarchi_book_1d}. The expectation values of the fields can be derived from integrability\cite{lukyanov_sinegordon_correlations}. In particular $m^x_a$ can be determined self-consistently.

\subsection{Quantum Monte Carlo (QMC)}

In order to take into account the detailled coupling structure of the BPCB compound shown in Fig.~\ref{fig:structure}, which is neglected in the mean-field approximation (Sec.~\ref{sec:mean-field}), we employ a stochastic series expansion implementation of the QMC technique with directed loop updates~\cite{SSE_directed_loops} provided with the ALPS libraries~\cite{ALPSSSEpaper,ALPSpaper}. Nevertheless due to the strong anisotropy of the couplings \eqref{equ:Jprimecond}, the temparatures at which the effects of the interladder coupling $J'$ become visible are not reachable with this method. The QMC results
for the transition temperature of the 3D-ordered phase, $T_c$, presented in Ref.~\onlinecite{Thielemann_ND_3Dladder}, Sec.~\ref{sec:transitiontemperature} and appendix~\ref{sec:qmc} 
 are then computed with a $J'$ of $\sim3$ times bigger than that extracted in Ref.~\onlinecite{Klanjsek_NMR_3Dladder} and Sec.~\ref{sec:transitiontemperature} making the 3D effects numerically accessible.

\section{Static properties and NMR relaxation rate}\label{sec:staticproperties}

We begin our analysis of the different phases of the coupled spin ladder system, Fig.~\ref{fig:phasediagramm}.b, by computing
thermodynamic quantities, such as the magnetization, the rung state density and the
specific heat. In particular, we test the LL low energy prediction of the latter and evaluate the related crossover to the quantum critical regime. Furthermore we discuss the effect of the 3D interladder coupling computing the staggered magnetization in the 3D-ordered phase and its critical temperature. We finally discuss the NMR relaxation rate in the gapless regime related to the low energy dynamics. In order to compare these physical quantities
to the experiments, all of them are computed
for the BPCB parameters (see Sec.~\ref{sec:bpcb}).

\subsection{Critical fields}\label{sec:criticalfields}

\begin{figure}
\begin{center}
\includegraphics[width=0.9\linewidth]{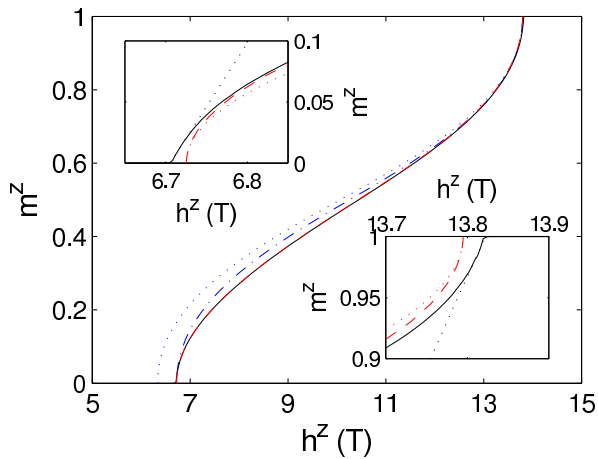}
\end{center}
\caption{(Color online): Dependence of the magnetization per rung $m^z$ on the magnetic field $h^z$ at zero temperature for the single ladder with the BPCB couplings (Sec.~\ref{sec:bpcb}) (dashed red line), the spin chain mapping (dotted blue line) rescaled to fit with the single ladder critical fields (dash-dotted blue line), and for the weakly coupled ladders treated by the mean-field approximation (solid black line). The insets emphasize the different behavior of the magnetization curves for the single (dashed red line) and coupled (solid black line) ladders close to the critical fields which are indistiguishable in the main part of the figure. The dotted lines in the insets correspond to the linear and square root like critical behaviour.
\label{fig:mz_zeroT}}
\end{figure}

The zero temperature magnetization contains extremely useful
information. Its behavior directly gives the critical
values of the magnetic fields $h_{c1}$ and $h_{c2}$ at which
the system enters and leaves the gapless regime, respectively (Fig.~\ref{fig:phasediagramm}.b). In
Fig.~\ref{fig:mz_zeroT} the dependence of the magnetization on the applied magnetic field is shown for a single
ladder and for the weakly coupled ladders. At low magnetic field,
$h^z<h_{c1}$, the system is in the gapped spin liquid regime
with zero magnetization, and spin singlets on the rungs
dominate the behavior of the system\footnote{The perturbative expression of the ground state in the spin liquid regime and the corresponding singlet and triplet densities are given in appendix~\ref{sec:tjmodelmapping}}, see Fig.~\ref{fig:tripletdensity}.
At $h^z=h_{c1}$, the
Zeemann interaction closes the spin gap to the rung triplet band
$\ket{t^+}$ (Fig.~\ref{fig:phasediagramm}).
Above $h^z>h_{c1}$ the triplet $\ket{t^+}$ band starts to be
populated leading to an increase of the magnetization with
$h^z$.  The lower
critical field in a 13th order expansion\cite{Weihong_spin_ladder} in $\gamma$ is $h_{c1} \approx 6.73~{\rm
T}$ for the BPCB parameters. At the same time the singlet and the high energy triplet occupation decreases, Fig.~\ref{fig:tripletdensity}. For $h^z>h_{c2}= J_\perp+2J_\parallel\approx13.79~{\rm
T}$ (for the compound BPCB), the $\ket{t^+}$ band is completely filled and the other bands are depopulated. The system becomes
fully polarized ($m^z=1$) and gapped. The two critical fields,
$h_{c1}$ and $h_{c2}$, are closely related to the two ladder
exchange couplings, $J_\perp$ and $J_\parallel$. As they are experimentally easily accessible, assuming that a ladder Hamiltonian is an accurate description of the system, these critical fields can be used to
determine the ladder couplings\cite{Klanjsek_NMR_3Dladder}.

Such a  general behavior of the magnetization is seen
for both the single ladder and the weakly coupled ladders in
Fig.~\ref{fig:mz_zeroT}. In particular, the effect of a small
coupling $J'$ between the ladders is completely negligible in the
central part of the curve. Only in the vicinity of the critical
fields, the single ladder and the coupled ladders show a
distinct behavior. The single ladder behaves like an empty
(filled) one-dimensional system of non-interacting fermions
which leads to a square-root behavior $m^z\propto(h^z-h_{c1})^{1/2}$ close to the lower critical field and  $1-m^z \propto(h_{c2}-h^z)^{1/2}$ close to the upper critical field. In contrast, in the system of weakly
coupled ladders, a 3D-ordered phase appears at low enough temperatures
in the gapless regime (see Sec.~\ref{sec:weakinterladderc} and~\ref{sec:mean-field}). The
magnetization dependence close to the critical fields becomes linear, $m^z\propto h^z-h_{c1}$, and $1-m^z \propto h_{c2}-h^z$, respectively\cite{giamarchi_ladder_coupled,sachdev_qaf_magfield}. In comparison with the single ladder, the critical
fields given above are shifted  by a value of the order of $J'$. This
behavior is in the universality class of the Bose-Einstein
condensation and is well reproduced by the mean-field
approximation shown in the insets of Fig.~\ref{fig:mz_zeroT}
close to the critical fields.

For comparison, the magnetization of a single ladder in the spin chain mapping is also plotted in
Fig.~\ref{fig:mz_zeroT}. This approximation reproduces well the
general behavior of the ladder magnetization discussed above.
However, note that for the exchange coupling constants considered here the
lower critical field in this approximation is different from the ladder one.
The lower critical field is
$h_{c1}^{\rm XXZ}=J_\perp-J_\parallel\approx 6.34 ~{\rm
T}<h_{c1}$. The upper critical field
$h_{c2}^{\rm XXZ}=J_\perp+2J_\parallel=h_{c2}$ is the same as for the ladder. If we rescale $h_{c1}^{\textrm{XXZ}}$ and $h_{c2}^{\textrm{XXZ}}$ to match the critical fields $h_{c1}$ and $h_{c2}$ ($\tilde h^z\rightarrow
\frac{(\tilde h^z-h_{c1}^{\textrm{XXZ}})(h_{c2}-h_{c1})}{h_{c2}^{\textrm{XXZ}}-h_{c1}^{\textrm{XXZ}}}+h_{c1}$), the magnetization curve gets very close to the one calculated for a ladder.
However, in contrast to the magnetization curve for the ladder, the corresponding curve in the spin chain mapping is symmetric with respect to its center at $h_m^{\textrm{XXZ}}=\frac{h_{c1}^{\textrm{XXZ}}+h_{c2}^{\textrm{XXZ}}}{2}=J_\perp+J_\parallel/2$ due to the absence of the high energy triplets.

\begin{figure}
\begin{center}
\includegraphics[width=0.85\linewidth]{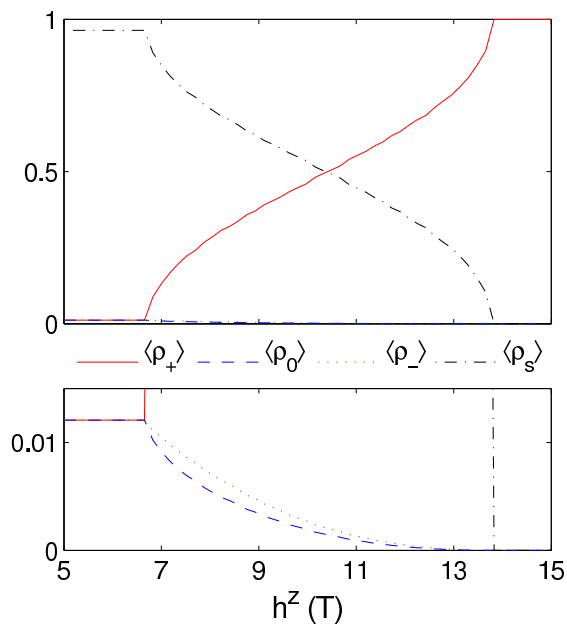}
\end{center}
\caption{(Color online): Rung state density versus the applied magnetic field $h^z$ at zero temperature for the single ladder with the BPCB couplings (zero temperature DMRG calculations averaging on the central sites of the ladder). The dash-dotted (black) lines correspond to the singlet density $\langle \rho_s\rangle$. The triplet densities are represented by the solid (red) lines for $\langle \rho_+\rangle$, the dashed (blue) lines for $\langle \rho_0 \rangle$ and the dotted (green) lines for $\langle\rho_-\rangle$.
\label{fig:tripletdensity}}
\end{figure}

\subsection{The Luttinger liquid regime and its crossover to the critical regime}

The thermodynamics of the spin-$1/2$ ladders has been studied in the
past\cite{Troyer_thermo_ladder,Wang_thermo_ladder,wessel01_spinliquid_bec,Ruegg_thermo_ladder}.
We here summarize the main interesting features of the
magnetization and the specific heat focussing on the crossover between
the LL regime and the quantum critical region using the BPCB
parameters (Sec.~\ref{sec:bpcb}).
As the interladder exchange coupling $J'$ is very small compared to the
ladder exchange couplings $J_\parallel$ and $J_\perp$, it is reasonable
to neglect it in this regime far from the
3D phase. Therefore we focus on a single ladder in the
following.

\subsubsection{Finite temperature magnetization}\label{sec:finiteTmagnetization}

We start the description of the temperature dependence of the magnetization in the two gapped regimes: the spin liquid phase and the fully polarized phase (not shown, cf.~Refs.~\onlinecite{Wang_thermo_ladder,normand_bond_spinladder}).
For small magnetic fields $h^z<h_{c1}$, the magnetization
vanishes exponentially at zero temperature and after a maximum
at intermediate temperatures it decreases to zero for large
temperatures. For large magnetic fields  $h^z>h_{c2}$ the
magnetization increases exponentially up to $m^z=1$ at
low temperature and decreases monotonously in the limit of
infinite temperature.

In the gapless regime, the magnetization at low temperature has
a non-trivial behavior that strongly depends on the applied
magnetic field. The temperature dependence of the magnetization
computed with the T-DMRG (see appendix~\ref{sec:temperatureDMRG})
is shown in Fig.~\ref{fig:Tmagnetization}.a for different
values of the magnetic field in the gapless phase, $h^z=9,10,11~{\rm
T}$ ($h_{c1}<h^z<h_{c2}$). In this regime new extrema appear in the magnetization at low temperature. This behavior can be
understood close to the critical fields where the ladder can be
described by a one-dimensional fermion model with negligible interaction between fermions. Indeed, in
this simplified picture\cite{giamarchi_book_1d} and in more
refined calculations\cite{Maeda_spinchain_magnetization,Wang_thermo_ladder,wessel01_spinliquid_bec}
the magnetization has an extremum where the temperature reaches
the chemical potential, i.e., at the temperature at which the energy of
excitations starts to feel the curvature of the energy band. This
specific behavior is illustrated in
Fig.~\ref{fig:Tmagnetization}.a with the curve for $h^z=11~{\rm T}$
($h_m=\frac{h_{c1}+h_{c2}}{2}<h^z<h_{c2}$). The low temperature maximum moves to higher temperature for $h^z<h_m$ and goes over to the already
discussed maximum for $h^z<h_{c1}$. Symmetrically with respect
to $h_m$, a low temperature minimum appears in the curve for $h^z=9~{\rm T}$
($h_{c1}<h^z<h_m$). This minimum slowly disappears for
$h^z\to h_m$ (the curve for $h^z=10~{\rm T}$ is close to that).

The location of the lowest extremum is a reasonable
criterion to characterize the crossover temperature between the
LL and the quantum critical regime\cite{Wang_thermo_ladder,wessel01_spinliquid_bec},
since the extremum occurs at temperatures of the order of the
chemical potential. A plot of this crossover temperature versus
the magnetic field is presented in
Fig.~\ref{fig:Tmagnetization}.c. Following this criterium, the
crossover has a continuous shape far from $h_m$. Nevertheless, close to $h_m$ both extrema are close to each other (since the maximum still exists for $h^z<h_m$ field at which the minimum appears). The criterium is thus not well defined. It presents a discontinuity at $h_m$ which is obviously an artefact. In the vicinity of $h_m$, we thus
use another crossover criterium based on the specific heat (see
Sec.~\ref{sec:specificheat}) that seems to give a more accurate
description. In Ref.~\onlinecite{Ruegg_thermo_ladder}, both criteria have been
applied on the magnetocaloric effect and specific heat measurements on the
compound BPCB, and give a crossover temperature in agreement with the computed ones.

The temperature dependence of the magnetization of the spin chain mapping, Fig.~\ref{fig:Tmagnetization}.b, exhibits a single low temperature
maximum if $h_m^{\textrm{XXZ}}<h^z<h_{c2}^{\textrm{XXZ}}$ (minimum
if $h_{c1}^{\textrm{XXZ}}<h^z<h_m^{\textrm{XXZ}}$). The appearance of a single extremum and its convergence to $m^z\rightarrow 0.5$ when $T\rightarrow\infty$
is due to the exact
symmetry with respect to the magnetic field $h_m^{\textrm{XXZ}}$. This approximation reproduces the main low energy features of the ladder but fails to describe the high energy behavior which strongly depends on the high energy triplets.

\begin{figure}
\begin{center}
\includegraphics[width=1\linewidth]{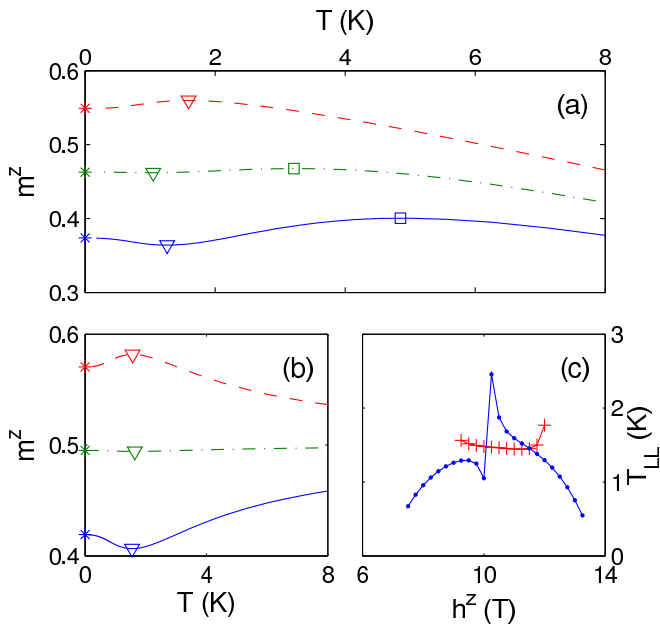}
\end{center}
\caption{(Color online): Temperature dependence of the magnetization per rung, $m^z(T)$, for (a) the ladder with the BPCB couplings~\eqref{equ:couplings}, and (b) the spin chain mapping at
  different applied magnetic fields $h^z=9~{\rm T}$ (solid blue lines), $h^z=10~{\rm T}$ (dash-dotted green lines) and $h^z=11~{\rm T}$ (dashed red lines).  The results were obtained using
  T-DMRG. The stars at $T=0~\mathrm{K}$ are the ground state magnetization per rung
  determined by zero temperature DMRG. The triangles (squares) marks the
  low (high) energy extrema. (c) Crossover temperature $T_{\rm LL}$of the LL to the quantum critical regime versus the applied magnetic field (blue circles for the extremum in $m^z(T)|_{h^z}$ criterium and red crosses for the maximum in $c(T)|_{h^z}$ criterium)\label{fig:Tmagnetization}.}
\end{figure}

\subsubsection{Specific heat}\label{sec:specificheat}

\begin{figure}
\begin{center}
\includegraphics[width=\linewidth]{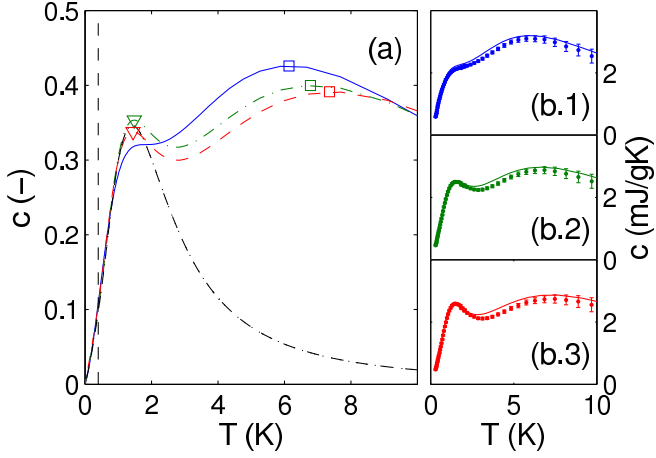}
\end{center}
\caption{(Color online): Specific heat per rung $c$ versus the temperature $T$ for different applied magnetic fields in the gapless regime. (a) Full ladder T-DMRG calculations with the BPCB couplings~\eqref{equ:couplings} for $h^z=9~{\rm T}$ (solid blue line), $h^z=10~{\rm T}$ (dash-dotted green line), and $h^z=11~{\rm T}$ (dashed red line).  Spin chain mapping at $h^z=10~{\rm T}$ solved by DMRG (dashed black line) and by Bethe ansatz (dotted black line). Note that the two lines are hardly distinguishable. The triangles (squares) mark the low (high) energy maxima of the specific heat versus temperature. The vertical dashed line marks the temperature $T=0.4~{\rm K}$ below which the DMRG results are extrapolated (see appendix \ref{sec:temperatureDMRG}). (b) Comparison between measurements on the compound BPCB from Ref.~\onlinecite{Ruegg_thermo_ladder} (dots) and the T-DMRG calculations\cite{\specificheatfactor}  (solid lines) at (b.1) $h^z=9~{\rm T}$, (b.2) $h^z=10~{\rm T}$ and (b.3) $h^z=11~{\rm T}$.\label{fig:cvsT}}
\end{figure}

The specific heat has been investigated for similar parameters as the ones considered in this paper in the
gapped and gapless regime in Refs.~\onlinecite{Wang_thermo_ladder,Ruegg_thermo_ladder,normand_bond_spinladder}.
Here we concentrate on the detailed analysis in the gapless
regime, in particular on the low temperature behavior and the
determination of a crossover temperature from the first
maximum.  We show in Fig.~\ref{fig:cvsT} the typical
temperature dependence of the specific heat for different
values of the magnetic field. Comparisons with actual
experimental data\cite{Ruegg_thermo_ladder} for BPCB are excellent (see Fig.~\ref{fig:cvsT}.b). For these comparisons the theoretical data are computed with $g=2.06$ related to the experimental orientation of the sample with respect to the magnetic field\cite{Ruegg_thermo_ladder} (see Sec.~\ref{sec:bpcb}) and rescaled by a factor $0.98$ in agreement with the global experimental uncertainties\namedfootnote{\specificheatfactor}{An additional scaling factor $7.47\ {\rm mJ/gK}$ has to be applied on the theoretical specific heat (per rung) to convert to the experimental units.}.

At low temperatures the specific heat
has a contribution due to the gapless spinon excitations. This
results in a peak around $T\sim 1.5~{\rm K}$. This peak is most
pronounced for the magnetic field values lying mid value between
the two critical fields. At higher temperatures the
contribution of the gapped triplet excitations leads to a second
peak whose position depends on the magnetic field. To separate
out the contribution from the low lying spinon excitations, we
compare the specific heat of the ladder to the results obtained
by the spin chain mapping in which we just keep the lowest two modes of the ladder (see Sec.~\ref{sec:spinchainmap} and appendix~\ref{sec:tjmodelmapping}).
The resulting effective chain model is solved using
Bethe-ansatz and
T-DMRG methods. The
agreement between these methods is excellent and the corresponding curves in Fig.~\ref{fig:cvsT} can
hardly be distinguished. However, a clear difference with the
full spin ladder results is revealed. While at low
temperatures the curves are very close, the first peak
in the spin chain mapping already lacks some weight, which stems
from higher modes of the ladder.

\begin{figure}
\begin{center}
\includegraphics[width=0.9\linewidth]{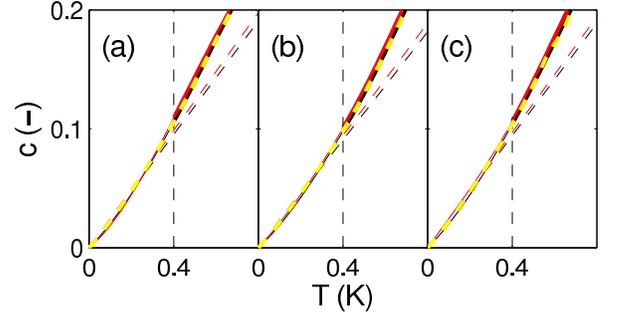}
\end{center}
\caption{(Color online): Low temperature dependence of the specific heat per rung versus the temperature, $c(T)$, for (a) $h^z=9~\mathrm{T}$, (b) $h^z=10~\mathrm{T}$, (c) $h^z=11~\mathrm{T}$. The T-DMRG calculations are in red thick lines for the ladder with the BPCB couplings~\eqref{equ:couplings} (black thick lines for the spin chain mapping). The two curves can hardly be distinguished.
Their low temperature polynomial extrapolation is plotted in thin lines below $T=0.4~{\rm K}$ (represented by a vertical dashed line). The linear low temperature behavior of the LL is represented by dashed lines (red for the ladder, black for the spin chain mapping). The dashed yellow lines correspond to the Bethe ansatz computation for the spin chain mapping.
\label{fig:cvsT_zoom}}
\end{figure}

In Fig.~\ref{fig:cvsT_zoom} the low temperature region is
analyzed in more detail. At very low temperatures the spinon
modes of the ladder can be described by the LL theory (see Sec.~\ref{sec:luttinger_liquid}) which
predicts a linear rise with temperature inversely proportional
to the spinon velocity\cite{giamarchi_book_1d,Suga_LL_spinchain}
\begin{equation}\label{equ:LLspec_heat}
c_{\textrm{LL}}(T)=\frac{T\pi}{3u}.
\end{equation}
In Fig. \ref{fig:cvsT_zoom} we compare the results of the LL,
the Bethe ansatz and the DMRG results for the effective spin chain
and the numerical DMRG results taking the full ladder into
account. The numerical results for the adaptive T-DMRG at finite
temperature are extrapolated to zero temperature by connecting
algebraically to zero
temperature DMRG results (see appendix~\ref{sec:temperatureDMRG}). A very good agreement between~\eqref{equ:LLspec_heat} and numerics is found for
low temperatures. However, at higher
temperatures, the slope of the $T\to 0$ LL
description slightly changes with respect to the curves calculated with other methods. This change of slope reflects the fact that the
curvature of the energy dispersion must be taken into account
when computing the finite temperature specific heat, and this
even when the temperature is quite small compared to the
effective energy bandwidth of the system. The effective
spin chain and the numerical results for the ladder agree 
for higher temperatures (depending on the magnetic field),
before the higher modes of the ladder cause deviations.

As for the magnetization (Sec.~\ref{sec:finiteTmagnetization}),
the location of the low temperature peak can be interpreted as the
crossover of the LL to the quantum critical regime.  Indeed, in
a free fermion description which is accurate close to the critical fields, this
peak appears at the temperature for which the excitations stem
from the bottom of the energy band. The corresponding temperature
crossover is compared in Fig.~\ref{fig:Tmagnetization}.c to the
crossover temperature extracted from the first magnetization
extremum (Sec.~\ref{sec:finiteTmagnetization}). The two
crossover criteria are complementary due to their domain of
validity. The first specific heat maximum is well pronounced
only in the center of the gapless phase. In contrast in this regime the presence of two extrema close to each other in the
magnetization renders the magnetization
criterium very imprecise (cf.~Sec.~\ref{sec:finiteTmagnetization}).

\subsection{Spin-lattice relaxation rate}\label{sec:relaxationtime}

As the spin-lattice relaxation in quantum spin systems is due to pure magnetic coupling between electronic and nuclear spins, the NMR spin-lattice relaxation rate
$T_1^{-1}$ is directly related to the local transverse correlation function\cite{slichter_NMR_book}
$\chi_a^{xx}(x=0,t)$ defined as in Eq.~\eqref{equ:xxacorrelationdef}
\begin{equation}\label{equ:T1init}
T_1^{-1}=-\frac{T\gamma_n^2A_\perp^2}{\omega_0}\Im\left(\int_{-\infty}^{\infty}dte^{i\omega_0t}\chi_a^{xx}(x=0,t)\right)
\end{equation}
where $T\gg\omega_0$ is assumed with the Larmor frequency  $\omega_0$.
$\gamma_n=19.3~\mathrm{MHz/T}$ is the nuclear gyromagnetic ratio
of the measured nucleus ($^{14}{\rm N}$ in
Ref.~\onlinecite{Klanjsek_NMR_3Dladder} for BPCB) and
$A_\perp$ is the transverse hyperfine coupling constant.

\begin{figure}
\begin{center}
\includegraphics[width=0.9\linewidth]{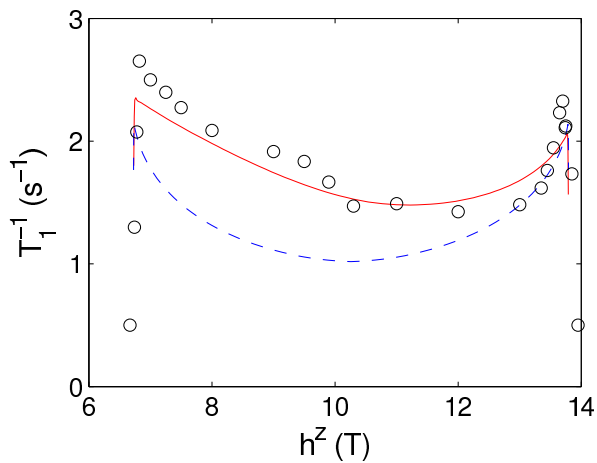}
\end{center}
\caption{(Color online): Magnetic field dependence of the NMR relaxation rate, $T_1^{-1}(h^z)$, at $T=250~\mathrm{mK}$. The solid red line is the bosonization determination using the ladder LL parameters for the BPCB couplings (the dashed blue line uses the LL parameters of the spin chain mapping). The black circles are the measurements from Ref.~\onlinecite{Klanjsek_NMR_3Dladder}.\label{fig:T1}}
\end{figure}

Assuming $J_\parallel\gg T$, the $T^{-1}_1$ due to the spin dynamics
can be computed in the gapless regime using the LL low energy description. Following Ref.~\onlinecite{giamarchi_ladder_coupled} we introduce the LL correlation~\eqref{equ:xxacorrelation} into Eq.~\eqref{equ:T1init}, and obtain
\begin{equation}\label{equ:T1}
T_1^{-1}=\frac{\gamma^2A_\perp^2A_x\cos\left(\frac{\pi }{4K}\right)}{u}\left(\frac{2\pi T}{ u}\right)^{\frac{1}{2K}-1}B\left(\frac{1}{4K},1-\frac{1}{2K}\right).
\end{equation}
According to the known LL
parameters 
(Fig.~\ref{fig:LLparameter})
the shape of
$T^{-1}_1(h_z)$ plotted in Fig. \ref{fig:T1} at
$T=250~\mathrm{mK}\gg T_c$ is strongly asymmetric with respect to the middle of the gapless phase. The only free (scaling) parameter, $A_\perp=0.057~\mathrm{T}$, is deduced
from the fit of Eq.~\eqref{equ:T1} to the experimental data\cite{Klanjsek_NMR_3Dladder} and
agrees with other measurements. For
comparison, the $T^{-1}_1$ obtained in the spin chain mapping
approximation is also plotted in  Fig.~\ref{fig:T1}. As for
other physical quantities, this description fails to
reproduce the non-symmetric shape.

\subsection{Properties of weakly coupled ladders}\label{sec:coupledladderproperties}

The interladder coupling $J'$ induces a low temperature ordered phase (the 3D-ordered phase
in Fig.~\ref{fig:phasediagramm}.b). Using the mean-field
approximation presented in Sec.~\ref{sec:mean-field} we
characterize the ordering and compute the critical temperature
and the order parameter related to this phase.

\subsubsection{3D order transition temperature}\label{sec:transitiontemperature}

\begin{figure}
\begin{center}
\includegraphics[width=0.9\linewidth]{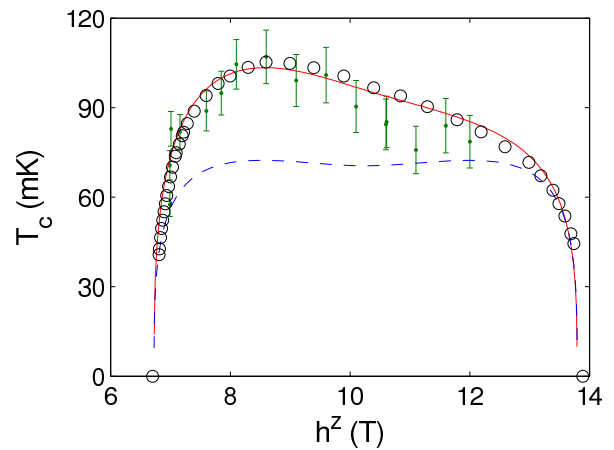}
\end{center}
\caption{(Color online): Magnetic field dependence of the transition temperature between the gapless regime and the 3D-ordered phase, $T_c(h^z)$, is plotted in solid red line for the ladder LL parameters of BPCB (in dashed blue line for the LL parameters of the spin chain mapping). The NMR measurements from Ref.~\onlinecite{Klanjsek_NMR_3Dladder} are represented by black circles and the neutron diffraction measurements from Ref.~\onlinecite{Thielemann_ND_3Dladder} by green dots.
\label{fig:criticaltemperature}}
\end{figure}

In order to compute the critical temperature of the 3D
transition, we follow
Ref.~\onlinecite{giamarchi_ladder_coupled} and treat the
staggered part of the mean-field Hamiltonian $H_{\textrm{MF}}$~\eqref{equ:mean-fieldhamiltonian} perturbatively using linear
response. The instability of the resulting mean-field
susceptibility, due to the 3D transition, appears at $T_c$
when\cite{scalapino_q1d}
\begin{equation}\label{equ:instabilitycondition}
\left.\chi_a^{xx}(q=0,\omega=0)\right|_{T_c}=-\frac{2}{n_cJ'}
\end{equation}
where $\chi_a^{xx}$ is the transverse staggered retarded correlation
function of an isolated single ladder system (appendix~\ref{sec:finiteluttingerliquidcorr}). This correlation can be computed
analytically (see
Eq.~\eqref{equ:xxacorrelationqo}) using the LL low energy description
of the isolated ladder (Eq.~\eqref{equ:luttingerliquid}) in the gapless regime. Applying the condition~\eqref{equ:instabilitycondition} to the LL correlation~\eqref{equ:xxacorrelationqo} leads to the critical temperature
\begin{equation}\label{equ:criticaltemperature}
T_c=\frac{u}{2\pi}\left(\frac{A_xJ'n_c\sin\left(\frac{\pi}{4K}\right)B^2\left(\frac{1}{8K},1-\frac{1}{4K}\right)}{2u}\right)^{\frac{2K}{4K-1}}.
\end{equation}
Introducing the computed LL parameters $u$, $K$ and $A_x$ (see
Fig.~\ref{fig:LLparameter}) in this expression, we get
the critical temperature\cite{Klanjsek_NMR_3Dladder} as a function of the magnetic field. This is
shown in Fig.~\ref{fig:criticaltemperature} together with the
experimental data. This allows us to extract the {\it mean-field} interladder
coupling $J'_{\textrm{MF}}\approx20~\mathrm{mK}$ for the experimental compound
BPCB (the \emph{only} free (scaling) parameter in Eq.~\eqref{equ:criticaltemperature}). The asymmetry of the LL
parameters induces a strong asymmetry of $T_c$ with respect
to the middle of the 3D phase.

As the mean-field approximation
neglects the quantum fluctuations between the ladders, the
critical temperature $T_c$ is overestimated for a given
$J'_{\textrm{MF}}$. We thus performed a Quantum Monte Carlo (QMC)
determination of this quantity based on the same 3D lattice structure.
Let us note that QMC  simulations of the coupled spin ladder Hamiltonian~\eqref{equ:coupledladdershamiltonian} are possible, since the 3D lattice structure, Fig.~\ref{fig:structure}, is unfrustrated. In appendix~\ref{sec:qmc} we present results on how to determine the critical temperatures for the 3D ordering transition using QMC.
This determination shows\cite{Thielemann_ND_3Dladder} that the real critical
temperature is well approximated by the mean-field
approximation, but with a rescaling of the real interladder
coupling $J'\approx27~{\rm mK}=\alpha^{-1} J'_{\textrm{MF}}$ with
$\alpha\approx0.74$.
The rescaling factor $\alpha$ is similar to the values obtained for other quasi one-dimensional antiferromagnets\cite{yasuda_neelorder,Todo_chain_MFT}.

\subsubsection{Zero temperature 3D order parameter}\label{sec:orderparameter}

\begin{figure}
\begin{center}
\includegraphics[width=0.9\linewidth]{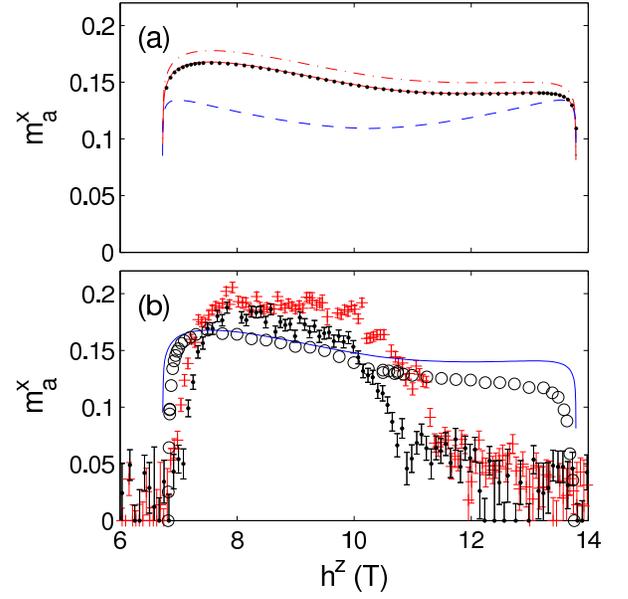}
\end{center}
\caption{(Color online): Magnetic field dependence of the staggered
  magnetization per spin, $m^x_a(h^z)$, in the 3D-ordered phase determined for
  the ground state. (a) Its computation with the analytical bosonization technique for the LL parameters of the compound BPCB and $J'=27~{\rm mK}$ is represented by the dash-dotted red line (dashed blue line for the LL parameters of
  the spin chain mapping). The DMRG result for $J'=20~{\rm
    mK}(=J'_{\textrm{MF}})$ is represented by black dots (as a comparison
  the bosonization result for $J'=20~{\rm mK}$ is plotted in solid red
  line). Note, that these two data are almost indistinguishable. 
(b) Comparison between NMR measurements (black cirles) done at $T=40~\mathrm{mK}$ from Ref.~\onlinecite{Klanjsek_NMR_3Dladder} and
  scaled on the theoretical results for $J' =27~{\rm mK}$,  neutron
  diffraction measurements on an absolute scale from Ref.~\onlinecite{Thielemann_ND_3Dladder} at
  $T=54~\mathrm{mK}$ ($T=75~\mathrm{mK}$) (red crosses (black dots)) and the computation for $J'=27~{\rm mK}$ (blue line as in panel (a)).  Recent neutron diffraction measurements as a function of temperature suggest that the data of Ref.~\onlinecite{Thielemann_ND_3Dladder} was taken at temperatures approximately $10~{\rm mK}$ higher than the nominal indicated temperature. \label{fig:orderparameter}}
\end{figure}

The staggered order parameter in the 3D-ordered phase, $m^x_a$, can  be
analytically determined at zero temperature using the mean-field approximation for the interladder coupling and the bosonization technique (see Sec.~\ref{sec:mean-field}). As
$m^x_a=\sqrt{A_x}\langle\cos(\theta(x))\rangle$ in the
bosonization description and the expectation value\cite{lukyanov_sinegordon_correlations} of the
operator $e^{i\theta(x)}$ is
\begin{equation}\label{equ:expsinegordon} \left\langle
e^{i\theta(x)}\right\rangle = F(K)\left(\frac{\pi
\sqrt{A_x} n_cJ'm^x_a}{2u}\right)^{\frac{1}{8K-1}}
\end{equation}
for the sine-Gordon Hamiltonian $H_{\textrm{SG}}$~\eqref{equ:sinegordonhamiltonian} with
\begin{equation}
F(K)=\frac{\frac{\pi^2}{\sin\left(\frac{\pi}{8K-1}\right)}
\frac{8K}{8K-1}  \left[\frac {\Gamma\left(1-\frac 1{8K}\right)}
{\Gamma\left(\frac{1}{8K}\right)}\right]^{\frac{8K}
{8K-1}}}{\left[\Gamma \left(\frac{4K}{8K-1} \right)  \Gamma
\left(\frac{16K-3}{16K-2} \right)\right]^2},
\end{equation}
we can extract
\begin{equation}\label{equ:orderparameter}
m_a^x=\sqrt{A_x}F(K)^{\frac{8K-1}{8K-2}}\left(\frac{\pi n_c
A_xJ'}{2u}\right)^{\frac{1}{8K-2}}.
\end{equation}
This can be evaluated in the 3D-ordered phase by introducing into~\eqref{equ:orderparameter} the LL parameters $u$, $K$ and $A_x$ from Fig.~\ref{fig:LLparameter}. Fig.~\ref{fig:orderparameter} shows the order parameter
versus the magnetic field determined analytically and
numerically by DMRG (see Sec.~\ref{sec:numericalmean-field}). The two
curves are almost indistinguishable and exhibit a strongly
asymmetric camel-like shape\cite{Klanjsek_NMR_3Dladder} with
two maxima close to the critical fields. The asymmetry of the curve
is again due to the
presence of the additional triplet states. This asymmetry disappears in the spin chain mapping. 

\section{Dynamical correlations of a single ladder}\label{sec:dynamicalcorrelation}

In the next paragraph, we discuss the possible excitations and the corresponding correlation functions created
by the spin operators. Such dynamical correlations allow us to
study the excitations of our system. They are also directly
related to many experimental measurements (NMR, INS ... ). We first focus on the gapped spin liquid
and then treat the gapless regime, for fields
$h_{c1}<h^z<h_{c2}$ (see Fig.~\ref{fig:phasediagramm}.b). All correlations are computed using the
t-DMRG at zero temperature for the single
ladder (appendix~\ref{sec:timeDMRG}) and are
compared to analytical results when such results exist. In particular we check the overlap with the LL description at low energy and use a strong coupling expansion (appendix~\ref{sec:tjmodelmapping}) to qualitatively characterize the obtained spectra. We start by a discussion of the correlations for the parameters of the compound BPCB (see Sec.~\ref{sec:bpcb}) and then turn to the evolution of the spectra with the coupling ratio $\gamma$ from the weak ($\gamma\rightarrow \infty$) to strong
coupling ($\gamma\approx0$).

\subsection{Zero temperature correlations and excitations}\label{par:zerocorrelations}

In a ladder system different types of correlations are
possible. We focus here on the quantities
\begin{equation}\label{equ:correlation1}
S^{\alpha\beta}_{q_y}(q,\omega)=\sum_l\int_{-\infty}^\infty dt\langle S^\alpha_{l,q_y}(t)S^\beta_{0,q_y}\rangle e^{i(\omega t-ql)}
\end{equation}
where $S^\alpha_{l,q_y}=S^\alpha_{l,1}\pm S^\alpha_{l,2}$ are
the symmetric ($+$) and antisymmetric ($-$) operators with rung momentum
$q_y=0,\pi$ and parity $P=+1,-1$ respectively, $\alpha,\beta=z,+,-$,
$S^\alpha_{l,q_y}(t)=e^{iHt}S^\alpha_{l,q_y}e^{-iHt}$ (for a single ladder $H$ corresponds to the Hamiltonian~\eqref{equ:spinladderhamiltonian}) and the
momentum $q$ is given in reciprocal lattice units $a^{-1}$. The rung momentum $q_y$ is a good quantum number. The dynamical correlations are directly related to INS measurements (see
Sec.~\ref{sec:INS}). They select different types of rung excitations (as
summarized in table~\ref{tab:excitations}). Using the
reflection and translation invariance of an infinite size system
($L\rightarrow\infty$), we can
rewrite the considered correlations~\eqref{equ:correlation1} in a more explicit form (at
zero temperature)\footnote{For the considered correlations ${S^\alpha}^\dagger=S^\beta$.}, i.e.~
\begin{equation}\label{equ:correlation2}
S^{\alpha\beta}_{q_y}(q,\omega)=\frac{2\pi}{L}\sum_\lambda|\langle\lambda|S_{q_y}^\beta(q)|0\rangle|^2\delta(\omega+E_0-E_\lambda)
\end{equation}
where $|0\rangle$ denotes the ground state of $H$ with energy
$E_0$, $S_{q_y}^\beta(q)=\sum_le^{-iql}S^\beta_{l,q_y}$ and
$\sum_\lambda$ is the sum over all eigenstates
$|\lambda\rangle$ of $H$ with energy $E_\lambda$. The form of
Eq.~\eqref{equ:correlation2} clearly shows that
$S^{\alpha\beta}_{q_y}(q,\omega)$ is non-zero if the operator
$S^\beta_{q_y}$ can create an excitation $|\lambda\rangle$ of
energy $E_0+\omega$ and momentum $q$ from the ground state. The
correlations $S_{q_y}^{\alpha\beta}$ are then direct probes of
the excitations $|\lambda\rangle$ in the system.
\begin{table}
\begin{center}
$$
\begin{array}{l||cccccc}
&S_0^z&S_\pi^z&S_0^+&S_\pi^+&S_0^-&S_\pi^-\\
\hline
\hline
 |s\rangle&0&|t^0\rangle&0&-\sqrt{2}|t^+\rangle&0&\sqrt{2}|t^-\rangle\\
 |t^+\rangle&|t^+\rangle&0&0&0&\sqrt{2}|t^0\rangle&-\sqrt{2}|s\rangle\\
 |t^0\rangle&0&|s\rangle&\sqrt{2}|t^+\rangle&0&\sqrt{2}|t^-\rangle&0\\
 |t^-\rangle&-|t^-\rangle&0&\sqrt{2}|t^0\rangle&\sqrt{2}|s\rangle&0&0\\
\hline
P&+1&-1&+1&-1&+1&-1\\
\Delta M^z&0&0&+1&+1&-1&-1\\
\end{array}
$$
\end{center}
\caption{Table of the rung excitations created by the symmetric and
antisymmetric operators in the decoupled bond limit. $P$ is the parity of the operators in
the rung direction and $\Delta M^z$ the change of the total
magnetization. \label{tab:excitations}}
\end{table}

Since the experimentally relevant case (compound BPCB) corresponds to a relatively
strong coupling situation ($\gamma\ll 1$, Eq. \ref{equ:couplingratio}), we
use the decoupled bond limit introduced in
Sec.~\ref{sec:spinchainmap} to present the expected
excitations $|t^+\rangle$, $|t^0\rangle$, $|t^-\rangle$ or
$|s\rangle$. In table \ref{tab:excitations}, we summarize
the rung excitations created by all the operators
$S^\beta_{q_y}$ and their properties. For example, the rung
parity $P$ is changed by applying an operator with rung momentum
$q_y=\pi$ and the $z$-magnetization is modified by $\Delta
M^z=\pm1$ by applying the operators $S^\pm_{q_y}$ respectively.

\subsection{Excitations in the spin liquid}\label{sec:spinliquidexcitations}

\begin{figure}
\begin{center}
\includegraphics[width=1\linewidth]{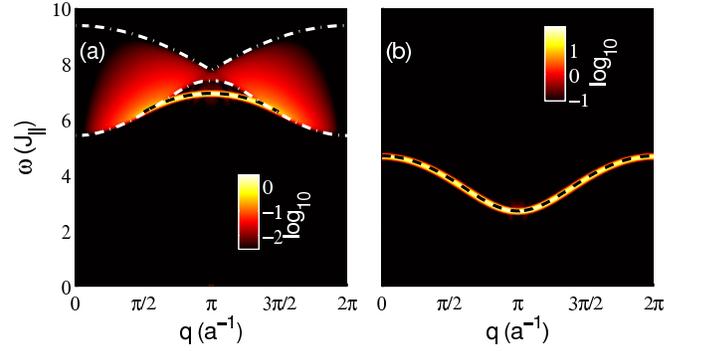}
\end{center}
\caption{(Color online): Momentum-energy dependent correlation functions at $h^z=0$ (a) Symmetric part $S_0^{\alpha\alpha}(q,\omega)$ with $\alpha=x,y,z$. The dashed (black) line marks the $(q,\omega)$ position of the two-triplet bound state, Eq.~(61b) in Ref.~\onlinecite{Zheng_bound_state_ladder}.
The dash-dotted (white) lines correspond to the boundaries of the two-triplet continuum.
(b) Antisymmetric part $S_\pi^{\alpha\alpha}(q,\omega)$.
The dashed (black) line corresponds to the predicted dispersion relation of a single triplet excitation (Eq.~(8) in Ref.~\onlinecite{reigrotzki_ladder_field}).\label{fig:spectrummz0}}
\end{figure}

Using the decoupled bond limit in the spin liquid phase, the excitations in the system
can be pictured as the excitation of rung singlets to rung
triplets. At zero magnetic field $h^z=0$, the system is spin
rotational symmetric and the different triplet excitations have
the same energy $\sim J_\perp$. It has been seen
previously that in this system both single triplet excitations
and two-triplet excitations play an important role \cite{barnes_ladder,reigrotzki_ladder_field,Zheng_bound_state_ladder,sushkov_ladder_boundstates,knetter_ladder}.
We discuss these excitations in the following focusing on the
ones that can be created by the symmetric
$S^{\alpha\alpha}_0=2S^{\pm\mp}_0$ and the antisymmetric
$S^{\alpha\alpha}_\pi=2S^{\pm\mp}_\pi$ correlations (see
Fig.~\ref{fig:spectrummz0}) for the
BPCB parameters~\eqref{equ:couplings}. Note that these correlations are
independent of the direction $\alpha=x,y,z$ due to the spin
rotation symmetry. Our results are in very good agreement to
previous findings \cite{reigrotzki_ladder_field,Zheng_bound_state_ladder,sushkov_ladder_boundstates,knetter_ladder}.

\subsubsection{Single triplet excitations}\label{sec:singletripletspinliquid}

At $h^z=0$, the system is in a global spin singlet state\cite{Auerbach_book_magnetism} ($S=0$). The $q_y=\pi$ correlation
couples then to states with an odd number of triplet excitations with rung parity $P=-1$ and
total spin $S=1,\ M^z=\pm1,0$ (see table~\ref{tab:excitations}). Nevertheless, only single triplet excitations are numerically resolved.
Their spectral weight is concentrated in a very sharp peak whose dispersion relation can be approximated using a \emph{strong coupling expansion} in
$\gamma$. Up to first order it is simply given
by a cosine dispersion\cite{barnes_ladder}, i.e. $\omega_t(q)/J_\perp \approx
1+\gamma \cos q$. Further corrections up to third
order in $\gamma$ have been determined in
Ref.~\onlinecite{reigrotzki_ladder_field}. In
Fig.~\ref{fig:spectrummz0}.b we compare the numerical results
for the BPCB parameters~\eqref{equ:couplings} to the expression up to third
order in $\gamma$. The strong
coupling expansion describes very well the position of the
numerically found excitations.

\subsubsection{Two-triplet excitations}\label{sec:zerofieldtwotriplets}

The structure of the $q_y=0$ correlation
is more complex (Fig.~\ref{fig:spectrummz0}.a). Due to the rung parity $P=1$ of the operators~$S^\alpha_0$, the
excitations correspond to an even number of
triplet excitations with total spin $S=1,\ M^z=\pm1,0$. We focus here on the two-triplet excitations that can be resolved numerically. These can be divided into a broad continuum and a very sharp triplet ($S=1$) bound state of
a pair of rung triplets. Since these excitations stem from the
coupling to triplets already present in the ground state (Fig.~\ref{fig:tripletdensity}), their
amplitude for the considered BPCB parameters~\eqref{equ:couplings} is
considerably smaller than the weight of the single triplet
excitations\cite{knetter_ladder}.

The dispersion relation of the bound states has been calculated
using a linked cluster series
expansion\cite{Zheng_bound_state_ladder}. The first terms of
the expansion have an inverse cosine form and the bound state
only exists in an interval around $q=\pi$ (cf. Ref.~\onlinecite{Zheng_bound_state_ladder,sushkov_ladder_boundstates}). The numerical
results for the BPCB parameters~\eqref{equ:couplings} agree very well with
the analytic form of the dispersion (Fig.~\ref{fig:spectrummz0}.a). The upper and lower limits of the continuum
can be determined by considering the boundary of the two non-interacting triplet continuum. They are numerically computed using the single triplet dispersion (Eq.~(8) in Ref.~\onlinecite{reigrotzki_ladder_field}) and shown in Fig.~\ref{fig:spectrummz0}.a. They agree very well with the numerically found results. The comparison with the known solutions serves as a check of the quality of our numerical results.

\subsection{Excitations in the gapless regime}

A small applied magnetic field ($h^z<h_{c1}$), at first order,
only smoothly translates the excitations shown in
Fig.~\ref{fig:spectrummz0} by an energy $-h^zM^z$ due to the
Zeeman effect. However, if the magnetic field exceeds
$h_{c1}$, the system enters into the gapless regime with a
continuum of excitations at low energy. For small values of $\gamma$
most features of this low energy continuum are qualitatively
well described by considering the lowest two modes of the
ladder only. Beside the low energy continuum, a complex
structure of high energy excitations exist. Contrarily to the
low energy sector this structure crucially depends on
the high energy triplet modes. In the following, we give a simple picture for
these excitations starting from the decoupled bond limit.
\begin{figure}
\begin{center}
\includegraphics[width=1\linewidth]{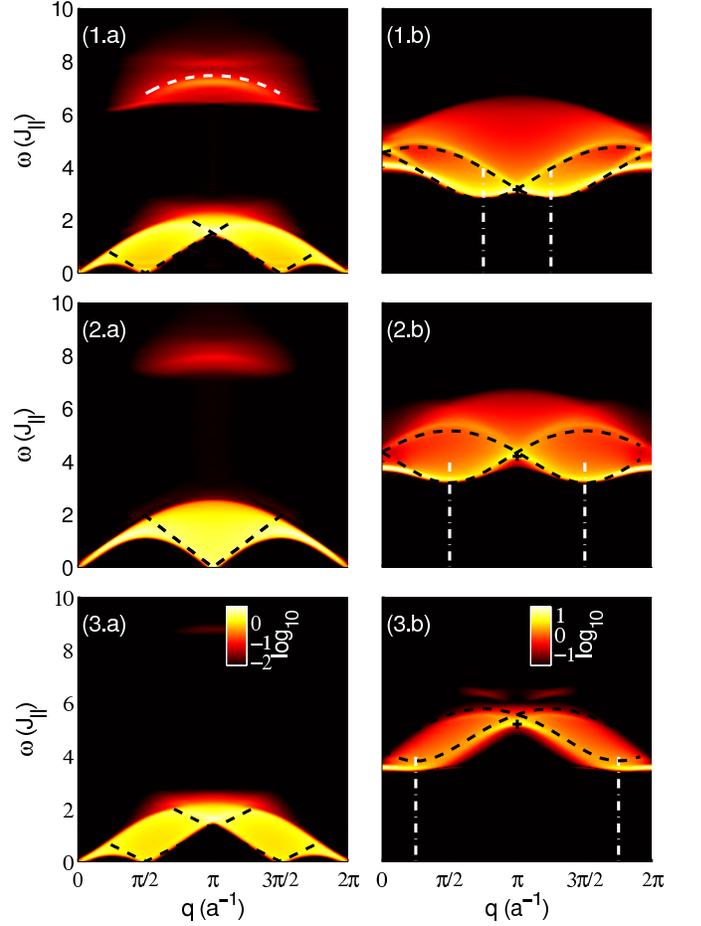}
\end{center}
\caption{(Color online): Momentum-energy dependent $zz$-correlation function (1)~at $m^z=0.25$ ($h^z=3.153~J_\parallel$), (2)~at $m^z=0.5$ ($h^z=4.194~J_\parallel$), and (3)~at $m^z=0.75$ ($h^z=5.192~J_\parallel$). (a)~Symmetric part $S_0^{zz}(q,\omega)$ without Bragg peak at $q=0$. The dashed black lines correspond to the location of the slow divergence at the lower edge of the continuum predicted by the LL theory. The dashed white curve corresponds to the predicted two-triplet bound state location. (b)~Antisymmetric part $S_\pi^{zz}(q,\omega)$. The dashed black lines correspond to the position of the high energy divergences or cusps predicted by the approximate mapping on the t-J model. The vertical white dash-dotted lines mark the momenta of the minimum energy of the high energy continuum and the black cross is the energy of its lower edge\cite{furusaki_correlations_ladder} at $q=\pi$.\label{fig:zzcorrelationmz}}
\end{figure}

\begin{figure}
\begin{center}
\includegraphics[width=1\linewidth]{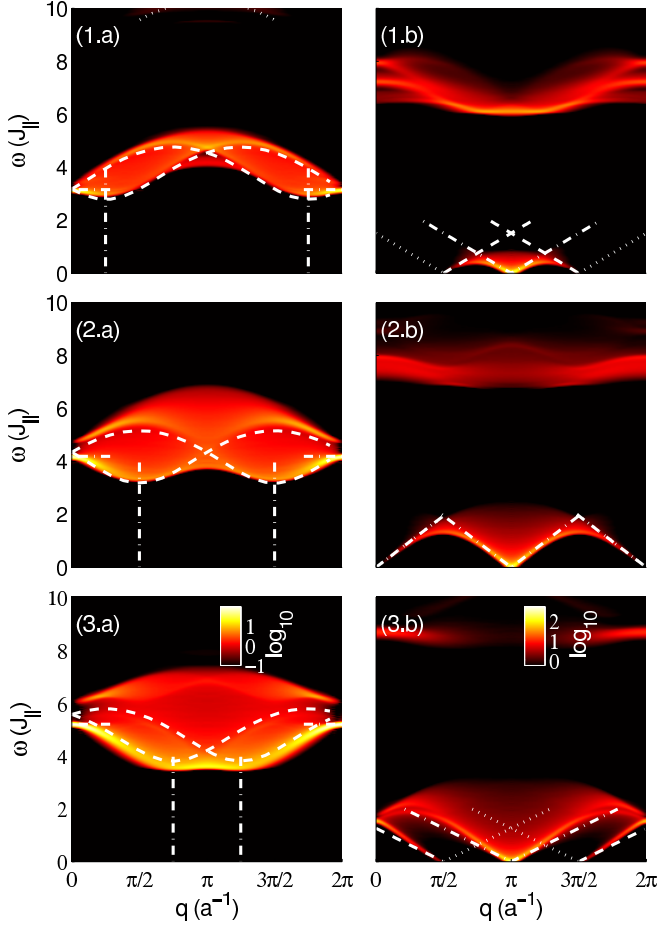}
\end{center}
\caption{(Color online): Momentum-energy dependent $+-$-correlation function (1)~at $m^z=0.25$ ($h^z=3.153~J_\parallel$), (2)~at $m^z=0.5$ ($h^z=4.194~J_\parallel$), and (3)~at $m^z=0.75$ ($h^z=5.192~J_\parallel$). (a)~Symmetric part $S_0^{+-}(q,\omega)$. The vertical dash-dotted white lines mark the momenta of the minimum energy of the high energy continuum and the horizontal ones the frequency of its lower edge\cite{furusaki_correlations_ladder} at $q=0,2\pi$. 
The dashed white lines correspond to the position of the high energy divergences or cusps predicted by the approximate mapping on the t-J model.
The dotted white curve corresponds to the predicted two-triplet bound state location. The high energy excitations at $\omega=3h_z$ are hardly visible.
(b)~Antisymmetric part $S_\pi^{+-}(q,\omega)$. The dashed and dash-dotted (dotted) white lines correspond to the location of the strong divergence (cusp) at the lower edge of the continuum predicted by the LL theory.
\label{fig:pmcorrelationmz}}
\end{figure}

\begin{figure}
\begin{center}
\includegraphics[width=1\linewidth]{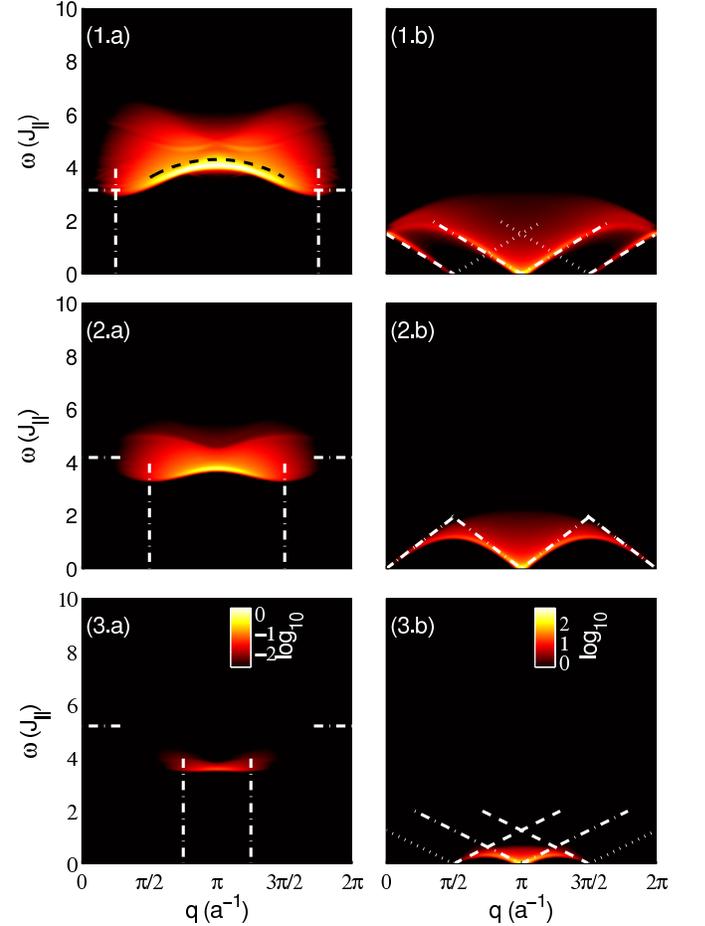}
\end{center}
\caption{(Color online): Momentum-energy dependent $-+$-correlation function (1)~at $m^z=0.25$ ($h^z=3.153~J_\parallel$), (2)~at $m^z=0.5$ ($h^z=4.194~J_\parallel$), and (3)~at $m^z=0.75$ ($h^z=5.192~J_\parallel$). (a)~Symmetric part $S_0^{-+}(q,\omega)$. The vertical dash-dotted white lines correspond to the momenta at which the minimum energy of the high energy continuum occurs and the horizontal line to the frequency of its lower edge\cite{furusaki_correlations_ladder} at $q=0,2\pi$. The dashed black curve corresponds to the predicted two-triplet bound state location. (b)~Antisymmetric part $S_\pi^{-+}(q,\omega)$. The dashed and dash-dotted (dotted) white lines correspond to the location of the strong divergence (cusp) at the lower edge of the continuum predicted by the LL theory.\label{fig:mpcorrelationmz}}
\end{figure}

\subsubsection{Characterization of the excitations in the decoupled bond limit}\label{sec:excitationcharacterization}
The evolution of the spectra for the BPCB parameters with
increasing magnetic field are presented in
Fig.~\ref{fig:zzcorrelationmz} for $S^{zz}_{q_y}$, in
Fig.~\ref{fig:pmcorrelationmz} for $S^{+-}_{q_y}$, and in
Fig.~\ref{fig:mpcorrelationmz} for $S^{-+}_{q_y}$.
Three different classes of excitations occur:
\begin{itemize}
\item[(i)] a continuum of excitations at low energy for $S^{zz}_0$ and $S^{\pm\mp}_\pi$
\item[(ii)] single triplet excitations at higher energy with a clear substructure for $S^{zz}_\pi$, $S^{+-}_0$, and $S^{+-}_\pi$
\item[(iii)] excitations at higher energy for $S^{zz}_0$ and $S^{+-}_0$ and $S^{-+}_0$ stemming from two-triplet excitations which have their main weight around $q\approx \pi$.
\end{itemize}
In the following we summarize some of the characteristic
features of these excitations, before we study them in more
detail in Secs.~\ref{sec:lowenergyexcitations} to~\ref{sec:singletripletLL}.

(i)  The continuum at low energy which does not exist in the
spin liquid is a characteristic signature of the gapless
regime. It stems from excitations within the low energy band which
corresponds to the $\ket{s}$ and $\ket{t_+}$ states in the decoupled bond limit (cf.~Fig.~\ref{fig:phasediagramm}.a and table \ref{tab:excitations}):
\begin{itemize}
\item[$S^{zz}_0$ :] excitations within the triplet $\ket{t_+}$ mode
\item[$S^{\mp\pm}_\pi$ :] excitations between the singlet $\ket{s}$ and the triplet $\ket{t_+}$ mode.
\end{itemize}
This continuum is smoothly connected to the spin liquid spectrum in the case of $S^{-+}_\pi$. It originates from the single triplet $\ket{t^+}$ branch (Sec.~\ref{sec:singletripletspinliquid}) when the latter reaches the ground state energy due to the Zeeman effect. Since two modes play the main role in the description of these low energy features,
many of them can already be explained qualitatively by the spin chain mapping. The excitations in the chain have been
studied previously using a Bethe ansatz description and exact
diagonalization calculations in
Ref.~\onlinecite{Muller_spinchain_dyncor}. More recently they
were computed in Ref.~\onlinecite{caux_heisenbergchaindyn} due
to recent progress in the Bethe ansatz method.
In particular, the boundary of the spectrum at low energy is well described by this approach, since the LL velocity determining it is hardly influenced by the higher modes (cf.~Fig.~\ref{fig:LLparameter}).
However, a more quantitative description requires to take into account the higher modes of the system as well. In Sec.~\ref{sec:lowenergyexcitations} we compare in detail our results with the LL theory and the spin chain mapping pointing out their corresponding ranges of validity.

(ii) The single high energy triplet excitations form a continuum with a clear substructure.
In the decoupled bond limit, these excitations correspond to
\begin{itemize}
\item[$S^{zz}_\pi$ :] Single triplet excitations $|t^0\rangle$ at energy $\sim h^z$
\item[$S^{+-}_0$ :] Single triplet excitations $|t^0\rangle$ at energy $\sim h^z$
\item[$S^{+-}_\pi$ :] Single triplet excitations $|t^-\rangle$ at energy $\sim 2h^z$.
\end{itemize}
Many of the features of these continua can be understood by
mapping the problem onto a mobile hole in a chain, as pointed out first in Ref.~\onlinecite{LaeuchliManep2007}.
We detail in Sec.~\ref{sec:singletripletLL} and appendix~\ref{sec:tjmodelmapping} this mapping. It opens the possibility to investigate the behavior of a single hole in a t-J like model using experiments in pure spin ladder compounds.

(iii) The high energy continuum, which has almost no weight close
to the Brillouin zone boundary ($q=0,2\pi$), is related to two-triplet excitations of the spin liquid (Sec.~\ref{sec:zerofieldtwotriplets}). They are
generated from high energy triplet components of the ground
state. Their weight therefore vanishes for $\gamma \to 0$ and the excitations correspond to
\begin{itemize}
\item[$S^{-+}_0$ :] Two-triplet excitations. $\frac{1}{\sqrt{2}}(|t^0\rangle|t^+\rangle-|t^+\rangle|t^0\rangle)$ at energy $\sim h^z$
\item[$S^{zz}_0$ :] Two-triplet excitations  $\frac{1}{\sqrt{2}}(|t^+\rangle|t^-\rangle-|t^-\rangle|t^+\rangle)$ at energy $\sim2h^z$
\item[$S^{+-}_0$ :] Two-triplet excitations $\frac{1}{\sqrt{2}}(|t^0\rangle|t^-\rangle-|t^-\rangle|t^0\rangle)$ at energy $\sim 3h^z$.
\end{itemize}

\subsubsection{Low energy continuum}\label{sec:lowenergyexcitations}

In this section we concentrate on the low energy excitations of type (i) discussing first their support and then comparing their spectral weight to the LL prediction.

\paragraph{Support of the low energy excitations}

\begin{figure}
\begin{center}
\includegraphics[width=1\linewidth]{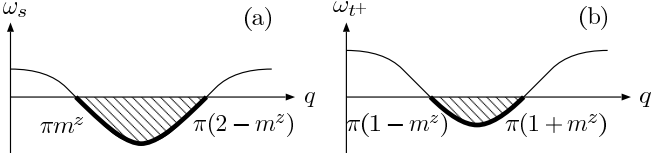}
\end{center}
\caption{Fermionic picture for the effect of the magnetic field: Filling of (a) the singlet band $|s\rangle$, (b) the triplet band $|t^+\rangle$ in the gapless phase for a given magnetization $m^z$.\label{fig:bandfilling}}
\end{figure}
The position of the soft modes in the low energy continuum can
be directly obtained from the bosonization representation
\cite{chitra_spinchains_field,giamarchi_ladder_coupled,furusaki_correlations_ladder} (see appendix~\ref{sec:luttingerliquidcorr}).
They can also be understood in a simple picture which we
outline in the following. The distribution of the rung state
population in the ground state depends on the magnetic field
$h^z$ (see Fig.~\ref{fig:tripletdensity}). Taking a fermionic point of view, the magnetic field
acts as a chemical potential that fixes the occupation of the
singlet and triplet rung states. Increasing the magnetic field
reduces the number of singlets, whereas at the same time the
number of triplets increases (see sketch in
Fig.~\ref{fig:bandfilling}). The Fermi level lies at the
momenta $q= \pi m^z, \pi(2-m^z)$ for the singlet states and at
the momenta $q= \pi(1-m^z), \pi(1+m^z)$ for the triplet states.
In this picture the soft modes correspond to excitations at the
Fermi levels. For transitions
$|t^+\rangle\leftrightarrow|t^+\rangle$ the transferred momenta
of these zero energy excitations are $q=0,2\pi m^z,2\pi(1-
m^z)$. In contrast the interspecies transitions
$|t^+\rangle\leftrightarrow|s\rangle$ allow the transfer of
$q=\pi(1-2m^z),\pi,\pi(1+2m^z)$. Therefore the positions of the
soft modes in the longitudinal correlation $S^{zz}_0$ which
allows transitions within the triplet states shift from the
boundaries of the Brillouin zone inwards towards $q=\pi$ when $m^z$ increases
(Fig.~\ref{fig:zzcorrelationmz}.a). In contrast the positions
of the soft modes in the transverse correlations
$S^{\pm\mp}_\pi$ which allow transitions between the singlets
and the triplets move with increasing magnetic field outwards
(Figs.~\ref{fig:pmcorrelationmz}.b and~\ref{fig:mpcorrelationmz}.b).

The top of these low energy continua are reached when the
excitations reach the boundaries of the energy band. In
particular, the maximum of the higher boundary lies at the
momentum $q=\pi$ which is easily understood within the simple
picture drawn above (cf.~Fig.~\ref{fig:bandfilling}). A more
detailed description of different parts of these low energy
continua is given in Ref.~\onlinecite{Muller_spinchain_dyncor}.

Let us compare the above findings with the predictions of the
LL theory for the dynamical correlations \cite{chitra_spinchains_field,furusaki_correlations_ladder,giamarchi_ladder_coupled}. Details on the LL description of the correlations
are given in appendix~\ref{sec:luttingerliquidcorr}. The LL theory predicts
a linear momentum-frequency dependence of the lower continuum
edges with a slope given by the LL velocity $\pm
u$ (Fig.~\ref{fig:LLparameter}). The position of
the soft modes are given by the ones outlined above (see
Fig.~\ref{fig:LLcorrschema}).  The predicted support at low
energy agrees very well with the numerical results (Fig.~\ref{fig:zzcorrelationmz}.a,~\ref{fig:pmcorrelationmz}.b, and~\ref{fig:mpcorrelationmz}.b). Of course
when one reaches energies of order $J_\parallel$ in the spectra one
cannot rely on the LL theory anymore. This is true in
particular for the upper limit of the spectra.

\paragraph{Spectral weight of the excitations}\label{sec:strongcouplingLLcomp}
Let us now focus on the distribution of the spectral
weight in the low energy continuum. In particular we compare our numerical findings to the Luttinger liquid description. Qualitatively, the LL theory predictions for the low energy spectra are well reproduced by the DMRG computations.

The Luttinger liquid predicts typically an
algebraic behavior of the correlations at the low energy
boundaries which can be a divergence or a cusp.
\begin{itemize}
\item[$S^{zz}_0$ :] The Luttinger liquid predicts peaks at the $q=0,2\pi$ branches and a slow divergence at the lower edge of the incommensurate branches $q=2\pi m^z,2\pi(1-m^z)$ (with exponent $1-K\approx0.2\ll1$). In the numerical results (Fig.~\ref{fig:zzcorrelationmz}.a) a slight increase of the weight towards the lower edge of the incommensurate branches can be seen.

\item[$S^{+-}_\pi$ :] A strong divergence at the lower edge of the $q=\pi$ branch (with exponent $1-1/4K\approx3/4\gg 0$) is obtained within the Luttinger liquid description. This is in good qualitative agreement with the strong increase of the spectral weight observed in the numerical data. A more interesting behavior is found close to the momenta $q=\pi(1\pm2m^z)$ in the incommensurate branches. Here a strong divergence is predicted for momenta higher (lower) than the soft mode $q=\pi(1- 2m^z)$ ($q=\pi(1+ 2m^z)$) with exponent $1-\eta_-\approx3/4\gg 0$. In contrast for momenta lower (higher) than the soft mode $q=\pi(1- 2m^z)$ ($q=\pi(1+ 2m^z)$) a cusp with exponent $1-\eta_+\approx-5/4\ll 0$ is expected\namedfootnote{\furusakibranches}{Note that the edge exponents in the incommensurate branches of the correlations \eqref{equ:LLtimepmcorrelation} are inverted compared to their expression in Ref.~\onlinecite{furusaki_correlations_ladder}.}. In the numerical results (Fig.~\ref{fig:pmcorrelationmz}.b) this very different behavior below and above the soft modes is evident. The divergence and cusp correspond to a large and invisible weight, respectively.

\item[$S^{-+}_\pi$ :] The same behavior as $S^{+-}_\pi$ replacing $m^z\rightarrow -m^z$ can be observed in Fig.~\ref{fig:mpcorrelationmz}.b.
\end{itemize}

\begin{figure}
\begin{center}
\includegraphics[width=0.9\linewidth]{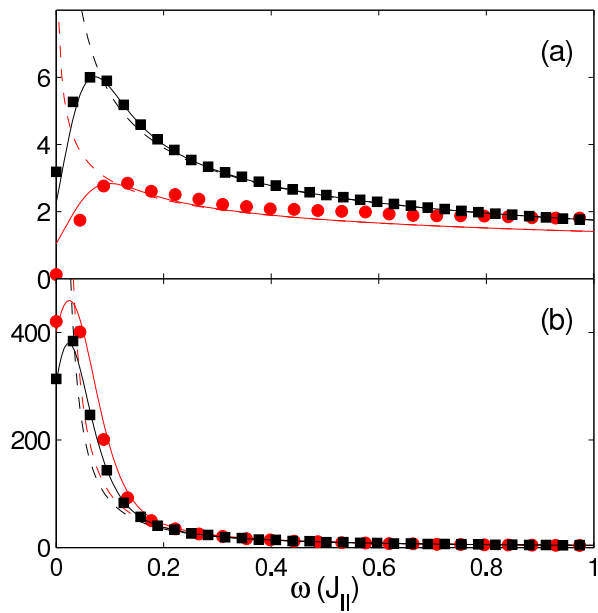}
\end{center}
\caption{(Color online): Cuts at fixed momentum $q=\pi$ and magnetization $m^z=0.5$ of the low energy spectrum (a) $S^{zz}_0(q=\pi,\omega)$, and (b) $S^{-+}_\pi(q=\pi,\omega)$.  The (red) circles and the (black) squares are the numerical results for the ladder and its spin chain mapping, respectively. The dashed lines correspond to the LL predictions and the solid lines are the latter convolved with the same Gaussian filter than the numerical data. The DMRG frequency numerical
limitation is of the order of the peak broadening of width $\delta\omega\approx
0.1~J_\parallel$ (see appendix~\ref{sec:tDMRGcorrelation}).
\label{fig:LLcomparisons}}
\end{figure}

To compare quantitatively the predictions of the LL to the numerical results
we show in Fig.~\ref{fig:LLcomparisons} different cuts of the
correlations at fixed momentum $q=\pi$ and magnetization $m^z=0.5$ for
the ladder and the spin chain mapping.
These plots show the DMRG results, the LL
description, and the latter convolved with the
Gaussian filter. The filter had been used in the numerical data to avoid effects due to the finite time-interval simulated (see
appendix~\ref{sec:tDMRGcorrelation}). Note that the amplitude of the LL results are inferred from the static correlation functions, such that the LL curve is fully determined and \emph{no fitting parameter} is left. Therefore the convolved LL results can directly be compared to the numerical results. Even though the presented numerical resolution might not be good enough to resolve the behavior close to the divergences (cusps), interesting information as the arising differences between the spin chain mapping and the full ladder calculations can already be extracted. In Fig.~\ref{fig:LLcomparisons}.a, we show a cut through the correlation at fixed momentum $S^{zz}_0(q=\pi,\omega)$.  The convolved LL and the numerical results compare very well. The difference between the
real ladder calculations and the spin chain mapping that neglect the effects of the higher triplet states
$|t^-\rangle$, $|t^0\rangle$ is obvious. From the LL description point of view, the shift of the spin chain correlation compared to the real ladder curve comes mainly from the prefactor $A_z$ and the algebraic exponent which are clearly modified by the effects of the high energy triplets (see Fig.~\ref{fig:LLparameter}).

For the transverse correlations, the LL theory predicts a strong
divergence (with an exponent $1-1/4K\approx3/4\gg0$ at the lower boundary of the
continuum branch at $q=\pi$.
A cut through the low energy continuum $S^{-+}_\pi(q=\pi,\omega)$ is shown in Fig.~\ref{fig:LLcomparisons}.b. The convolved Luttinger liquid reproduces well the numerical results.

\subsubsection{High energy excitations}

\paragraph{Weak coupling description of the high energy excitations}\label{sec:weakcouplinghighenergy}

Before looking in detail at the two kinds of high energy excitations presented in Sec.~\ref{sec:excitationcharacterization} we compare our computed high energy spectra with the weak coupling
description ($\gamma\gg1$). In this limit information on the spectrum
can be extracted from the bosonization
description\cite{chitra_spinchains_field,giamarchi_ladder_coupled,furusaki_correlations_ladder}.
In particular one expects a power law singularity at the lower
edge continuum with a minimal position at $q=\pi(1\pm m^z)$
(for $S^{zz}_\pi$), $q=\pi m^z, \pi(2-m^z)$ (for
$S^{\pm\mp}_0$) and an energy $h^z$ at momentum $q=\pi$ (for
$S^{zz}_\pi$), $q=0$ (for $S^{\pm\mp}_0$).
Except for $S^{-+}_0$ in which the spectral weight is
too low for a good visualization, our computed spectra
reproduces well the predictions for the minimal positions even though the coupling strength considered is not in the weak coupling limit (cf.~Figs.~\ref{fig:zzcorrelationmz}.b,~\ref{fig:pmcorrelationmz}.a and~\ref{fig:mpcorrelationmz}.a).

\paragraph{High energy single triplet excitations}\label{sec:singletripletLL}

The high energy single triplet continua originate from the transition of the low energy rung states $\ket{s}$ and $\ket{t^+}$ to the high energy triplets $\ket{t^0}$ and $\ket{t^-}$. The excitations coming from the singlets $\ket{s}$ (in $S^{zz}_\pi$ and $S^{+-}_\pi$) are already present in the spin liquid phase (cf.~Sec.~\ref{sec:singletripletspinliquid}) in which they have the shape of a sharp peak centered on the triplet dispersion. The transition between the gapped spin liquid and the gapless regime is smooth and consists in a splitting and a broadening of the triplet branch that generates a broad continuum of new excitations. Contrarily to the latter the excitations coming from the low energy triplets $\ket{t^+}$ (in $S^{+-}_0$) are not present in the spin liquid phase. The corresponding spectral weight appears when $h^z>h_{c1}$.

An interpretation of the complex structure of these high energy continua can be obtained in terms of itinerant quantum chains.
 Using a strong coupling expansion of the Hamiltonian~\eqref{equ:spinladderhamiltonian} (appendix~\ref{sec:tjmodelmapping}) it is possible to map the high energy single triplet excitations $\ket{t^0}$ to a single hole in a system populated by two types of particles with
pseudo spin $|\tilde\uparrow\rangle=|t^+\rangle$,
$|\tilde\downarrow\rangle=|s\rangle$ (with the
Sec.~\ref{sec:spinchainmap} notation).

In this picture the effective Hamiltonian of the $J_\perp$ energy sector is approximately equivalent to the half
filled anisotropic 1D t-J model with one hole (see appendix~\ref{sec:sec_sector_Jperp}). The effective Hamiltonian is given by
\begin{equation}\label{equ:t-J}
H_\text{t-J}= H_\text{XXZ} + H_\text{t} + H_\text{s-h}+\epsilon.
\end{equation}
where $\epsilon=(J_\perp+h^z)/2 $ is an energy shift and $
H_\text{t}=J_\parallel/2\sum_{l,\sigma}(c_{l,\sigma}^\dagger c_{l+1,\sigma}^\phd +h.c.)$ is the usual hopping term. Here $c^\dagger_{l,\sigma}$ ($c_{l,\sigma}$) is the creation (annihilation) operator of a fermion with pseudo spin $\sigma=\tilde \uparrow,\tilde\downarrow$ at the site $l$. Note that although we are dealing here with spin states, it is possible to 
faithfully represent the three states of each site's Hilbert space ($\ket{s}$, $\ket{t^+}$, $\ket{t^0}$) using a fermion representation. 

Additionally to the usual terms of the t-J model a nearest neighbor interaction term between one of the spins and the hole arises
\begin{equation}
H_\text{s-h}=-\frac{J_\parallel}{4}\sum_l\left[n_{l,h}n_{l+1,\tilde\uparrow}+n_{l,\tilde\uparrow}n_{l+1,h}\right].
\end{equation}
Here $n_{l,h}$ is the density operator of the hole on the site $l$.
In this language the spectral weight of $S^{zz}_\pi$ and $S^{+-}_0$ 
corresponding to the single high energy triplet
excitations is equivalent to the single particle spectral functions of the up-spin and down-spin particle respectively:
\begin{equation}
\begin{array}{lllll}
S^{zz}_\pi&\propto&\langle c^\dagger_{\tilde\downarrow} c_{\tilde\downarrow}^\phd\rangle&\mbox{ with hole of type }&|s\rangle\rightarrow|t^0\rangle\\
S^{+-}_0&\propto&\langle c^\dagger_{\tilde\uparrow} c_{\tilde\uparrow}^\phd\rangle&\mbox{ with hole of type }&|t^+\rangle\rightarrow|t^0\rangle.
\end{array}
\end{equation}
Here $\langle c^\dagger_\sigma
c_\sigma^\phd\rangle(q,\omega)=\sum_\lambda|\langle\lambda|c_{q,\sigma}|0\rangle|^2\delta(\omega+E_0-E_\lambda)$.

For the standard t-J model (for $SU(2)$ invariant XXX spin background and without the anisotropic term $H_\text{s-h}$ in Eq.~\eqref{equ:t-J}), these spectral functions have been studied in
Refs.~\onlinecite{sorella_spectrum_hubbard_1D,sorella_spectrum_hubbard_1D_long}. The
presence of singularities of the form
\begin{equation}\label{equ:highenergysingularity}
\langle c^\dagger_\sigma c_\sigma^\phd\rangle(q,\omega)\propto[\omega-\omega_{t^0}(q-q_\nu)]^{2X_\nu(q)-1}
\end{equation}
were found. Here $\omega_{t^0}(q)$ is the $\ket{t^0}$ triplet
dispersion relation, $q_\nu$ the spinon momentum at the Fermi
level and $X_\nu$ the algebraic decay exponent at the
singularity. This exponent is not known in our case and depends
on the magnetization $m^z$ and the momentum $q$. It generates a
peak or a cusp at the energy
$\omega=\omega_{t^0}(q-q_\nu)$. The spinon momentum $q_\nu$ depends on the type of the rung state before excitation
($\nu=s,t^+$). For an excitation created from a singlet state
$q_s=\pm\pi m^z$ (for $S^{zz}_\pi$) and 
from the triplet state $q_{t^+}=\pi(1\pm m^z)$ (for $S^{+-}_0$) (Fig.~\ref{fig:bandfilling}). At $h^z=0$, a series expansion of
$\omega_{t^0}(q)$ can be performed (Eq.~8 in Ref.~\onlinecite{reigrotzki_ladder_field}). To extend it into the gapless phase ($h_{c1}<h^z<h_{c2}$), we approximate
$\omega_{t^0}(q)$ by shifting the value $\omega_t(q)$ at $h^z=0$ by the Zeeman shift, i.e.
\begin{equation}
\omega_{t^0}(q)=\omega_t(q)+\Delta E_0(h^z).
\end{equation}
Here we used the shift of the
ground state energy per rung $\Delta
E_0(h^z)=E_0(h^z)-E_0(0)$. $\Delta E_0$ was determined by
DMRG calculations (Fig.~\ref{fig:decay_exp}.b for the BPCB
parameters). The resulting momentum-frequency positions
$\omega=\omega_{t^0}(q-q_\nu)$ of the high energy
singularities (cusps or divergencies) are plotted 
on the spectrum
Fig.~\ref{fig:zzcorrelationmz}.b and
Fig.~\ref{fig:pmcorrelationmz}.a. They agree remarkably well with
the shape of the computed spectra in particular for small magnetic field \footnote{For the correlations
$S^{zz}_\pi$ and $S^{+-}_0$, some of these singularities
correspond to the lower edge description in
Ref.~\onlinecite{furusaki_correlations_ladder} and discussed in Sec.~\ref{sec:weakcouplinghighenergy}.}.  Neglecting the additional interaction term $H_\text{s-h}$ the t-J model Hamiltonian would lead to a symmetry of these excitations with respect to half magnetization. 
However in the numerical spectra the effect of the interaction shows up in a clear asymmetry of these excitations (Figs.~\ref{fig:zzcorrelationmz}.1.b  and~\ref{fig:pmcorrelationmz}.3.a).
In particular  in the $S^{+-}_0$ correlation some of the weight is seemingly detaching and pushed towards the upper boundary of the continuum (Fig.~\ref{fig:pmcorrelationmz}.3.a) for large magnetization. 
A more detailed account of the spectra can be found in Ref.~\onlinecite{LaeuchliRuegg}.

A similar mapping can be performed for the single $\ket{t^-}$ excitation. 
In contrast to the $J_\perp$ sector in the $2J_\perp$ sector not only the $\ket{t^-}$ excitation occurs, but the effective Hamiltonian mixes also $\ket{t^0}$ triplets into the description. Therefore the description by a single hole in a spin-$1/2$ chain breaks down and more local degrees of freedom are required.
This results in a more complex structure as seen in Fig.~\ref{fig:pmcorrelationmz}.1.b. Previously high-energy excitations in dimerized antiferromagnets have been described rather generally by a mapping to an X-ray edge singularity problem~\cite{Kolezhuk_ESR_high_fields,Kolezhuk_response_high_fields,Friedrich_edge_haldane}. It is interesting though that in the present setup these excitations
can be understood as  t-J hole spectral functions, which display a much richer structure than anticipated.

\paragraph{High energy two-triplet excitations}\label{sec:twotripletLL}

The two-triplet continua and bound states already discussed in the spin liquid phase (cf.~Sec.~\ref{sec:zerofieldtwotriplets}) are still visible in the gapless regime in the symmetric
correlations ($S^{zz}_0$ and $S^{\pm\mp}_0$). At low magnetic field the location of their maximal spectral weight can
be approximated by the expression of the bound state dispersion at zero field (Eq.~(61b) in Ref.~\onlinecite{Zheng_bound_state_ladder}) shifted by the Zeeman energy\footnote{The Zeeman shift includes both the shift of the ground state (Fig.~\ref{fig:decay_exp}.b) and the shift of the excited state.}.
The two-triplet excitation location obtained in this way agrees to a good extent with the location found in the numerical calculations
(cf.~Figs.~\ref{fig:zzcorrelationmz}.1.a,~\ref{fig:pmcorrelationmz}.1.a and~\ref{fig:mpcorrelationmz}.1.a). Since these excitations
are generated from the high energy triplet components in the
ground state and these vanish with increasing magnetic field (cf.~Fig.~\ref{fig:tripletdensity}),
their residual spectral weight slowly disappears with
increasing magnetization.

\subsection{Weak to strong coupling evolution}

\begin{figure*}
\begin{center}
\includegraphics[width=1\linewidth]{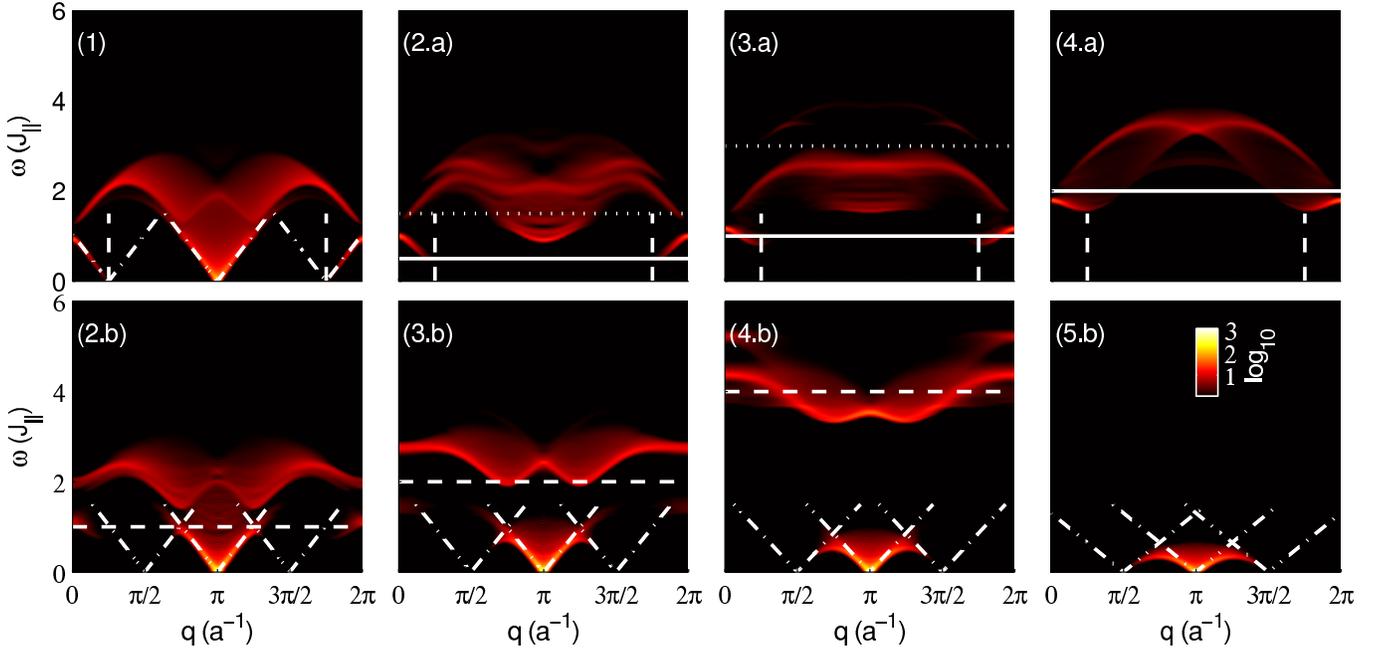}
\end{center}
\caption{(Color online):
Momentum-energy dependent $+-$-correlations ($S_{q_y}^{+-}(q,\omega)$) at $m^z=0.25$ for different ladder couplings ($\gamma=J_{\parallel}/J_{\perp}$) (1) $\gamma\rightarrow\infty$, (2) $\gamma=2$, (3) $\gamma=1$, (4) $\gamma=0.5$, (5) $\gamma\rightarrow0$. The symmetric (antisymmetric) correlations with $q_y=0$ ($q_y=\pi$) are presented in the figures labeled by a (b). In (1,2-4.a) the vertical dashed lines represent the incommensurate momenta of the low energy branches of the single spin chain at $q=\pm\pi m^z=\pm\pi/4$ (they also correspond to the predicted momenta of the lowest energy excitations of the symmetric correlations\cite{furusaki_correlations_ladder}). The horizontal solid (dotted) horizontal lines in (2-4.a) correspond to the approximate energy $J_\perp$ ($3J_\perp$) of the single triplet excitations of type (ii) (two-triplet excitations of type (iii)). The horizontal dashed lines in (2-4.b) correspond to the approximate energy $2J_\perp$ of the excitations of type (ii). The dash-dotted lines in (1) correspond to the linear low energy boundaries of the continuum of excitations given by the LL theory applied on single Heisenberg chains $\gamma\rightarrow \infty$. The dash-dotted lines in (2-5.b) correspond to the linear low energy boundaries of the continuum of excitations given by the LL theory on spin ladder with finite $\gamma$ (appendix~\ref{sec:luttingerliquidcorr}).
\label{fig:weaktostrong}}
\end{figure*}

For all the excitation spectra presented above the intrachain coupling ratio of BPCB $\gamma =J_\parallel/J_\perp\approx1/3.55\ll 1$ was taken. For this chosen value of $\gamma$, a strong coupling approach gives a reasonable
description of the physics. In this section we discuss the evolution of the spectra from weak ($\gamma\rightarrow\infty$) to strong coupling ($\gamma\rightarrow0$).
To illustrate this behavior, we show in
Fig.~\ref{fig:weaktostrong} the symmetric  and antisymmetric
parts of the correlations $S_{q_y}^{+-}$ at $m^z=0.25$ for
different coupling ratios $\gamma =\infty,2,1,0.5,0$.

At $\gamma\rightarrow\infty$ (Fig.~\ref{fig:weaktostrong}.1), the chains forming the ladder correspond to two decoupled Heisenberg chains. In this case the symmetric and antisymmetric correlations are
identical $S_0^{+-}=S_\pi^{+-}$ and are equivalent to the correlation $2S^{+-}$ of the single chain\cite{Muller_spinchain_dyncor} with magnetization per spin $m^z/2=0.125$. A complex low energy continuum exists with zero energy branches\cite{giamarchi_book_1d,chitra_spinchains_field,Muller_spinchain_dyncor} at momenta $q=\pm\pi m^z,\pi$ similar to that discussed in Sec.~\ref{sec:lowenergyexcitations}.
In contrast, in the strong coupling limit ($\gamma \rightarrow0$) (Fig.~\ref{fig:weaktostrong}.5.b)
the symmetric correlations vanish and the antisymmetric
part corresponds to the single chain correlation $2S^{+-}$ with
anisotropy $\Delta=1/2$ and magnetization per spin $m^z-1/2$ (see the spin chain mapping in Sec.~\ref{sec:spinchainmap}). The antisymmetric part consists of a low energy continuum with branches at momenta $q=(1\pm2 m^z)\pi,\pi$ (Sec.~\ref{sec:lowenergyexcitations}). Note, that a bosonization description of the low energy
sectors of both extreme regimes can be formulated \cite{chitra_spinchains_field,giamarchi_ladder_coupled,furusaki_correlations_ladder} (appendix~\ref{sec:luttingerliquidcorr}).

In the following we discuss the evolution between these two limits.
In the antisymmetric correlation (cf.~Fig.~\ref{fig:weaktostrong}.2-5.b) a low energy continuum exists  at all couplings with a zero energy excitation branch at $q=\pi$. These low energy excitations correspond mainly to the excitations with $\Delta S=\Delta M^z=-1$. This has been pointed out for the weak coupling limit\cite{Muller_spinchain_dyncor,chitra_spinchains_field}. They become the transitions $|t^+\rangle\rightarrow|s\rangle$ with the same quantum numbers in the decoupled bond limit.
Additionally the upper part of the excitation spectrum at weak coupling, which mainly corresponds to excitations with \cite{Muller_spinchain_dyncor} $\Delta S=0,1$ and $\Delta
M^z=-1$ splits from the lower part of the spectrum and moves to higher energy while increasing the coupling. It evolves to a high energy excitation branch which corresponds in the decoupled bond limit to the $|s\rangle\rightarrow|t^-\rangle$ transition, i.e.~single triplet excitations of type (ii) (Sec.~\ref{sec:singletripletLL}) approximately at\namedfootnote{\Jperptohz}{Note that in the strong coupling limit the energy scales set by $J_\perp$ and $h^z$ become very close, such that this position is in agreement with the ones previously discussed for the strong coupling limit. }
$2J_\perp$.

The properties of the zero energy excitation branch at $q=\pi$ show a smooth transition between the two limits\cite{giamarchi_ladder_coupled,chitra_spinchains_field}.
For example the slope of the lower edge continuum which is determined by the LL velocity $u$ decreases smoothly from its value for the Heisenberg chain to the lower value for the anisotropic spin chain with $\Delta=1/2$ in the strong coupling limit.
In contrast to this smooth change, the presence of a finite value of $J_\perp$ leads to the formation of a gap in the
incommensurate low energy branches\cite{chitra_spinchains_field} at $q=\pm\pi m^z$. With increasing coupling strength $J_\perp$ new low energy branches at momenta $q=\pi(1\pm 2m^z)$ become visible\cite{giamarchi_ladder_coupled,furusaki_correlations_ladder}. The weight of these gapless branches is very small for small coupling and increases with stronger coupling \cite{giamarchi_ladder_coupled}.

In contrast to the antisymmetric part, the
symmetric part $S_0^{+-}$ becomes gapped when the interladder coupling $J_\perp$ is turned on. The lowest energy excitations remain close to the momenta $q=\pm\pi m^z$ in agreement with Ref.~\onlinecite{furusaki_correlations_ladder}. They connect to the single triplet excitations of type (ii) (Sec.~\ref{sec:singletripletLL}) which are approximately at an energy $J_\perp$. While increasing $\gamma$ the higher part of the spectrum starts to separate from the main part and evolves to a branch of high energy two-triplet excitations of type (iii) (Sec.~\ref{sec:twotripletLL}). These are located at approximately $3J_\perp$.
Our computed spectra for $\gamma=2,1,0.5$ presented in
Fig.~\ref{fig:weaktostrong}.2-4.a clearly show this behavior. In Fig.~\ref{fig:weaktostrong}.4.a the highest two-triplet excitations cannot be seen anymore since their spectral weight is too low.

\subsection{Influence of the weak interladder coupling on the excitation spectrum}

Up to now we only discussed the excitations of a single spin ladder and neglected the weak interladder coupling $J'$ usually present in real compounds.

Deep inside of the spin liquid phase, the correlations for a single ladder are dominated by high energy single or multi triplet excitations as discussed in Sec.~\ref{sec:spinliquidexcitations}. The presence of a small interladder coupling $J'$ causes a dispersion in the interladder direction with an amplitude of order $J'$. This effect can be evaluated for independent triplet excitations using a single mode approximation\cite{Auerbach_book_magnetism}. However, for the compound BPCB the interladder coupling is so small that present day experiments do not resolve this small broadening\cite{Thielemann_INS_ladder,Savici_BPCB_INS}.

In contrast in the gapless phase the effect of the interladder coupling can change considerably the excitations.
In particular below the transition temperature to the 3D-ordered phase, a
Bragg peak appears at $q=\pi$ in the transverse dynamical functions
$S^{\pm\mp}_\pi$. As discussed in Ref.~\onlinecite{schulz_coupled_spinchains},
this Bragg peak is surrounded by gapless Goldstone modes and it has been
measured in the compound BPCB\cite{Thielemann_ND_3Dladder}. Additional high energy modes are predicted to occur in the transverse $S^{\pm\mp}_\pi$ and longitudinal $S^{zz}_0$\cite{schulz_coupled_spinchains}.
It would be interesting to compute the excitations using random phase approximation analogously to Ref.~\onlinecite{schulz_coupled_spinchains} in combination with the computed dynamical correlations for the single ladder in order to investigate the effect of a weak interladder coupling in more detail. However, this goes beyond the scope of the present work and
will be left for a future study.	

\section{Experimental measurements}\label{sec:experimental}

In this section we summarize the results of experiments on the compound BPCB and compare them to the theoretical predictions. First we introduce the structure of the BCPB compound. Then we focus on static and low energy results of the system. Finally we discuss the detection of excitations by INS\cite{Thielemann_INS_ladder,Savici_BPCB_INS} measurements. In particular we give a prediction of the INS cross section in a full range of energy and compare its low energy part with the measured spectra on the compound BPCB.

\subsection{Structure of the compound $\mathrm{(C}_5\mathrm{H}_{12}\mathrm{N)}_2\mathrm{CuBr}_4$}\label{sec:bpcb}

The compound $\mathrm{(C}_5\mathrm{H}_{12}\mathrm{N)}_2\mathrm{CuBr}_4$, customarily called BPCB or (Hpip)$_2$CuBr$_4$, has been intensively
investigated using different experimental methods such as nuclear magnetic
resonance\cite{Klanjsek_NMR_3Dladder} (NMR), neutron
diffraction\cite{Thielemann_ND_3Dladder} (ND), inelastic
neutron scattering\cite{Thielemann_INS_ladder,Savici_BPCB_INS}
(INS), calorimetry\cite{Ruegg_thermo_ladder}, magnetometry\cite{watson_bpcb},
magnetostriction\cite{Anfuso_BPCB_magnetostriction,lorenz_thermalexp_magnetostriction}, and electron spin resonance spectroscopy\cite{cizmar_esr_bpcb} (ESR).

\begin{figure}
\begin{center}
\includegraphics[width=1\linewidth]{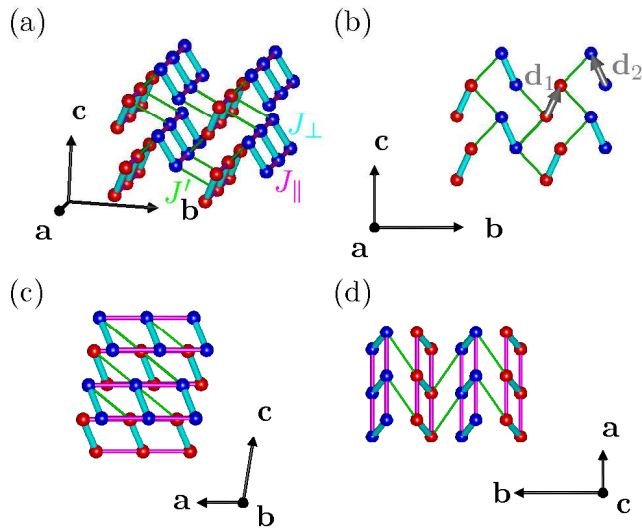}
\end{center}
\caption{(Color online): Coupling structure of BPCB where the unpaired electron spins of the $\mathrm{Cu^{2+}}$ atoms in the first (second) type of ladders are pictured by red (blue) spheres. The $J_\perp$, $J_\parallel$ and $J'$ coupling paths are represented in turquoise, pink, and green, respectively. ${\bf a}$, ${\bf b}$ and ${\bf c}$ are the three unit cell vectors of the structure. Gray arrows are the rung vectors of the two types of ladders ${\bf d}_{1,2}$.
(a) 3D structure. (b) Projection of the 3D structure onto the ${\bf bc}$-plane.
(c) Projection of the 3D structure onto the ${\bf ac}$-plane.
(d) Projection of the 3D structure onto the ${\bf ab}$-plane.
\label{fig:structure}}
\end{figure}

The magnetic properties of the compound are related to the
unpaired highest energy orbital of the $\mathrm{Cu^{2+}}$
ions. Thus the corresponding spin structure (Fig.~\ref{fig:structure}) matches with the
$\mathrm{Cu^{2+}}$ location\cite{Patyal_BPCB,Klanjsek_NMR_3Dladder,Thielemann_INS_ladder}. The unpaired spins form two types
of inequivalent ladders (Fig.~\ref{fig:structure}) along the ${\bf a}$ axis (${\bf a}$,
${\bf b}$ and ${\bf c}$ are the unit cell vectors of BPCB).
The direction of the rung vectors of these ladders are $\mathbf{d}_{1,2}=(0.3904,\pm0.1598,0.4842)$ in the
primitive vector coordinates (Fig.~\ref{fig:structure}.b).
As one can
see from the projection of the spin structure onto the ${\bf
bc}$-plane (Fig.~\ref{fig:structure}.b), each rung has $n_c=4$
interladder neighboring spins.

The BPCB structure has been identified as a good experimental realization of
the system of weakly coupled spin-$1/2$ ladders\cite{Patyal_BPCB} described by the Hamiltonian~\eqref{equ:coupledladdershamiltonian}. As we will explain in the next section, the interladder coupling $J'$ has been experimentally determined to be\cite{Klanjsek_NMR_3Dladder,Thielemann_ND_3Dladder}
\begin{equation}\label{equ:jprime}
J'\approx20-100~\mbox{mK}.
\end{equation}
The intraladder couplings from Eq.~\eqref{equ:spinladderhamiltonian} were determined to be $J_\perp\approx12.6-13.3~{\rm K}$, $J_\parallel\approx3.3-3.8~{\rm K}$ with different experimental techniques and at different experimental conditions\cite{Klanjsek_NMR_3Dladder,Thielemann_ND_3Dladder,Thielemann_INS_ladder,Savici_BPCB_INS,Ruegg_thermo_ladder,watson_bpcb,Anfuso_BPCB_magnetostriction,lorenz_thermalexp_magnetostriction,Patyal_BPCB}. In this paper, we use the values\namedfootnote{\NMRparameters}{These  parameters were the first outputs from the NMR measurements, which were later refined to the values from Ref.~\onlinecite{Klanjsek_NMR_3Dladder}. Note that small changes in these values do not affect the main results of the calculations.}
\begin{equation}\label{equ:couplings}
J_{\perp}\approx12.6~\mathrm{K},\quad J_\parallel\approx3.55~\mathrm{K}.
\end{equation}

Recently a slight anisotropy of the order of $5\%$ of $J_\perp$ has been discovered by ESR\cite{cizmar_esr_bpcb} measurements. This anisotropy could explain the small discrepancies between the couplings found in different experiments.
The magnetic field in Tesla is related to $h^z$ replacing
\begin{equation}\label{equ:experimentalhz}
h^z\rightarrow g\mu_B h^z
\end{equation}
in Eq.~\eqref{equ:spinladderhamiltonian} with $\mu_B$ being the Bohr magneton and $g$ being the Land\'e factor
of the unpaired copper electron spins. The latter depends on the orientation of the sample with respect to the magnetic field. For the orientation chosen in the NMR measurements\cite{Klanjsek_NMR_3Dladder}, it amounts to\cite{\NMRparameters} $g\approx2.126$. It can vary up to $\sim 10\%$ for other experimental setups\cite{Patyal_BPCB}.

\subsection{Thermodynamic measurements and low energy properties}

Many interesting thermodynamic measurements have been performed on BPCB. We select in the following some of these experiments and compare them to the theoretical predictions.

The longitudinal magnetization that can be measured very
precisely by NMR (see Ref.~\onlinecite{Klanjsek_NMR_3Dladder}) was shown to agree remarkably with the one computed using the weakly coupled ladder model (Fig.~\ref{fig:mz_zeroT}). Unfortunately, the magnetization is not very sensitive to the underlying model (Sec.~\ref{sec:criticalfields}). Thus it cannot be used to select between various models, but once the model is fixed, it can
be used to fix precisely the parameters given the high accuracy
of the experimental data. In particular, the position of the critical fields are very sensitive to the values of the intraladder couplings (Sec.~\ref{sec:criticalfields}). The couplings determined by this method are
$J_{\perp}\approx12.6~\mathrm{K}$ and $J_\parallel\approx3.55~\mathrm{K}$.

A more selective test to distinguish between various models is provided by
the specific heat. This is due to the fact that the specific heat contains information on high energy excitations which are characteristic for the underlying model.  As shown in Fig.~\ref{fig:cvsT} the
experimental data are remarkably described, up to an accuracy of
a few percent, by a simple Heisenberg ladder Hamiltonian with the parameters extracted from the magnetization. In particular, not only the low temperature behavior is covered by the ladder description, but also the higher maxima. This indicates that the ladder
Hamiltonian is an adequate description of the compound.
The small discrepancies between the specific heat data and the calculation
which is essentially exact can have various sources. First of all,
the substraction of the non-magnetic term in the experimental data
can account for some of the deviations.
Furthermore the interladder coupling can induce slight changes in the behavior of the specific heat.
Finally deviations from the simple ladder Hamiltonian can be present.
Small anisotropy of the couplings
can exist and indeed are necessary to interpret recent ESR experiments \cite{cizmar_esr_bpcb}. Other
terms such as longer range exchanges or Dzyaloshinskii-Moryia
(DM) terms might occur along the legs even if the latter is forbidden
by symmetry along the dominant rung coupling. Clearly all these
deviations from the Heisenberg model cannot be larger than a
few percents. They will not lead to any sizeable deviation
for the Luttinger parameters (Fig.~\ref{fig:LLparameter}) in the one dimensional regime.
Close to the critical points they can, however, play a more important role. It
would thus be interesting in subsequent studies to refine the model
to take such deviations into account.

After having fixed the model and the intraladder couplings up to a few percents we use it to compute other experimentally accessible quantities such as the magnetostriction, thermal expansion, the NMR relaxation rate, the transition temperature to the ordered
phase and its order parameter. In Refs.~\onlinecite{Anfuso_BPCB_magnetostriction,lorenz_thermalexp_magnetostriction} the magnetostriction and thermal expansion were compared to the theoretical results using the described ladder model. A very good agreement was found  in a broad range of temperature (not shown here). Note that only a full ladder model allows a global quantitative description of the magnetostriction effect which provides an additional confirmation of the applicability of the model. The quality of the determination of the model and its intraladder parameters becomes even more evident in the
comparison of the NMR data for the relaxation rate $T_1^{-1}$ with the theoretical results of the Luttinger liquid theory as
shown in Fig.~\ref{fig:T1}. Only \emph{one}
adjustable parameter is left, namely the hyperfine coupling constant (see Sec.~\ref{sec:relaxationtime}).
This parameter allows for a global expansion of the
theoretical curve, but not for a change of its shape.
The agreement between the theory and the experimental data is very good over
the whole range of the magnetic field and only small deviations can be
seen. The compound BPCB
thus allows to \emph{quantitatively} test the Luttinger liquid
universality class. Even though the Luttinger liquid description is restricted to low energies,
in BPCB its range of validity is rather large. Indeed
at high energy, its breakdown is approximately signaled by the first peak of the specific
heat\cite{Ruegg_thermo_ladder} (see Sec.~\ref{sec:specificheat}). Here this has a maximum scale of
about $T \sim 1.5~{\rm K}$ at midpoint between $h_{c1}$ and $h_{c2}$ (see Fig.~\ref{fig:Tmagnetization}.c).
Given the low ordering temperature which has a maximum at about
$T \sim 100~{\rm mK}$ this leaves a rather large Luttinger regime for
this compound.

Taking now the coupling between ladders into account, one can induce a transition to a three-dimensional ordered phase.
The transition temperature is shown in
Fig.~\ref{fig:criticaltemperature}. Experimentally it is determined by
NMR\cite{Klanjsek_NMR_3Dladder} and neutron diffraction
measurements \cite{Thielemann_ND_3Dladder}.
Theoretically the ladders are described by Luttinger liquid theory and their interladder coupling is treated in a mean-field approximation (Secs.~\ref{sec:luttinger_liquid} and~\ref{sec:mean-field}).
As shown in Fig.~\ref{fig:criticaltemperature}, the Luttinger liquid theory provides a remarkable description
of the transition to the transverse antiferromagnetic order at low
temperatures. The shape of $T_c(h^z)$ is almost perfectly reproduced, in agreement with both the
NMR\cite{Klanjsek_NMR_3Dladder} and the ND data\cite{Thielemann_ND_3Dladder}.
The comparison with the experiments determines the interladder coupling $J'$, the only adjustable parameter.
The simple mean-field approximation would
give a value of $J' \sim 20~{\rm mK}$. As discussed in Sec.~\ref{sec:transitiontemperature} mean-field tends to
underestimate the coupling and it should be corrected by
an essentially field independent factor. Taking this into account we obtain a coupling
of the order of $J' = 27~{\rm mK}$.

The order parameter in the antiferromagnetic phase can also be observed by
experiments. It shows a very interesting shape. At a pure experimental level
neutron diffraction and
NMR have some discrepancies as shown in
Fig.~\ref{fig:orderparameter}. These discrepancies can be attributed to the different temperatures at which the data has been taken,
and a probable underestimate of the temperature in the neutron diffraction experiments\cite{Thielemann_ND_3Dladder}. Indeed
the order parameter close to the critical magnetic field $h_{c2}$ is very sensitive to temperature, since the transition temperature drops steeply in this regime. Note that although
the NMR allows clearly for a more precise measurement of the
transverse staggered magnetization it cannot give its absolute
value. Thus the amplitude of the order parameter is fixed from the
neutron diffraction measurement. Even though a good agreement between the theoretical results and the experimental results is obtained, several questions concerning the deviations remain to be addressed.

First, the theoretical curve does not fully follow the shape of the experimental data. Particularly at high fields the experimental data shows a stronger decrease. A simple
explanation for this effect most likely comes from the fact that the
calculation is performed at zero temperature, while the
measurement is done at $40~{\rm mK}$. This is not a negligible temperature with
respect to $T_c$, in particular at magnetic fields close to $h_{c2}$. Extrapolation of the experimental data
to zero temperature\cite{Klanjsek_NMR_3Dladder} improves the
agreement. However, for a detailed comparison either lower temperature measurements or
a calculation of the transverse staggered magnetization at finite temperature would
be required. Both are quite difficult to perform and will clearly require further studies.

The second question comes from the amplitude of the
staggered magnetization. Indeed the experimental data seem to be slightly
above the theoretical curve, even if one uses the value $J' =
27~{\rm mK}$ for the interladder coupling. This is surprising since
one would expect that going beyond the mean-field approximation
could only reduce the order parameter. Naively, one would thus need
a larger coupling, perhaps of the order of $J' \sim
60-80~{\rm mK}$ to explain the amplitude of the order parameter. This is a much
larger value than the one extracted from the comparison of $T_c$. How to reconcile
these two values remains open. The various anisotropies and additional
small perturbations in the ladder Hamiltonian could
resolve part of this discrepancy. However, it seems unlikely that they result in a correction of $J'$ by a factor of about 2-3.
Another origin might be the presence of some level of frustration present in the
interladder coupling. Clearly more experimental and theoretical
studies are needed on that point.

\subsection{Inelastic neutron scattering}\label{sec:INS}

The inelastic neutron scattering (INS) technique is
a direct probe for dynamical spin-spin correlation functions. Measurements have been performed on the compound BPCB in the spin liquid
phase\cite{Savici_BPCB_INS,Thielemann_INS_ladder} (low magnetic
field) and in the gapless regime\cite{Thielemann_INS_ladder}.
Modeling the compound BPCB by two inequivalent uncoupled ladders  oriented along the two rung
vectors ${\bf d}_{1,2}$ (see Fig.~\ref{fig:structure}) the magnetic INS cross section\cite{lovesey_neutron_scattering} is given by the formula
\begin{multline}\label{equ:laddercross_section}
\frac{d^2\sigma}{d\Omega dE'}\propto \frac{q'}{q}|F(\mathbf{Q})|^2\sum_{i=1,2}\left\{\frac{1}{2}\left(1+\frac{{Q^z}^2}{\mathbf{Q}^2}\right)\right.
\\
\left[\cos(\mathbf{Q}\cdot \mathbf{d}_i)(S^{+-}_{21}+S^{-+}_{21})+S^{+-}_{11}+S^{-+}_{11}\right] \\
\left.+2\left(1-\frac{{Q^z}^2}{\mathbf{Q}^2}\right)
\left[\cos(\mathbf{Q}\cdot \mathbf{d}_i)S^{zz}_{21}+S^{zz}_{11}\right]\right\}.
\end{multline}
Here $\mathbf{Q}=(Q^x,Q^y,Q^z)=\mathbf{q}-\mathbf{q}'$ is the momentum transferred to the sample ($\mathbf{q}$, $\mathbf{q}'$ are the incident, scattered neutron momenta) and $\omega=E-E'$ is the transferred energy  ($E$, $E'$ are the incident, scattered neutron energies). The correlations $S_{ij}^{\alpha\beta}$ are understood to be evaluated at the momentum $\mathbf{Q}\cdot\mathbf{a}$ and frequency $\omega$, and are defined by
\begin{equation}
S_{ij}^{\alpha\beta}(\mathbf{Q}\cdot\mathbf{a},\omega)=\frac{ S_0^{\alpha\beta}(\mathbf{Q}\cdot\mathbf{a},\omega)\pm S_\pi^{\alpha\beta}(\mathbf{Q}\cdot\mathbf{a},\omega)}{2}
\end{equation}
($+$ sign if $i=j$ and $-$ sign if $i\neq j$) with $S^{\alpha\beta}_{q_y}$ defined at zero temperature in Eq.~(\ref{equ:correlation1}) and evaluated for a momentum $q={\bf Q}\cdot {\bf a}$ along the ${\bf a}$ unit cell vector (momentum along the ladder direction). The magnetic form factor $F({\bf{Q}})$ of the $\mathrm{Cu^{2+}}$ and the ratio $q'/q$ are corrected in the experimental data.

The INS cross section (\ref{equ:laddercross_section}) is
directly related to a combination of different correlation functions
$S_{q_y}^{\alpha\beta}$ with weights depending on the transferred momentum ${\bf Q}$ and the
magnetic field orientation. In the model definition (see
Sec.~\ref{sec:model}), the magnetic field ${\bf h}$ is pointing
along the $z$ direction. Aligning it to the ${\bf b}$ unit cell
vector and tuning ${\bf Q}$ in the ${\bf a^\star c^\star}$-plane
(${\bf a^\star}$, ${\bf b^\star}$ and ${\bf c^\star}$ are the
reciprocal vectors of ${\bf a}$, ${\bf b}$ and ${\bf c}$) allows one
to keep constant the prefactors in Eq.~\eqref{equ:laddercross_section} scanning the ${\bf
a^\star}$-momenta with the condition ${\bf Q}\cdot{\bf
d_i}=0\text{ or }\pi$ to target the symmetric or
antisymmetric part, respectively.

\begin{figure}
\begin{center}
\includegraphics[width=1\linewidth]{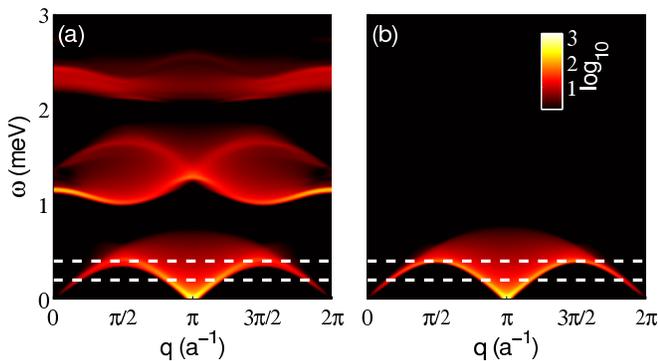}
\end{center}
\caption{(Color online): Theoretical momentum-energy dependent INS cross section for BPCB with $\mathbf{Q}\cdot\mathbf{d}_i=\pi$ ($i=1,2$) and $q={\bf Q}\cdot {\bf a}$ at $m^z=0.5$ in (a) a ladder system and (b) the spin chain mapping. The horizontal dashed lines correspond to the constant energy scans at $\omega=0.2,0.4 ~{\rm meV}$ shown in Fig.~\ref{fig:experimentalINS1}.\label{fig:theoreticalINS1}}
\end{figure}

\begin{figure*}
\begin{center}
\hspace{-1cm}
\includegraphics[width=0.75\linewidth]{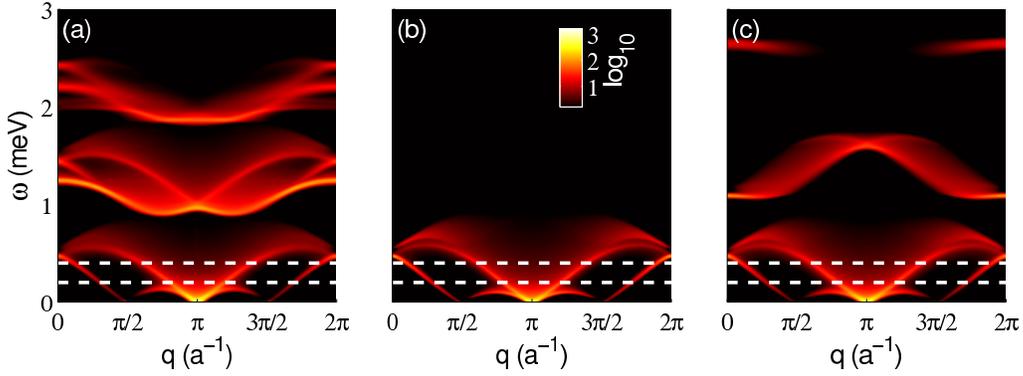}
\end{center}
\caption{(Color online): Theoretical momentum-energy dependent INS cross section for BPCB with $\mathbf{Q}\cdot\mathbf{d}_i=\pi$ ($i=1,2$) and $q={\bf Q}\cdot {\bf a}$, (a) at $m^z=0.25$ (b) at $m^z=0.25, 0.75$ in the spin chain mapping, (c) at $m^z=0.75$. The horizontal dashed lines correspond to the constant energy scans at $\omega=0.2,0.4 ~{\rm meV}$ plotted in Fig.~\ref{fig:experimentalINS2}.\label{fig:theoreticalINS2}}
\end{figure*}

We focus here on the antisymmetric
part for which the low energy spectra have already been
studied experimentally and theoretically\cite{Thielemann_INS_ladder}.
Theoretically the focus so far lay on the description by the spin chain mapping.
We compute here the INS cross section (\ref{equ:laddercross_section}) for the full ladder at $m^z=0.25,0.5,0.75$ using the correlations presented in Sec.~\ref{sec:dynamicalcorrelation}. The results are shown in Figs.~\ref{fig:theoreticalINS1}~and~\ref{fig:theoreticalINS2} and are compared to the results from the spin chain mapping.

As expected from expression (\ref{equ:laddercross_section}), it contains the different excitations
present in the spectra of $S_\pi^{zz}$ and $S_\pi^{\pm\mp}$ (cf.~Figs.~\ref{fig:zzcorrelationmz}.b, \ref{fig:pmcorrelationmz}.b and~\ref{fig:mpcorrelationmz}.b):
\begin{itemize}
 \item[(a)] The low energy continuum originates from the transversal correlations $S_\pi^{\pm\mp}$. It is qualitatively well described by the spin chain mapping that presents a symmetry with respect to half magnetization.
 \item[(b)] The continuum of excitations at energy\cite{\Jperptohz} $\sim J_\perp$ comes from the longitudinal correlation $S_\pi^{zz}$ and is not present in the spin chain mapping.
 \item[(c)] The continuum of excitations at energy\cite{\Jperptohz} $\sim 2J_\perp$ comes from the transversal correlation $S_\pi^{+-}$ and is not present in the spin chain mapping.
\end{itemize}

\begin{figure}
\begin{center}
\includegraphics[width=1\linewidth]{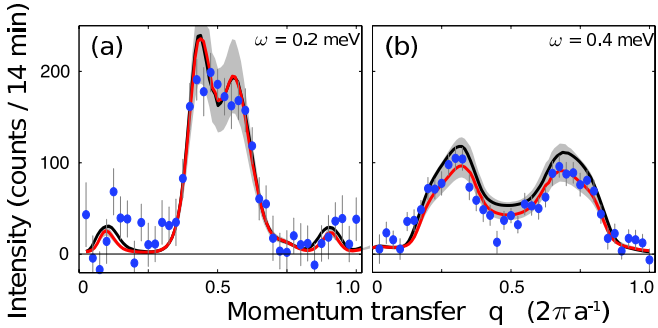}
\end{center}
\caption{(Color online): Inelastic neutron scattering intensity measured along
  ${\bf a^\star}$ of BPCB\cite{Thielemann_INS_ladder} with the momentum $\pi$ in the rung direction (${\bf Q}\cdot{\bf
d_i}=\pi$) at $h^z=10.1~{\rm T}$ ($m^z\approx0.5$) and $T=250~{\rm mK}$ after substraction of the zero-field background. In each panel, fixed energy scans (shown by white dashed lines in Fig.~\ref{fig:theoreticalINS1}) are plotted: (a) $\omega=0.2~{\rm meV}$, (b) $\omega=0.4~{\rm meV}$. The circles correspond to the experimental data. The red (black) solid lines are the $m^z=0.5$ theoretical data for the ladder (the spin chain mapping) convolved with the instrumental resolution. The shaded bands indicate the error bar in the experimental
determination of a single proportionality constant valid for all fields, energies, and wave vectors. The width of these areas combines the statistics of all our scans with uncertainties
in the exact magnetization values at the chosen
fields and in the convolution procedure.
\label{fig:experimentalINS1}}
\end{figure}

\begin{figure}
\begin{center}
\includegraphics[width=1\linewidth]{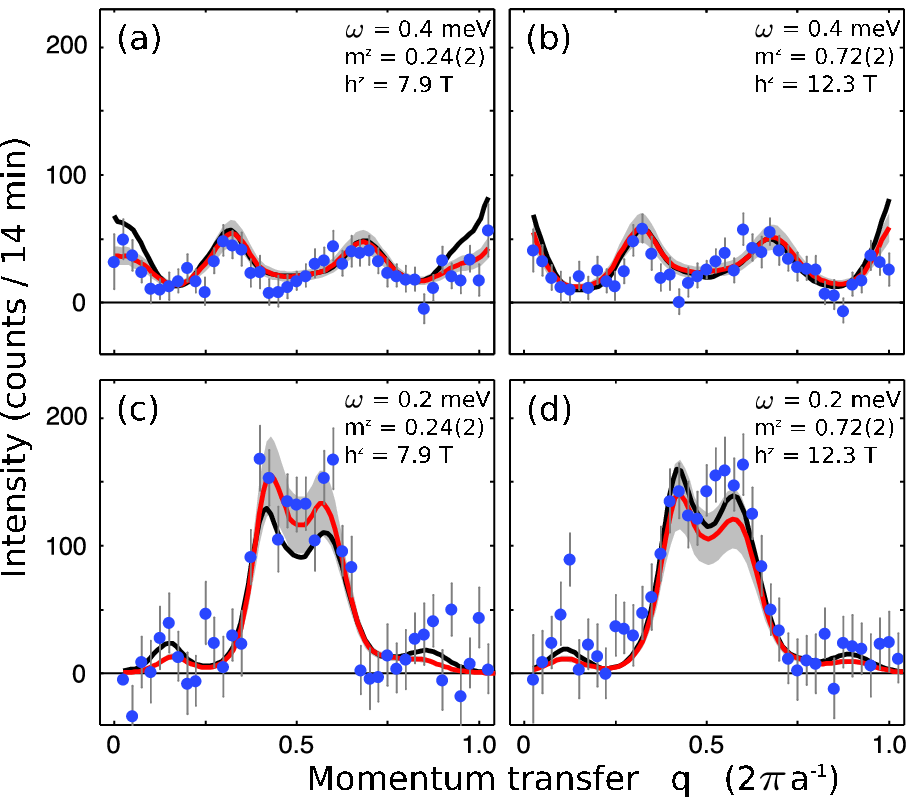}
\end{center}
\caption{(Color online): Inelastic neutron scattering intensity measured along $\bf a^\star$ of BPCB\cite{Thielemann_INS_ladder} with a $\pi$ momentum in the rungs (${\bf Q}\cdot{\bf
d_i}=\pi$) at $T=250~{\rm mK}$ after substraction of the zero-field background. In each panel, cuts at fixed energy (shown by white dashed lines in Fig. \ref{fig:theoreticalINS2}) are plotted: (a) $\omega=0.4~\mathrm{meV}$ and $m^z = 0.24$,
(b) $\omega=0.4~\mathrm{meV}$ and $m^z = 0.72$,
(c) $\omega=0.2~\mathrm{meV}$ and $m^z = 0.24$,
(d) $\omega=0.2~\mathrm{meV}$ and $m^z = 0.72$. The circles correspond to the experimental data. The solid red (black) curves are the theoretical data for the ladder (the spin chain mapping) convolved with the instrumental resolution. The shaded bands indicate the error bar in the experimental
determination of a single proportionality constant valid for all fields, energies, and wave vectors. The width of these areas combines the statistics of all our scans with uncertainties
in the exact magnetization values at the chosen
fields and in the convolution procedure.\label{fig:experimentalINS2}}
\end{figure}

The main features of the low energy continuum (a) are well covered by the spin chain mapping\cite{Thielemann_INS_ladder}. However, slight differences between the low
energy excitations in the spin ladder and the spin chain are still visible (cf.~also Sec.~\ref{sec:lowenergyexcitations}). They can even be distinguished in the experimental data as shown in Figs.~\ref{fig:experimentalINS1}~and~\ref{fig:experimentalINS2} where some cuts at fixed energy $\omega=0.2,0.4{\rm
meV}$ are plotted. The
INS measured intensity is directly compared to the theoretical
cross section (\ref{equ:laddercross_section}) computed for the
ladder and the spin chain mapping at $m^z=0.24,0.5,
0.72$ convolved with the instrumental resolution. The amplitude is fixed by fitting one proportionality constant for all fields, energies, and wave vectors.

These scans at fixed energy present peaks when the lower edge of the continua (related to the correlations $S_\pi^{\pm\mp}$) is crossed (see dashed white lines in Figs.~\ref{fig:theoreticalINS1}~and~\ref{fig:theoreticalINS2}). As one can see, the theoretical curves for the ladder
and the spin chain both reproduce well the main features in the experimental data and only small differences are present:
\begin{itemize}
 \item[-] The spectral weight intensity at $m^z=0.5$ and $\omega=0.4~{\rm meV}$ (in Fig.~\ref{fig:experimentalINS1}.b) is slightly overestimated by the spin chain mapping.
\item[-] The height of the two central peaks at $m^z=0.24$ and $\omega=0.2~{\rm meV}$ (in Fig.~\ref{fig:experimentalINS2}.c) is underestimated by the spin chain mapping.
\end{itemize}

Whereas the low energy excitations (a) only showed a slight asymmetry with respect to the magnetization, a very different behavior can be seen in the high energy part (b)-(c). Indeed, the
high energy part of the INS cross section (Fig.~\ref{fig:theoreticalINS2}) is very asymmetric with respect to half magnetization. As we discussed in
Sec.~\ref{sec:dynamicalcorrelation}, these excitations are due
to the high energy triplets that can be excited in $S_\pi^{zz}$
and $S_\pi^{+-}$ (see Fig.~\ref{fig:zzcorrelationmz}.b and
Fig.~\ref{fig:pmcorrelationmz}.b) and are totally neglected in
the spin chain mapping. Their corresponding spectral
weight is of the same order than the low energy spectra, and
thus should be accessible experimentally. It would be
very interesting to have an experimental determination of this
part of the spectrum, since as we have seen it contains characteristic information on the system itself and related to itinerant systems via the various mappings.

\section{Conclusions}\label{sec:conclusions}

In this paper we have looked at the thermodynamic and dynamical
properties of weakly coupled spin ladders under a magnetic field. This was
done by a combination of analytical techniques, such as Bethe
ansatz, bosonization and Luttinger liquid theory, and numerical
techniques such as DMRG and quantum Monte Carlo.
Using this combination of techniques we were able to explore
the physical properties in the three main regions of the phase
diagram of such spin ladders under a magnetic field, namely: a)
a gapped spin liquid at low fields; b) a massless phase at
intermediate fields $h_{c1} < h^z < h_{c2}$; c) a saturated phase
at larger fields. In addition to the theoretical analysis we compared our findings to the experimental results on
the compound BPCB ($\mathrm{(C}_5\mathrm{H}_{12}\mathrm{N)}_2\mathrm{CuBr}_4$) which is
an excellent realization of such ladder systems.

For thermodynamics we computed the magnetization and specific
heat of the system as a function of temperature and magnetic
field. The extension of the DMRG technique to finite
temperature allows us to compute these quantities with an
excellent accuracy. In the gapless phase the low energy part
of the specific heat agrees well with the prediction of the
Luttinger liquid theory, which is the low energy theory
describing most of massless one dimensional systems. At higher
temperatures, the numerical solution is needed to capture the
precise structure of the peaks in the specific heat, that
reflect the presence of the excited states in the ladder.
Comparison of the theoretical calculations with the measured
magnetization and specific heat proves to be remarkable. This good agreement confirms that the ladder model is indeed
a faithful description of the compound BPCB. It also gives
direct access, via the maxima in magnetization and peaks in the
specific heat to the approximate region of applicability of the Luttinger
liquid description. For the low energy dynamics we
used a combination of the numerical techniques to determine the
Luttinger liquid parameters and then the analytical description
based on Luttinger liquids to compute the dynamical spin-spin
correlation functions. This allowed to extract the NMR
relaxation rate and the one dimensional antiferromagnetic
transverse susceptibility. If the ladders are very weakly
coupled, which is the case in the considered material, the divergence
of the susceptibility leads to a three dimensional
antiferromagnetic order at low temperatures in the direction
transverse to the applied magnetic field. We computed this
transition temperature and the order parameter at zero
temperature. Comparison with the measured experimental
quantities both by NMR and neutron diffraction proved again to be
remarkable. This excellent agreement between theory and
experiment for these quantities as a function of the magnetic
field allows to \emph{quantitatively} test the Luttinger liquid
theory. Indeed it shows that several different correlations are
indeed totally described by the knowledge of the two Luttinger
liquid parameters (and the amplitudes relating the microscopic
operators to the field theory one). This is something that could not
really be tested previously since either the microscopic
interactions were not known in details, leaving the Luttinger
parameters as adjustable parameters, or only one correlation
function could be measured in a given experiment, not allowing
to test for universality of the description.

We also gave a detailed description of the dynamical spin-spin
correlations, for $T=0$ using the time dependent DMRG method,
for a wide range of energies and all momenta. The excitations reveal many important informations on the system and are well suited to characterize it. In particular we show the interesting evolution of the excitations in the system with the magnetic field and with different coupling strengths. Quite interestingly the intermediate energy
part can be related to the excitations of a t-J model and shows thereby features of itinerant systems. We also showed that the
dynamical correlations of the ladder posses characteristic high
energy features that are clearly distinct from the
corresponding spectrum for spin chains.

The numerical calculation is efficient for the high and intermediate energy
part of the spectrum for which the Luttinger liquid description
cannot be applied. We showed that the two methods, numerics and
LL have enough overlap, given the accuracy of our calculation
so that we can have a full description of the dynamical
properties at all energies. This allowed us to use each of the method in
the regime where it is efficient. In particular, in the present study
we did not push the numerical calculations to try to obtain the exact behavior at low energies,
but focus on the high and intermediate energy regime. We used the analytical description coupled
to the numerical determination of the Luttinger parameters to obtain a very accurate low
energy description. We made the connection between our
results and several analytical predictions. In particular at
intermediate-low energy our calculation agrees with the
Luttinger liquid prediction of incommensurate points and
behavior (divergence, convergence) of the correlations.

We compared our numerical results with existing INS data on
the compound BPCB and found excellent agreement. It is
rewarding to note that the resolution of our dynamical
calculation is, in energy and momentum, at the moment better
than the one of the experiment. The comparison between theory
and experiment is thus essentially free of numerical error bars.
Given the current resolution of the INS
experiment it is difficult to distinguish in the low energy
part of the spectrum the difference between the dynamical
correlation of the true ladder and the one of an anisotropic
spin $1/2$ system, which corresponds to the strong rung
exchange limit. More accurate experiments would be
desirable in this respect. An alternative route is to probe experimentally the
high energy part of the spectrum, since these high energy excitations contain many characteristic features of the underlying model.

On the conceptual side and also in connection with the
compound BPCB several points remain to be investigated.
An extension of the dynamical results to finite temperature would be desirable. This could be used to study different
effects such as the interesting shifts and damping of the triplet modes
with temperature that have been observed in three dimensionnal gapped system\cite{Ruegg_quantum_statistics_triplons}.
A second important
step would be to improve the description of the quasi one
dimensional systems, by including in a mean-field way the
effect of the other ladders in the numerical study. This is
something we already partly performed, but the extension to the dynamical
quantities remains to be done. This is specially important close
to the quantum critical points $h_{c1}$ and $h_{c2}$ where the
interladder coupling becomes crucial and the system undergoes a
dimensional crossover between a one dimensional and a higher
dimensional (three dimensional typically) behavior.
Understanding such a crossover is a particularly challenging
question since the system goes from a description for which an
image of essentially free fermions applies (in the one
dimensional regime) to one for which a description in terms of
essentially free bosons (the three dimensional regime) applies.

One step further would be the extension of the investigation of the combined numerical-analytical methods we used here to other systems including ladder systems with certain asymmetries or in the presence of disorder.

\acknowledgments

We thank T.~Barthel, C.~Berthod, J.-S.~Caux, V.~V.~Cheianov, A.~Furusaki, U.~Schollw\"ock, S.~Trebst, and  S.~White for useful discussions and specially J.-S.~Caux for the data exchange. We acknowledge support of the 'Triangle de la Physique', the ANR 'FAMOUS', and of the Swiss NSF, under MaNEP and Division II. T.~Giamarchi would like to thank the Aspen Center for Physics, where part of this work was completed. M. Klanj\v{s}ek acknowledges support from the Centre of Excellence EN$\rightarrow$FIST, Dunajska 156, SI-1000 Ljubljana, Slovenia.

\appendix

\section{Strong coupling expansion of a single spin ladder}\label{sec:tjmodelmapping}

In this appendix we show how the spin ladder Hamiltonian~\eqref{equ:spinladderhamiltonian} at strong coupling ($\gamma\ll1$) can be expressed in bosonic operators acting on single bonds introduced in Ref.~\onlinecite{sachdev_bot}. This representation allows a classification of the excitations due to their position in energy. We first derive perturbatively an effective system based on this Hilbert space organization by energy sectors. We introduce the Schrieffer-Wolff transformation that maps the physical system to the effective, and approximate it using a strong coupling expansion. Thus, we evaluate the rung densities of the ground state in the spin liquid, and derive an effective theory for the gapless regime. Furthermore we evaluate the deviation of the LL parameters from the spin chain mapping.

\subsection{Strong coupling expansion}

The four-dimensional Hilbert space on each rung $l$ is spanned by the states $\ket{s},$ $\ket{t^+},$ $\ket{t^0},$ and $\ket{t^-},$ (cf.~Eqs.~\eqref{equ:smult} and~\eqref{equ:tmult}), obtained by applying the boson creation operators $s^\dagger_l,$ $t^\dagger_{l,+},$ $t^\dagger_{l,0},$ and $t^\dagger_{l,-}$ to a vacuum state. A hardcore boson constraint applies on each rung $l$, i.e.
\begin{equation}\label{equ:hcfull}
\varrho_{l,s}+\varrho_{l,+}+\varrho_{l,0}+\varrho_{l,-}=1
\end{equation}
where $\varrho_{l,s}=s^\dagger_ls_l$ and $\varrho_{l,k}=t^\dagger_{l,k}t_{l,k}$, $k=\pm,0$.

While the Hamiltonian on the rung $H_\perp$~\eqref{equ:Hperp} is quadratic in the boson operators
\begin{equation}\label{equ:Hbosonperp}
H_\perp=\sum_{l=1}^L \left[(1-h^z/J_\perp)\varrho_{l,+} + \varrho_{l,0}+ (1+h^z/J_\perp)\varrho_{l,-}\right] -\frac34L,
\end{equation}
the chain Hamiltonian $H_\parallel$~\eqref{equ:Hparallel} is quartic, and its structure is quite complex. The advantage of the boson representation reveals when considering the case of small $\gamma$. In that case we perform a Schrieffer-Wolff transformation of the spin ladder Hamiltonian~\eqref{equ:spinladderhamiltonian}
\begin{equation}\label{equ:heff}
H_\text{eff}=e^{i\gamma A}He^{-i\gamma A}.
\end{equation}
The Hermitian operator $A$ can be expanded in powers of $\gamma$
\begin{equation}\label{equ:Aexpansion}
A= A_1+\gamma A_2+\cdots.
\end{equation}
Thus $H_\text{eff}$ can be written in orders of $\gamma$ as
\begin{equation}\label{equ:heffexp}
J_\perp^{-1} H_\text{eff} = H_\perp+\gamma  H^{(1)} + \gamma^2  H^{(2)}+ \cdots,
\end{equation}
where
\begin{align}\label{equ:H1_def}
&H^{(1)} = H_\parallel +i[A_1,H_\perp], \\
&H^{(2)} = i[A_2,H_\perp] -\frac12 [A_1,[A_1,H_\perp]]+i [A_1,H_\parallel],\label{equ:H2_def}
\end{align}
etc. Through this expansion, the unitary transform $e^{i\gamma A}$ can be perturbatively determined computing the $A_k$ recursively in order to eliminate the transitions between the energy sectors of excitations in $H_{\text{eff}}$. Since the first term $J_\perp H_\perp$ in Eq.~\eqref{equ:heffexp} leads to a separation of excitations on the order of the energy scale $J_\perp$ (cf.~Fig.~\ref{fig:phasediagramm}.a) the decoupled bond limit provides the effective Hilbert space that contributes to each energy sector. The second term $J_\parallel H^{(1)}$ causes broadening of these bands on the order of $J_\parallel$ and can induce a complex structure within the energy bands. To obtain the desired expansion up to the first order in $\gamma$ we choose
\begin{multline}\label{equ:A1bos}
A_1=\frac{i}{4}\sum_{l}
s_l^\dagger s_{l+1}^\dagger \left(t_{l,0} t_{l+1,0} - t_{l,+}t_{l+1,-} \right.\\
\left. - t_{l,-} t_{l+1,+}\right)+\text{h.c.},
\end{multline}
where h.c.\ stands for the Hermitian conjugation.

\subsection{Rung state density in the spin liquid}

In the spin liquid phase, the decoupled bond limit provides the effective ground state
$\ket{0_{\text{eff}}}=\ket{s\cdots s}$ which is related to the physical ground state by
\begin{equation}
\ket{0}=e^{-i\gamma A}\ket{0_{\text{eff}}}.
\end{equation}
So the triplet density of the ground state $\langle \rho_k\rangle$ (with $k=\pm,0$) is given by
\begin{equation}\label{equ:tripletdensity}
\langle \rho_k\rangle=\langle\varrho_{l,k}\rangle=\bra{s\cdots s}e^{i\gamma A}\varrho_{l,k}e^{-i\gamma A}\ket{s\cdots s}.
\end{equation}
Using Eq.~\eqref{equ:A1bos}, and keeping only the non-vanishing corresponding terms in~\eqref{equ:tripletdensity} up to second order we get
\begin{equation}\label{equ:tripletdensity2}
\langle \rho_k\rangle\cong \gamma^2\bra{s\cdots s}A_1\varrho_{l,k} A_1\ket{s\cdots s}=\frac{\gamma^2}{8}.
\end{equation}
In the case of the compound BPCB (see Sec.~\ref{sec:bpcb}) this expansion gives $\langle \rho_k\rangle\cong0.01$, and due to the hardcore boson constraint (Eq.~\eqref{equ:hcfull}) $\langle \rho_s \rangle=\langle\varrho_{l,s}\rangle\cong0.97$. Even though we took into account only the first order term for $A$ in Eq.~\eqref{equ:Aexpansion}, this approximation of the triplet density differ from the direct numerical computations (in Fig.~\ref{fig:tripletdensity}) of only $\sim20\%$.

\subsection{Effective Hamiltonian in the gapless regime}

The first order term $H^{(1)}$ of the effective Hamiltonian~\eqref{equ:heffexp} is computed substituting~\eqref{equ:A1bos} into~\eqref{equ:H1_def}. This leads to $H^{(1)}$ in the form
\begin{equation}\label{equ:1storderH}
H^{(1)} =\sum_{k=0}^4 \underbrace{\frac12\sum_{l} \mathcal{H}_{k,l}^{(1)}}_{=H_k^{(1)}}
\end{equation}
where
\begin{equation} \label{equ:H0}
\mathcal{H}_{0,l}^{(1)}=s_{l+1}^\dagger t_{l,+}^\dagger   t_{l+1,+}s_{l}+ \frac12 \varrho_{l+1,+} \varrho_{l,+} + \text{h.c.},
\end{equation}
\begin{equation}\label{equ:HJperp}
\mathcal{H}_{1,l}^{(1)}=s_{l+1}^\dagger t_{l,0}^\dagger t_{l+1,0} s_{l}
+  t_{l+1,+}^\dagger t_{l,0}^\dagger t_{l+1,0} t_{l,+} + \text{h.c.},
\end{equation}
\begin{multline}\label{equ:H2Jperp}
\mathcal{H}_{2,l}^{(1)}=s_{l+1}^\dagger  t_{l,-}^\dagger t_{l+1,-} s_{l}  - \frac12\left( \varrho_{l+1,+}\varrho_{l,-}
+ \varrho_{l+1,-}\varrho_{l,+}\right)\\
+ t_{l+1,0}^\dagger t_{l,0}^\dagger t_{l+1,-}t_{l,+}
+ t_{l+1,0}^\dagger t_{l,0}^\dagger t_{l+1,+}t_{l,-} +\text{h.c.},
\end{multline}
\begin{equation}\label{equ:H3Jperp}
\mathcal{H}_{3,l}^{(1)}= t_{l+1,0}^\dagger t_{l,-}^\dagger t_{l+1,-}t_{l,0} +\text{h.c.},
\end{equation}
and
\begin{equation}\label{equ:H4Jperp}
\mathcal{H}_{4,l}^{(1)}=\varrho_{l+1,-}\varrho_{l,-}.
\end{equation}
Here we regrouped the terms such that each $J_\perp H_k^{(1)}$ acts on  the corresponding energy sector $kJ_\perp$, $k=0,1,\ldots,4$ in the gapless regime.  Note that in each sector $A_1=0$ such that to the considered order in $\gamma$ the Hamiltonian~\eqref{equ:spinladderhamiltonian} corresponds to the effective Hamiltonian.

\subsubsection{Low energy sector}\label{sec:sec_sector_0}
When focusing on the low energy sector, the $\ket{s}$ and the $\ket{t_+}$ modes dominate the behavior and we can assume a vanishing density of $\ket{t_0}$ and $\ket{t_-}$ triplets. Thus the hardcore boson constraint~\eqref{equ:hcfull} simplifies to
\begin{equation}
\varrho_{l,s} +\varrho_{l,+}  =1
\end{equation}
and the rung Hamiltonian~\eqref{equ:Hbosonperp} to
\begin{equation}\label{equ:Hbpreduced}
H_\perp = (1-h^z/J_\perp) \sum_{l} \varrho_{l,+} -\frac34L.
\end{equation}
Further the only contribution to the first order term in $\gamma$ comes from $H_0^{(1)}$. Taking this into account we obtain from Eq.~\eqref{equ:heffexp} for the Hamiltonian~\eqref{equ:spinladderhamiltonian} in the lowest energy sector
\begin{equation}\label{equ:H_sector_0_init}
 H= J_\perp H_\perp +J_\parallel H_0^{(1)},
\end{equation}
where $H_\perp$ is given by Eq.~\eqref{equ:Hbpreduced} and $H_0^{(1)}$ by Eq.~\eqref{equ:H0}. Following Ref.~\onlinecite{mila_ladder_strongcoupling}, we map the two low energy modes on the two states of a pseudo spin (Eq.~\eqref{equ:Hsreduced}) and replace the boson operators $s^\dagger$ and $t_+^\dagger$ with the spin-$1/2$ operators
\begin{equation}\label{equ:operatormapping}
\tilde S_l^+=t^\dagger_{l,+}s_{l}, \quad \tilde S_l^- = s^\dagger_{l} t_{l,+}, \quad \tilde S_l^z = \varrho_{l,+} -\frac12.
\end{equation}
The effective Hamiltonian is the spin-$1/2$ XXZ Heisenberg Hamiltonian, Eq.~\eqref{equ:strongcouplinghamiltonian}.

\subsubsection{Sector of energy $J_\perp$}\label{sec:sec_sector_Jperp}
The effective Hilbert space of the $J_\perp$ energy sector corresponds to a single $\ket{t^0}$ triplet excitation lying in a sea of singlets $\ket{s}$ and triplets $\ket{t^+}$. The effective Hamiltonian up to first order in $\gamma$ is given by
\begin{equation}\label{equ:H_sector_Jperp_init}
H= J_\perp H_\perp +J_\parallel \left(H_0^{(1)}+H_1^{(1)}\right).
\end{equation}
The excitation $\ket{t^0}$ can be interpreted as a single hole excitation in a spin chain formed by $|s\rangle$ and $|t^+\rangle$. Each rung state of this sector is identified with
\begin{equation}
|\tilde\downarrow\rangle =|s\rangle,  \quad |\tilde\uparrow\rangle = |t^+\rangle,\quad |0\rangle = |t^0\rangle. \label{equ:Hsreduced1}
\end{equation}

In this picture the Hamiltonian \eqref{equ:H_sector_Jperp_init} can be mapped onto the anisotropic t-J model
\begin{equation}
H_\text{t-J}= H_\text{XXZ} + H_\text{t} + H_\text{s-h}+\epsilon.
\end{equation}
where $\epsilon=(J_\perp+h^z)/2$ is an energy shift.

The hopping term
\begin{equation}
H_\text{t}=\frac{J_\parallel}{2}\sum_{l,\sigma}\left(c_{l,\sigma}^\dagger c_{l+1,\sigma}^\phd +c_{l+1,\sigma}^\dagger c_{l,\sigma}^\phd \right)
\end{equation}
stems from the term $J_\parallel H_1^{(1)}$ in Eq.~\eqref{equ:H_sector_Jperp_init}. Here $c^\dagger_{l,\sigma}$ ($c_{l,\sigma}$) is the creation (annihilation) operator of a fermion with pseudo spin $\sigma=\tilde\uparrow,\tilde\downarrow$ at the site $l$. Note that although we are dealing here with spin states, it is possible to 
faithfully represent the three states of each site's Hilbert space ($\ket{s}$, $\ket{t^+}$, $\ket{t^0}$) using a fermion representation. 

Additionally a nearest neighbor density-density term between the up spin and the hole arises
\begin{equation}
H_\text{s-h}=-\frac{J_\parallel}{4}\sum_l\left[n_{l,h}n_{l+1,\tilde\uparrow}+n_{l,\tilde\uparrow}n_{l+1,h}\right].
\end{equation}
Here $n_{l,h}$ is the density operator of the hole at site $l$.
This term stems from the nearest-neighbour interaction between the $\ket{t_+}$ triplets, i.e.~the second term in Equ.~\eqref{equ:H0}.
Mapping this term onto a spin chain in the presence of a hole leads to an interaction term between the hole and the spin-up state. Note that this is in contrast to the usual mapping on a spin chain without a hole in which case it would only causes a shift in energy.

\subsection{Second order perturbation and Luttinger liquid parameters}\label{sec:secondorder_LL}

The second order term $H^{(2)}$~\eqref{equ:H2_def} of the expansion~\eqref{equ:heffexp}, contains a huge amount of terms. Nevertheless considering the low energy sector, only the following terms
\begin{multline}\label{equ:H2}
H^{(2)}_0=-\frac{3}{8}\sum_{l}\varrho_{l,s}\varrho_{l+1,s}\\
-\frac{1}{8}\sum_{l}\left(t_{l-1,+}s_{l-1}^\dagger\varrho_{l,s}t^\dagger_{l+1,+}s_{l+1}+\text{h.c.}\right)
\end{multline}
are important. The first term in Eq.~\eqref{equ:H2} is a singlet density-density interaction which can be absorbed into the coupling of the XXZ chain, and the second term is a conditionnal hopping\cite{picon_spindimer}. In order to study the effects of $H^{(2)}_0$ on the LL parameters $u$ and $K$ (see Fig.~\ref{fig:LLparameter}), we first replace the boson operators with the spin-$1/2$ operators (Eq.~\eqref{equ:operatormapping}). So the Hamiltonian~\eqref{equ:spinladderhamiltonian} in the low energy sector becomes
\begin{multline}\label{equ:Hxxz_2nd_order}
H=H_\text{XXZ}-\frac{1}{8}\sum_{l}\left[\tilde S_{l-1}^- \left(\frac{1}{2}-\tilde S^z_l\right)\tilde S_{l+1}^++{\rm h.c.}\right]+\text{const}
\end{multline}
where $H_\text{XXZ}$ is the XXZ Heisenberg chain Hamiltonian (Eq.~\eqref{equ:strongcouplinghamiltonian}) with the corrected parameters
\begin{equation}\label{equ:Hxxz_newparam}
\Delta^{(2)} =  \frac{1}{2}-\frac{3}{8}\gamma\quad,\quad \tilde {h^z}^{(2)}=\tilde h^z-\frac{3}{8}J_\perp\gamma^2
\end{equation}
up to the second order in $\gamma$. For the BPCB parameters (Eq.~\ref{equ:couplingratio2}) $\Delta^{(2)}\cong 0.4$ instead of $\Delta=0.5$ for the spin chain mapping (first order approximation). The LL parameters $u$, $K$ and $A_x$ of $H_\text{XXZ}$ with the anisotropy $\Delta^{(2)}$ are computed, and we treat the three terms interaction (conditional hopping) approximating $1/2-\tilde S^z_l\cong 1/2-\tilde m^z$ (mean-field approximation). The remaining term is then bosonized using the expression~\eqref{equ:luttingeroperator1} for $\tilde S^{\pm}(x=l)$. It leads to the corrected LL parameter $\tilde u$ and $\tilde K$ of the Hamiltonian~\eqref{equ:Hxxz_2nd_order} through the relations
\begin{equation}\label{equ:newLLparam2}
\left\{
\begin{array}{lll}
\tilde u\tilde K&=&u K + 2\pi\gamma^2J_\perp A_x\left(1/2-\tilde m^z\right)\\
\tilde u/\tilde K&=&u/K
\end{array}
\right..
\end{equation}
The corrected $\tilde u$ and $\tilde K$ are plotted in Fig.~\ref{fig:LLparameter} and clearly show the asymmetric signature of the full ladder parameters induced by the conditional hopping term in~\eqref{equ:Hxxz_2nd_order}. Note that the lack of convergence $K\rightarrow 1$  when $m^z=\tilde m^z+1/2\rightarrow0$ is obviously an artefact of the mean-field approximation $1/2-\tilde S^z_l\cong 1/2-\tilde m^z$ and the big sensitivity in the $K$ determination with the ratio of Eq.~\eqref{equ:newLLparam2}.

\section{Luttinger liquid}\label{sec:LLappendix}

In the following, we present several properties of the Luttinger liquid providing a quantitative description of the low energy physics of the gapless regime in the spin ladder model~\eqref{equ:spinladderhamiltonian} and the spin chain mapping~\eqref{equ:strongcouplinghamiltonian}. We start with the determination of the parameters that totally characterize the LL description. Then we summarize the spin-spin correlation functions deduced from the LL theory at zero and finite temperature.

\subsection{Luttinger liquid parameter determination}\label{sec:LLparameters}

In this appendix we detail the determination of the LL parameters $u$, $K$ and the prefactors $A_x$, $B_x$ and $A_z$ (see Eqs.~\eqref{equ:luttingerliquid}, \eqref{equ:luttingeroperator1} and~\eqref{equ:luttingeroperator2}). The parameters $K$, $A_x$, $B_x$ and $A_z$ and their dependence on the magnetic field have been previously determined in Refs.~\onlinecite{hikihara_LL_ladder_magneticfield,Usami_LL_parameter,furusaki_correlations_xxz_magneticfield} for different values of the couplings than considered here.  We obtain these parameters in two steps\cite{Klanjsek_NMR_3Dladder,Klanjsek_3Dladder_Ax}:
\begin{itemize}
 \item[(i)] We determine the ratio $u/K$ from its relation to the static susceptibility, Eq.~\eqref{equ:uoverKcalculation}, which is a quantity numerically easily accessible with DMRG or Bethe ansatz (for the spin chain only).
 \item[(ii)] The parameters $K$ and the prefactors $A_x$, $B_x$ and $A_z$ are extracted by fitting numerical results for the static correlation functions obtained by DMRG  with their analytical LL expression. For the spin chain, it is also possible to deduce the product $uK$ from the magnetic stiffness computed by Bethe ansatz\cite{haldane_luttinger,giamarchi_book_1d}.
\end{itemize}
We give the relations used for the spin chains only, since from these the relations for the spin ladders can be easily inferred using the following relations
\begin{multline*}
 m^z\rightarrow \tilde m^z +\frac{1}{2}\quad ,\quad\frac{\partial m^z}{\partial h^z}\rightarrow \frac{\partial \tilde m^z}{\partial \tilde h^z}\\
\langle S^x_{l,\pi} S^x_{l',\pi}\rangle\rightarrow2\langle \tilde S^x_l \tilde S^x_{l'}\rangle  \quad ,\quad
\langle S^z_{l,0} \rangle\rightarrow\langle \tilde S^z_l \rangle+\frac{1}{2}\\
 \langle S^z_{l,0} S^z_{l,0'}\rangle\rightarrow\langle \tilde S^z_l \tilde S^z_{l'}\rangle+\frac{1}{2}\left(\langle \tilde S^z_l\rangle+\langle \tilde S^z_{l'}\rangle\right)+\frac{1}{4}.
\end{multline*}

\subsubsection{Susceptibility and $u/K$}\label{sec:uK}
The LL theory predicts that the static susceptibility $\frac{\partial \tilde m^z}{\partial \tilde h^z}$ is related to the ratio $u/K$ through the relation\cite{haldane_luttinger,giamarchi_book_1d}
\begin{equation}\label{equ:uoverKcalculation}
\frac{u}{K}=\frac{1}{\pi \frac{\partial \tilde m^z}{\partial \tilde h^z}},
\end{equation}
We numerically compute the static susceptibility using DMRG or Bethe ansatz (for the spin chain only) and infer the ratio $u/K$ with a negligible error.

\subsubsection{Static correlation functions}
For the extraction of the parameters $K$, $A_x$, $B_x$ and $A_z$, we fit numerical DMRG results for the correlation functions $\langle \tilde S^x_l\tilde S^x_{l'}\rangle$, $\langle \tilde S^z_l\tilde S^z_{l'}\rangle$ and the local magnetization $\langle \tilde S^z_l\rangle$ with their analytical expression for finite system size, Eqs.~(11), (12) and~(13) in Ref.~\onlinecite{hikihara_LL_ladder_magneticfield} respectively. In the limit of infinite system size and far from the boundaries these LL correlations simplify to the well known power law decay
\begin{equation}
\label{equ:xxcorrelationsimplify}
\langle \tilde S^x_l\tilde S^x_{l'}\rangle=A_x\frac{(-1)^{l-l'}}{|l-l'|^{\frac{1}{2K}}}- B_x(-1)^{l-l'}\frac{\cos[q(l-l')]}{|l-l'|^{2K+\frac{1}{2K}}}
\end{equation}
\begin{equation}\label{equ:zzcorrelationsimplify}
\langle \tilde S^z_l\tilde S^z_{l'}\rangle=\tilde m^{z\,2}+A_z(-1)^{l-l'}\frac{\cos[q(l-l')]}{|l-l'|^{2K}}-\frac{K}{2\pi^2|l-l'|^2}
\end{equation}
and the local magnetization becomes constant,
$\langle \tilde S^z_l\rangle= \tilde m^z$.

\begin{figure}
\begin{center}
\includegraphics[width=0.8\linewidth]{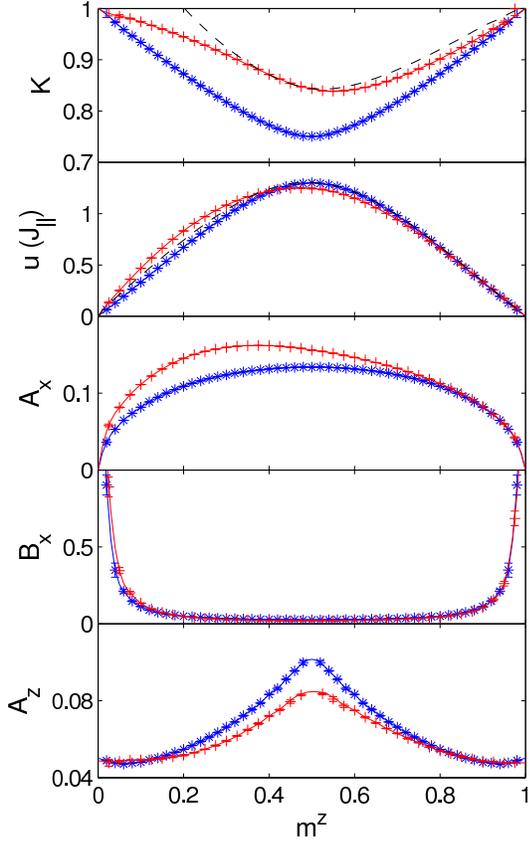}
\end{center}
\caption{(Color online): LL parameters $u$, $K$ and the prefactors of the spin operators $A_x$, $B_x$, $A_z$ versus
the magnetization per rung $m^z$ computed for a spin ladder with the BPCB couplings~\eqref{equ:couplings}~(red crosses) and for the spin chain mapping
(blue stars). The strong coupling expansion of $u$ and $K$ up to the second order in $\gamma$ (discussed in appendix~\ref{sec:secondorder_LL}) is plotted in black dashed lines.\label{fig:LLparameter}}
\end{figure}

We first fit the transverse correlation ($xx$-correlation $\langle \tilde S^x_l\tilde S^x_{l'}\rangle$) to extract the parameters $K$, $A_x$, and $B_x$. Then we use the previously extracted value for $K$ to fit the longitudinal correlation ($zz$-correlation $\langle \tilde S^z_l\tilde S^z_{l'}\rangle$) and the magnetization, $\langle \tilde S^z_l\rangle$, which allow us to determine $A_z$. The values determined by both fits are very close and in Fig.~\ref{fig:LLparameter} the average value of both is shown.

All the results presented in Fig.~\ref{fig:LLparameter} were done for $L=200$ and several hundred DMRG states after an average on the four sets of used data points in the fit $10<l,l'<170$, $30<l,l'<170$,  $10<l,l'<190$, $30<l,l'<190$. The error bars correspond to the maximum discrepancy of these four fits from the average. We further checked that different system lengths lead to similar results.

The LL parameters of the ladder system~\eqref{equ:spinladderhamiltonian} for the BPCB couplings~\eqref{equ:couplings} are
presented in Fig.~\ref{fig:LLparameter} as a function of the
magnetization per rung. Additionally we show the parameters of the spin chain mapping (computed for the spin chain Hamiltonian~\eqref{equ:strongcouplinghamiltonian}) for comparison.
When the ladder is just getting magnetized, or when the ladder is almost fully polarized,
$K\rightarrow 1$ (free fermion limit) and $u\rightarrow 0$  (because of the low density of
triplons in the first case, and low density of singlets in the second case).
For the spin chain mapping,
the reflection symmetry around $m^z=0.5$ arises from the
symmetry under $\pi$ rotation around the $x$ or $y-$axis of the spin chain.
This symmetry has no reason to be present in the original ladder model,
and is an artefact of the strong coupling limit, when truncated to the lowest order term as shown in appendix~\ref{sec:tjmodelmapping}.
The values for the spin ladder
with the compound BPCB parameters can deviate strongly from this
symmetry. The velocity $u$ and the prefactor $B_x$ remain very close to the values for the spin chain mapping. In contrast, the
prefactors $A_z$, $A_x$ and the exponent $K$ deviate
considerably and $A_x$ and $K$ become strongly asymmetric. The origin of the asymmetry lies in the contribution of the higher triplet states~\cite{giamarchi_ladder_coupled}, and can be understood using a strong coupling expansion of the Hamiltonian~\eqref{equ:spinladderhamiltonian} up to the second order in $\gamma$ (see appendix~\ref{sec:secondorder_LL}). This asymmetry has
consequences for many experimentally relevant quantities and it
was found to cause for example strong asymmetries in the 3D
order parameter, its transition temperature and the NMR
relaxation rate (see Fig.~\ref{fig:orderparameter},
Fig.~\ref{fig:criticaltemperature} and Fig.~\ref{fig:T1}).

\subsection{Dynamical Luttinger liquid correlations}\label{sec:luttingerliquidcorr}

In this appendix we summarize previous results of the Luttinger liquid description of the dynamical correlation functions\cite{chitra_spinchains_field,furusaki_correlations_ladder,Sato_NMR_frustrated_ladder} at zero temperature. Note that the weak and strong coupling limits which have been studied separately \cite{chitra_spinchains_field,furusaki_correlations_ladder} can be connected \cite{giamarchi_ladder_coupled}. The expression of the correlations in the whole regime is given by \cite{\furusakibranches}
\begin{widetext}
\begin{multline}\label{equ:LLtimezzcorrelation}
S^{zz}_0(q,\omega)=(2\pi{m^z})^2\delta(q)\delta(\omega)
+\frac{\pi^2A_z}{u\Gamma(K)^2}
\left[\Theta(\omega-u|q-2\pi m^z|)\left(\frac{4u^2}{\omega^2-u^2(q-2\pi m^z)^2}\right)^{1-K}\right.\\
+\left.\Theta(\omega-u|q-2\pi(1- m^z)|)\left(\frac{4u^2}{\omega^2-u^2(q-2\pi(1- m^z))^2}\right)^{1-K}\right]+ \frac{K\omega}{u}\Theta(\omega)\left[\delta(\omega-uq)+\delta(\omega+uq)\right]
\end{multline}
\begin{multline}\label{equ:LLtimepmcorrelation}
S^{+-}_\pi(q,\omega)=\frac{8\pi^2A_x}{u\Gamma(1/4K)^2}\Theta(\omega-u|q-\pi|)\left(\frac{4u^2}{\omega^2-u^2(q-\pi)^2}\right)^{1-1/4K}\\
+\frac{4\pi^2B_x}{u\Gamma(\eta_+)\Gamma(\eta_-)}\left[\Theta(\omega-u|q-\pi(1-2m^z)|)\left(\frac{2u}{\omega-u[q-\pi(1-2m^z)]}\right)^{1-\eta_-}\left(\frac{2u}{\omega+u[q-\pi(1-2m^z)]}\right)^{1-\eta_+}\right.\\
\left.+\Theta(\omega-u|q-\pi(1+2m^z)|)\left(\frac{2u}{\omega-u[q-\pi(1+2m^z)]}\right)^{1-\eta_+}\left(\frac{2u}{\omega+u[q-\pi(1+2m^z)]}\right)^{1-\eta_-}\right]
\end{multline}
\end{widetext}
with the Heaviside function $\Theta(x)$ and $\eta_{\pm}=1/4K\pm1+K$. The correlation $S^{-+}_\pi$ is obtained replacing $m^z\rightarrow-m^z$ in the $S^{+-}_\pi$ expression Eq.~\eqref{equ:LLtimepmcorrelation}.

\begin{figure}
\begin{center}
\includegraphics[width=\linewidth]{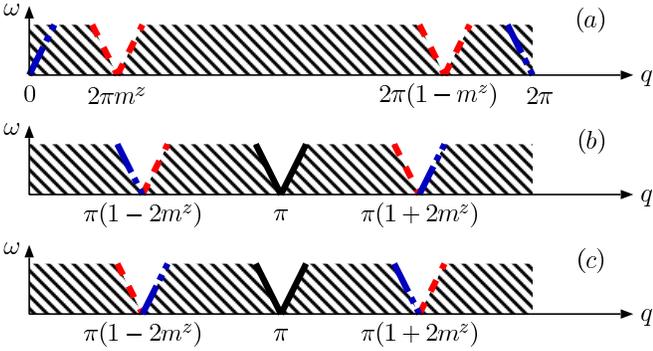}
\end{center}
\caption{\label{fig:LLcorrschema}(Color online): Map of the momentum-frequency low energy correlations of the LL model where the white areas represent the continuum of excitations. In the striped areas no excitations are possible. (a) $S^{zz}_0(q,\omega)$: the dash-dotted lines (blue) are the excitation peaks close to $q=0,2\pi$ and the dashed lines (red) are the continuum lower boundary with edge exponent $1-K$ close to $q=2\pi m^z,2\pi(1- m^z)$. (b) $S^{+-}_\pi(q,\omega)$, (c) $S^{-+}_\pi(q,\omega)$: the continuum lower boundary close to $q=\pi,\pi(1\pm 2m^z)$ is represented by solid lines (black) (edge exponent $1-1/4K$), dashed lines (red) (edge exponent $1-\eta_-=2-1/4K-K$) and dash-dotted lines (blue) (edge exponent $1-\eta_+=-1/4K-K$).}
\end{figure}

\begin{figure}
\begin{center}
\includegraphics[width=0.9\linewidth]{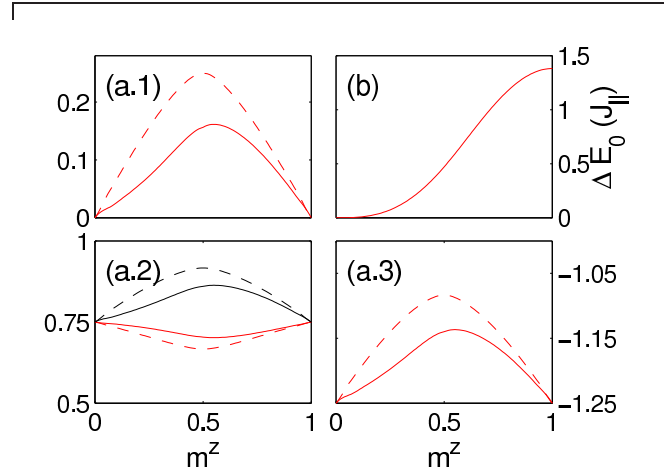}
\end{center}
\caption{(Color online): (a) Different exponents that appear in the LL correlation functions, Eqs.~\eqref{equ:LLtimezzcorrelation} and~\eqref{equ:LLtimepmcorrelation}, versus the magnetization $m^z$. The solid (dashed) lines are determined from the ladder (spin chain mapping) exponent $K$ in Fig.~\ref{fig:LLparameter}. The exponent $1-K$  of the $S^{zz}_0$ correlations is shown in (a.1), and the exponent  $1-1/4K$ of the $S^{\pm\mp}_\pi$ correlations at the $q=\pi$ branch in (a.2) (lower red curves). The exponents $1-\eta_-=2-1/4K-K$ (upper black curves) in (a.2) and $1-\eta_+=-1/4K-K$ in (a.3) correspond of both sides of the incommensurate branches of the $S^{\pm\mp}_\pi$ (see Fig.~\ref{fig:LLcorrschema}). (b) Shift of the ground state energy per rung versus the magnetization $\Delta E_0(m^z)=E_0(0)-E_0(m^z)$.\label{fig:decay_exp}}
\end{figure}

The expressions Eq.~\eqref{equ:LLtimezzcorrelation} and Eq.~\eqref{equ:LLtimepmcorrelation} exhibit the typical behavior of the frequency-momentum LL correlations: a continuum of low energy excitations exists with a linear dispersion with a slope given by the Luttinger velocity $\pm u$.
The spectral weight at the lower boundary of the continuum displays an algebraic singularity with the exponents related to the Luttinger parameter $K$. A summary of this behavior is sketched in Fig.~\ref{fig:LLcorrschema}.
For the considered system the longitudinal
correlation $S^{zz}_0$ is predicted to diverge with the
exponent $1-K$ at its lower edge. As shown in
Fig.~\ref{fig:decay_exp}.a.1 the exponent of this divergence is
very weak $<~0.2$ for the parameters of BCPB. The transverse
correlations $S^{\pm\mp}_\pi$ exhibit a distinct behavior
depending on the considered soft mode. Close to $q=\pi$ the
weight diverges with an exponent given by $1-1/4K$. This
divergence is strong for the considered parameters
($1-1/4K\approx3/4\gg0$ in Fig.~\ref{fig:decay_exp}.a.2). In
contrast at the soft mode  $q=\pi(1-2m^z),\pi(1+2m^z)$ a
divergence (cusp) is predicted at the lower edge with the
exponent $2-1/4K-K\approx 3/4$ in
Fig.~\ref{fig:decay_exp}.a.2 ($-1/4K-K\approx -5/4$ in
Fig.~\ref{fig:decay_exp}.a.3).

\subsection{Finite temperature transverse staggered Luttinger liquid correlation}\label{sec:finiteluttingerliquidcorr}

The determination of the relaxation time $T^{-1}_1$ (see Sec.~\ref{sec:relaxationtime}) and of the transition temperature $T_c$ to the 3D-ordered phase (see Sec.~\ref{sec:transitiontemperature}) requires the staggered transverse retarded correlation function
\begin{equation}\label{equ:xxacorrelationdef}
\chi_a^{xx}(x,t)=-i \Theta(t)(-1)^x\left\langle\left[\tilde S^x(x,t),\tilde S^x(0,0)\right]\right\rangle
\end{equation}
written in term of the spin chain mapping operators~\eqref{equ:spinchainmaping} with $\Theta(t)$ the Heaviside function. In the gapless regime, using the bosonization formalism~\eqref{equ:luttingeroperator1}, and taking into account only the most relevant term, we can compute it as described in Ref.~\onlinecite{giamarchi_book_1d} for the LL Hamiltonian~\eqref{equ:luttingerliquid}:
\begin{widetext}
\begin{equation}\label{equ:xxacorrelation}
\chi_a^{xx}(x,t)=-\Theta(t)\Theta(ut-x)\Theta(ut+x)\left(\frac{\pi}{\beta u}\right)^{\frac{1}{2K}}\frac{2A_x\sin\left(\frac{\pi }{4K}\right)}
{\left|\sinh\left(\frac{\pi}{\beta}\left(\frac{x}{u}+t\right)\right)\sinh\left(\frac{\pi}{\beta}\left(\frac{x}{u}-t\right)\right)\right|^{\frac{1}{4K}}}
\end{equation}
and its Fourrier transform:
\begin{equation}\label{equ:xxacorrelationqo}
\chi^{xx}_a(q,\omega)=-\frac{A_x\sin\left(\frac{\pi}{4K}\right)}{ u}\left(\frac{2\pi}{\beta u}\right)^{\frac{1}{2K}-2}
B\left(-i\frac{\beta(\omega-uq)}{4\pi}+\frac{1}{8K},1-\frac{1}{4K}\right)\\B\left(-i\frac{\beta(\omega+uq)}{4\pi}+\frac{1}{8K},1-\frac{1}{4K}\right)
\end{equation}
\end{widetext}
where $B(x,y)=\frac{\Gamma(x)\Gamma(y)}{\Gamma(x+y)}$.

\section{DMRG method}\label{sec:DMRG}

In this appendix, we describe first the time dependent DMRG method and its extension to finite temperature. Then we give the technical details related to the computation of the momentum-frequency correlations.

\subsection{Time dependent DMRG}\label{sec:timeDMRG}

The
t-DMRG\cite{Vidal_time_DMRG,daley_time_DMRG,white_time_DMRG,Schollwoeck_tDMRG}
(time dependent DMRG) method is based on the principle of the original DMRG to choose an effective reduced Hilbert space to describe the physics one is interested in. Instead of choosing only once the effective description for the
evolution of the state, the t-DMRG adapts its effective description at
each time-step. The implementation of this idea can be
performed using different time-evolution algorithms. Here we use the second
order Trotter-Suzuki expansion for the time-evolution operator
using a rung as a unit\cite{daley_time_DMRG,white_time_DMRG}. 
The errors arising in this method are the so-called truncation error and the Trotter-Suzuki error. Both are very well controlled (see Ref.~\onlinecite{Gobert_spindynamics} for a detailed discussion). 

\subsection{Finite temperature DMRG}\label{sec:temperatureDMRG}

The main idea of the finite temperature DMRG
\cite{Verstraete_finiteT_DMRG,Zwolak_finiteT_DMRG,White_finT}
(T-DMRG) is to represent the density matrix of the physical
state as a pure state in an artificially enlarged Hilbert
space. The auxiliary system is constructed by simply doubling
the physical system. Starting from the infinite temperature
limit the finite temperature is reached evolving down in
imaginary time. The infinite temperature state in this
auxiliary system corresponds to the totally mixed Bell state
$|\psi(0)\rangle=\frac{1}{{N_\lambda}^{L/2}}\prod_{l=1}^L\sum_{\lambda_l}|\lambda_l\lambda_l\rangle$,
where $|\lambda_l\lambda_l\rangle$ is the state at the bond $l$
of the auxiliary system that has the same value $\lambda_l$ on the
two sites of the bond (physical and its copy). The sum
$\sum_{\lambda_l}$ is done on all these $N_\lambda$ states
$|\lambda_l\rangle$. We evolve the physical part of
$|\psi(0)\rangle$ in imaginary time to obtain
\begin{equation}
|\psi(\beta)\rangle=e^{-\beta H/2}|\psi(0)\rangle
\end{equation}
using the t-DMRG algorithm presented in appendix~\ref{sec:timeDMRG}
with imaginary time. We
renormalize this state at each step of the imaginary time
evolution. Thus the expectation value of an operator $O$ acting in the physical system with respect to the normalized state $|\psi(\beta)\rangle$
is directly related to its
thermodynamic average, i.e.~
\begin{equation}
\langle O\rangle_\beta=\frac{\mathrm{Tr}[Oe^{-\beta H}]}{\mathrm{Tr}[e^{-\beta H}]}=\langle\psi(\beta)|O |\psi(\beta)\rangle
\end{equation}
at the temperature $T=1/\beta$.
We use this method to compute the average value of the local rung
magnetization $m^z(T)$ and energy per rung $E(T)$ in the center of
the system. Additionally we extract the
specific heat $c(T)$ by
\begin{equation}
c(\beta+\delta\beta/2) \approx -\frac{(\beta+\delta\beta/2)^2}{2\delta\beta} \left(\langle E\rangle_{\beta+\delta\beta}-\langle E\rangle_{\beta}\right)
\end{equation}
where $\delta\beta$ is the imaginary time-step used in the T-DMRG.

To reach very low temperatures $T\rightarrow 0$ for the specific heat,
we approximate the energy by its expansion in $T$
\begin{equation}\label{equ:lowTEexpansion}
E(T)\approx E_0 + \sum_{i=2}^{n}\alpha_iT^i
\end{equation}
up to $n=4$. The energy at zero temperature $E_0$ is
determined by a standard $T=0$ DMRG method. Since
$E(T)$ has a minimum at $T=0$ the linear term in the expansion~\eqref{equ:lowTEexpansion} does not exist. The numbers
$\alpha_i$ ($i=2,3,4$) are obtained fitting the expansion on the
low $T$ values of the numerically computed $E(T)$.

Typical system lengths used for the finite temperature
calculations are $L=80$ ($L=100$ for the spin chain mapping)
keeping a few hundred DMRG states and choosing a temperature
step of $\delta\beta=0.02~{\rm K}^{-1}$ ($\delta\beta=0.01~{\rm K}^{-1}$ for the spin chain mapping).

Let us note that recently a new method has been developed to treat finite temperatures \cite{White_TDMRG_thermal_state,Stoudenmire_TDMRG_thermal_state} which is very promising to reach even lower temperatures.

\subsection{Momentum-frequency correlation functions}\label{sec:tDMRGcorrelation}

To obtain the desired frequency-momentum spin-spin correlations
($S^{\alpha\beta}_{q_y}(q,\omega)$ in~\eqref{equ:correlation1}), we first compute the correlations in
space and time
\begin{equation}\label{equ:DMRGcorrelation}
S^{\alpha\beta}_{l,k}(t_n) =\langle 0 |e^{it_nH} S^\alpha_{l+L/2,k}e^{-it_nH}S^\beta_{L/2,1}|0\rangle
\end{equation}
with $l=-L/2+1,-L/2+2,\ldots,L/2$, $k=1,2$, and $t_n=n\delta t$ ($n=0,1,\ldots,N_t$) is the discrete time used. These
correlations are calculated by time-evolving the ground state
$\ket{\psi_0}=|0\rangle$ and the excited state
$\ket{\psi_1}=S^\beta_{L/2,1}|0\rangle$ using the t-DMRG (see
appendix~\ref{sec:timeDMRG}).

Afterwards the overlap of
$S^\alpha_{l+L/2,k}\ket{\psi_1(t)}$ and $\ket{\psi_0(t)}$ is
evaluated to obtain the correlation function~\eqref{equ:DMRGcorrelation}.

In an infinite system reflection symmetry would be fulfilled.
To minimize the finite system corrections and to recover the
reflection symmetry of the correlations, we average them
\begin{equation}
 \frac{1}{2}\left(S^{\alpha\beta}_{-l,k}(t_n)+S^{\alpha\beta}_{l,k}(t_n)\right)\rightarrow S^{\alpha\beta}_{l,k}(t_n).
\end{equation}
We then compute the symmetric (antisymmetric) correlations (upon leg exchange)
(see Sec.~\ref{par:zerocorrelations})
\begin{equation}
 S^{\alpha\beta}_{l,q_y}(t_n)=2(S^{\alpha\beta}_{l,1}(t_n)\pm S^{\alpha\beta}_{l,2}(t_n))
\end{equation}
with the rung momentum $q_y=0,\pi$ respectively. Finally, we perform a numerical Fourrier transform\footnote{In order to delete the numerical artefacts appearing in the zero frequency component of $S^{zz}_0(q,\omega)$ due to the boundary effects and the limitation in the numerical precision, we compute the Fourrier transform of $S^{zz}_{l,0}(t_n)$ only with its imaginary part that has no zero frequency component as proposed in Ref.~\onlinecite{pereira_spectrum_fermion1D}.}
\begin{equation}\label{equ:numFourrier}
S^{\alpha\beta}_{q_y}(q,\omega)\approx\delta t\sum_{n=-N_t+1}^{N_t}\sum_{l=-L/2+1}^{L/2}e^{i(\omega t_n -ql)}S^{\alpha\beta}_{l,q_y}(t_n)
\end{equation}
for discrete momenta $q=2\pi k/L$ ($k=0,1,\ldots,L-1$) and
frequencies $\omega$\footnote{The negative time correlations in the sum for $n<0$ are deduced from their value at positive time
since $S^{\alpha\beta}_{q_y}(q,-t_n)={ S^{\alpha\beta}_{q_y}}^\dagger(q,t_n)$, with $S^{\alpha\beta}_{q_y}(q,t_n)=\sum_le^{-iql}S^{\alpha\beta}_{l,q_y}(t_n)
$, for translation invariant systems, and for correlations such as ${S^\alpha}^\dagger=S^\beta$ with $\alpha,\beta=z,+,-$.}. The momentum $q$ has the reciprocal units of the interrung
spacing $a$ ($a=1$ is used if not mentioned otherwise). Due to the finite time step $\delta t$, our
computed $S^{\alpha\beta}_{q_y}(q,\omega)$ are limited to the
frequencies from $-\pi/\delta t$ to $\pi/\delta t$.
The finite calculation time $t_f=N_t\delta t$ induces artificial
oscillations of frequency $2\pi/t_f$ in
$S^{\alpha\beta}_{q_y}(q,\omega)$. To eliminate these artefacts
and reduce the effects of the finite system length, we apply
a filter to the time-space correlations before the numerical
Fourrier transform~\eqref{equ:numFourrier}, i.e.
\begin{equation}
S^{\alpha\beta}_{l,q_y}(t_n)f(l,t_n)\rightarrow
S^{\alpha\beta}_{l,q_y}(t_n).
\end{equation}
We tried different functional
forms for the filter $f(l,t_n)$ (cf.
Ref.~\onlinecite{white_time_DMRG} as well). In the following
the results are obtained by a Gaussian filter
$f(l,t_n)=e^{-\left(4l/L\right)^2-\left(2t_n/t_f\right)^2}$ if not stated
otherwise. As the effect of this filtering on the
momentum-energy correlations consists to convolve them by a
Gaussian function $f(q,\omega)=t_f
L/(32\pi)e^{-\left(\omega
t_f/4\right)^2-\left(qL/8\right)^2}$,
it minimizes the numerical artefacts but further reduces the momentum-frequency resolution.

Typical values we used in the simulations are system lengths of up to $L=160$
sites while keeping a few hundreds DMRG states. We limited
the final time $t_f$ to be smaller than the time necessary
for the excitations to reach the boundaries ($t_f\sim L/2u$ with
$u$ the LL velocity in Fig.~\ref{fig:LLparameter}) in order to
minimize the boundary effects. The computations for the BPCB parameters, Eq.~\eqref{equ:couplings}, were typically done with
a time step of $\delta t =0.0355 ~{J_\parallel}^{-1}$ up to $t_f=71~{J_\parallel}^{-1}$ (but calculating
the correlations only every second time step).
The
momentum-frequency limitations are then
$\delta\omega\approx0.11 ~J_\parallel$ and $\delta q\approx
0.1~a^{-1}$. Concerning the other couplings and the spin chain
calculations, we used a time step $\delta t =0.1 ~{J_\parallel}^{-1}$
up to $t_f=100 ~{J_\parallel}^{-1}$ (also with the correlation
evaluations every second time steps) for a momentum-frequency
precision $\delta\omega\approx0.08 ~J_\parallel$ and $\delta
q\approx 0.1~a^{-1}$.

Different techniques of extrapolation in time (using linear prediction or fitting the long time evolution with a guess asymptotic form cf.~Refs.~\onlinecite{white_spectral_heisenberg_spin_1,Barthel_spectrum_1D}) were recently used to improve the frequency resolution of the computed correlations. Nevertheless, as none of them can be apply systematically for our ladder system due to the presence of the high energy triplet excitations (which result in very high frequency oscillations), we decided not to use them.

\section{Quantum Monte Carlo determination of the 3D-ordering transition $T_c$}\label{sec:qmc}

The three dimensional network of couplings of the coupled ladder Hamiltonian Eq.~\ref{equ:coupledladdershamiltonian} and shown in Fig.~\ref{fig:structure} is not frustrated, and can
therefore be simulated using Quantum Monte Carlo algorithms.
We employ a stochastic series expansion implementation with directed loop updates~\cite{SSE_directed_loops} provided with the ALPS
libraries~\cite{ALPSSSEpaper,ALPSpaper}.
This numerical exact approach is complementary to mean-field approaches,
because the later tend not to be quantitatively accurate due to the neglect of quantum fluctuations in the interladder coupling.

In order to detect the transition temperature we measure the spin stiffness $\rho_s$ based on winding numbers in the three spatial directions. As pointed out in
Ref.~\onlinecite{Sandvik_tc_3D_heisenberg}, when plotting $L \rho_s$ for different system sizes $L$, the different curves cross at $T_c$. Alternatively we
measured the squared order parameter $m_x^2$. The quantity $L^{1+\eta} m_x^2$ (assuming the 3D XY universality class value of $\eta \approx 0.04$) also crosses
at $T_c$ when plotted for different $L$.

Previously the 3D
ordering temperatures of coupled spin ladders in a magnetic field have been determined using a specific feature of the magnetization $m(T)$~\cite{wessel01_spinliquid_bec}.
It turns out, that while the feature in $m(T)$ indeed locates $T_c$ accurately for simple coupled dimer systems~\cite{Nohadani_Tc_dimers}, the same does not hold for coupled
ladders. In the ladder case one has to resort to the spin stiffness crossing or order parameter measurements to locate $T_c$ accurately.

When simulating the coupled ladder Hamiltonian, a suitable finite size sample setup is required. Due to the spatial anisotropies in the problem -
the ladder direction being singled out over the two transverse spatial directions - an appropriate aspect ratio needs to be kept~\cite{yasuda_neelorder}.
In our simulations we chose an aspect ratio of about 12, i.e. the samples were 12 times longer along the ladder direction than in the transverse directions.

\begin{figure}
\begin{center}
\includegraphics*[width=0.95\linewidth]{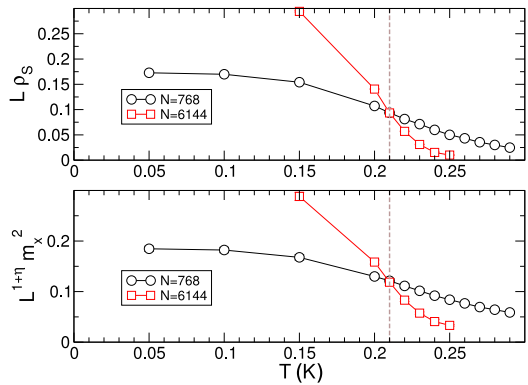}
\end{center}
\caption{(Color online): Quantum Monte Carlo simulation results for the spin stiffness scaling (upper panel), and the order parameter scaling (lower panel).
The vertical dashed line denotes the coinciding estimate of $T_c$ from the crossing curves in both panels.
\label{fig:qmc_tc}}
\end{figure}

In Fig.~\ref{fig:qmc_tc} we show simulation results for one particular set of couplings: the rung and leg couplings were set to $J_\perp =12.9~{\rm K}$ and $J_\parallel=3.3~{\rm K}$ respectively,
the $g$-factor was $2.17$, the magnetic field amounted to $8.9~{\rm T}$ and the interladder coupling $J'$ was set to $80~{\rm mK}$.  These couplings are identical to those
used in Ref.~\onlinecite{Thielemann_ND_3Dladder}. In the upper panel we show the rescaled spin stiffness $L\rho_s$ for two different system sizes (768 versus 6144 spins).
One locates a crossing at $210~{\rm mK}$ for this observable. In the lower panel we show the rescaled squared order parameter $L^{1+\eta} m_x^2$, which also exhibits a crossing at
the same temperature. These matching estimates for the critical temperature render us confident that we accurately locate the critical temperature.

Based on this and subsequent simulations either with an identical $J'$ but a higher magnetic field of $11.9~{\rm T}$, or the same field and a smaller $J'=50~{\rm mK}$, we are able to
determine and verify the use of single rescaling factor $\alpha=0.74(1)$ relating the real and effective mean-field interladder coupling~\cite{yasuda_neelorder}, as presented
in Sec.~\ref{sec:transitiontemperature}.

\end{document}